%% file: Current_current_deformation_notes.tex
\documentclass[11pt,a4paper,twoside]{article}
\usepackage{graphicx} 

\usepackage[headheight=25mm,top=30mm,inner=35mm,outer=35mm,marginparwidth=50pt,centering]{geometry}

\usepackage{amsmath,amsfonts,mathrsfs,graphicx,bbm}
\usepackage{fixltx2e}
\usepackage{cancel}
\usepackage{slashed}
\usepackage{braket}

\usepackage{tabularray}

\usepackage{comment}
\usepackage{color}
\usepackage[colorlinks=true]{hyperref}
\hypersetup{
    colorlinks,
    linkcolor={black},
    citecolor={black},
    urlcolor={black}
}

\numberwithin{equation}{section} 

\usepackage{tikz} 
\usetikzlibrary{calc}
\usetikzlibrary{arrows}
\usetikzlibrary{trees}
\usetikzlibrary{decorations.pathmorphing}
\usetikzlibrary{decorations.markings}

\newcommand{\Ad}{\mathrm{Ad}}
\newcommand{\ad}{\mathrm{ad}}

\newcommand{\Trh}{\mathrm{\hat{T}r}}

\newcommand{\figI}{\begin{tikzpicture}[scale=0.6, every node/.style={scale=0.8}]
\node (a) at (0,0) {$z_1$};
\node (b) at (2,0) {$z_2$};
\node (c) at (1,-1) {$z_1$};
\node (d) at (3,-1) {$z$};
\node (e) at (2,-2) {$z$};
\node (f) at (4,-2) {$w$};
\node (g) at (3,-3) {$\mathbf 1$};
    \draw   (a) -- (c) -- (b);
    \draw   (c) -- (e) -- (d);
    \draw   (e) -- (g) -- (f);
\end{tikzpicture}}

\newcommand{\figII}{\begin{tikzpicture}[scale=0.6, every node/.style={scale=0.8}]
\node (a) at (0,0) {$z_1$};
\node (b) at (2,0) {$z_2$};
\node (c) at (1,-1) {$z_1$};
\node (d) at (3,-1) {$w$};
\node (e) at (2,-2) {$w$};
\node (f) at (4,-2) {$z$};
\node (g) at (3,-3) {$\mathbf 1$};
    \draw   (a) -- (c) -- (b);
    \draw   (c) -- (e) -- (d);
    \draw   (e) -- (g) -- (f);
\end{tikzpicture}}

\newcommand{\figIII}{\begin{tikzpicture}[scale=0.6, every node/.style={scale=0.8}]
\node (a) at (0,0) {$w$};
\node (b) at (2,0) {$z_2$};
\node (c) at (1,-1) {$w$};
\node (d) at (3,-1) {$z_1$};
\node (e) at (2,-2) {$w$};
\node (f) at (4,-2) {$z$};
\node (g) at (3,-3) {$\mathbf 1$};
    \draw   (a) -- (c) -- (b);
    \draw   (c) -- (e) -- (d);
    \draw   (e) -- (g) -- (f);
\end{tikzpicture}}

\newcommand{\figIV}{\begin{tikzpicture}[scale=0.6, every node/.style={scale=0.8}]
\node (a) at (0,0) {$z$};
\node (b) at (2,0) {$z_2$};
\node (c) at (1,-1) {$z$};
\node (d) at (3,-1) {$z_1$};
\node (e) at (2,-2) {$z$};
\node (f) at (4,-2) {$w$};
\node (g) at (3,-3) {$\mathbf 1$};
    \draw   (a) -- (c) -- (b);
    \draw   (c) -- (e) -- (d);
    \draw   (e) -- (g) -- (f);
\end{tikzpicture}}

\newcommand{\figV}{\begin{tikzpicture}[scale=0.6, every node/.style={scale=0.8}]
\node (a) at (0,0) {$w$};
\node (b) at (2,0) {$z_2$};
\node (c) at (1,-1) {$w$};
\node (d) at (3,0) {$z$};
\node (e) at (5,0) {$z_1$};
\node (f) at (4,-1) {$z$};
\node (g) at (2.5,-2) {$\mathbf 1$};
    \draw   (a) -- (c) -- (b);
    \draw   (d) -- (f) -- (e);
    \draw   (c) -- (g) -- (f);
\end{tikzpicture}}

\newcommand{\figVI}{\begin{tikzpicture}[scale=0.6, every node/.style={scale=0.8}]
\node (a) at (0,0) {$z$};
\node (b) at (2,0) {$z_2$};
\node (c) at (1,-1) {$z$};
\node (d) at (3,0) {$w$};
\node (e) at (5,0) {$z_1$};
\node (f) at (4,-1) {$w$};
\node (g) at (2.5,-2) {$\mathbf 1$};
    \draw   (a) -- (c) -- (b);
    \draw   (d) -- (f) -- (e);
    \draw   (c) -- (g) -- (f);
\end{tikzpicture}}

\begin{document}


\vspace{24pt}

\begin{center}
{\huge{\bf Lecture notes on\\
\vspace{8pt}
current-current 
deformations}} 

\vspace{24pt}

Riccardo Borsato 

\vspace{15pt}

{
\small {\it 
Instituto Galego de F\'isica de Altas Enerx\'ias (IGFAE),\\ Universidade de  Santiago de Compostela, Spain}\\
\vspace{12pt}
\texttt{riccardo.borsato@usc.es}}\\

\vspace{24pt}

{\bf Abstract}
\end{center}
\noindent
These are pedagogical lecture notes discussing current-current deformations of 2-dimensional field theories. The deformations that are considered here are generated infinitesimally by bilinears of Noether currents corresponding to internal global symmetries of the ``seed'' theory. When the seed theory is conformal, these deformations are marginal and are often known as $J\bar J$-deformations. In this context,  we review the criterion for  marginal operators due to Chaudhuri and Schwartz. When the seed theory is an integrable $\sigma$-model (in the sense that it possesses a Lax connection), these deformations preserve the integrability. Here we review this fact by viewing the deformations as maps that leave  the equations of motion and the Poisson brackets of the 2-dimensional $\sigma$-models invariant. The reinterpretation as undeformed theories with twisted boundary conditions is also discussed, as well as the effect of the deformation at the level of the S-matrix of the quantum theory. The finite (or integrated) form of the deformations is equivalent to sequences of T-duality--shift--T-duality transformations (TsT's), and here we review the $O(d,d)$-covariant formalism that is useful to describe them.

The presentation starts with pedagogical examples of deformations of free massless scalars in 2 dimensions, and minimal prerequisites on conformal field theories or integrability are needed to understand later sections. Moreover, guided exercises are proposed to the reader.
These notes were prepared for the Young Researchers Integrability School and Workshop (YRISW) held in Durham from 17 to 21 July 2023.

\newpage 


\tableofcontents


\section{Introduction}\label{sec:prel}

In physics we are often confronted with problems that are difficult to describe and to  solve. A strategy that we often employ is to first understand simpler setups, where explicit and exact calculations can be carried out, and only later add the complications that we find hard to treat. A typical example is the way we approach the subject of quantum field theory in our undergraduate studies: we first learn about theories of free fields, and only later we add interactions. Because of the difficulty in working with them, actually, we typically deal with interactions only order by order around the free theory in a perturbative fashion.

In these notes we will discuss continuous deformations of 2-dimensional field theories. The construction of continuous deformations may be seen as a strategy to add ``complicated features'' to a theory that is assumed to be well understood. The deformations that we consider are characterised by the fact that they are  controlled by the \emph{symmetries} of the original theory (which we may call the ``seed'' theory). In particular, given the symmetries (in particular the \emph{global} symmetries) of a theory that we want to deform, we may compute the corresponding conserved currents. The deformations that we will consider may be understood as perturbations 
 of the original  action functional by current bilinears (i.e.~products of two currents), and for this reason they are normally called ``current-current deformations''. On the one hand, as already mentioned, the symmetries of the seed theory are the guiding principle to define and construct these deformations. On the other hand, the deformations may generally break the symmetries of the seed theory (like a squashed sphere that is no longer invariant under generic rotations). There is, therefore, an interesting interplay between the deformations and the symmetries.

We will discuss the deformations in relation to two aspects of central importance in the study of 2-dimensional field theories: the \emph{conformal invariance} and the \emph{integrability}. This does not mean that these two aspects are essential in order to be able to construct the deformations. In fact, these current-current deformations can be applied to any 2-dimensional field theory, even those that are neither conformal nor integrable. The only necessary feature of the original theory  is that it must have some symmetries that we can use to define and construct the deformation. Nevertheless, these deformations turn out to have an interesting interplay with the conformal invariance and the integrability of the theories, in the case when the seed theory is conformal or integrable. Actually, both the conformal invariance and the integrability may be seen as ``guiding principles'' to come up with the construction of these deformations.

\vspace{12pt}

Conformal field theories (CFTs) are invariant not only under the Poincar\'e group, but under an extension of this group called the conformal group, see for example~\cite{DiFrancesco:1997nk,Ribault:2016sla} for useful introductions. In two dimensions something special happens, because the conformal group becomes infinite dimensional. CFTs are an important subject firstly because they are fixed points of the renormalisation-group flow: because of the scale symmetry, the parameters in the action of a CFT do not depend on the scale. Moreover, the conformal invariance is very restricting, and for example  it severely fixes the allowed forms of the correlation functions of CFTs, therefore helping a great deal to solve the theory.
When deforming CFTs, we may consider deformations of three types:  relevant, irrelevant or marginal. Relevant deformations modify the Lagrangian by operators that are important in the infrared: they change the nature of the original theory, which flows to a new one. Irrelevant deformations, on the other hand, are not important in the infrared, but may be  understood as an attempt to ``climb up'' the renormalisation-group flow and find a modification of the theory in the ultraviolet. Both relevant and irrelevant deformations break the conformal invariance. Marginal deformations, instead, by definition do not break conformal symmetry. If the deformation is controlled by a continuous parameter, then this parameter does not depend on the scale (it is the same in the infrared and in the ultraviolet) and the deformed theory is still conformal. The case of marginal deformations is, of course, the one that is more under control, because even for the deformed theory we still have at our disposal the methods that we use to solve CFTs. Constructing marginal deformations of CFTs may be interesting for various reasons, for example to understand how large is the family of theories that enjoy conformal invariance.
As anticipated, in these notes we will consider marginal deformations whose infinitesimal version is generated by current bilinears. We will, however, also discuss when the infinitesimal marginal deformation can be promoted to a \emph{finite} marginal deformation. These deformations will be called ``exactly'' or ``integrably'' marginal. Here ``integrably'' is meant in the sense of the deformation being ``integrated'', and it shouldn't be confused with the different notion of ``integrability'' of 2-dimensional models.

\vspace{12pt}

Typically, we refer to 2-dimensional models as being \emph{integrable} when they possess a number of conserved quantities that is often enough to reach an exact solution of the theory, at the classical or quantum level. Techniques for integrable models range from the inverse scattering method, to exact S-matrices, Bethe ansatz methods, etc. Integrable models may be seen as non-trivial interacting theories that admit an exact formulation, and they turn out to have interesting applications from condensed matter physics to string theory. Integrability is a very large subject and many books may be consulted. For lecture notes on integrability see for example~\cite{Arutyunov:2021ygo,Torrielli:2022byn}, and~\cite{Bombardelli:2016rwb,Beisert:2010jr} for reviews with an emphasis on integrability in the context of the AdS/CFT correspondence. Although the concept of integrability has various incarnations (e.g.~integrable spin-chains), in these lecture notes we will only discuss integrable $\sigma$-models. The current-current deformations that we will consider here have the crucial feature of preserving the integrability of the original theory, and that means that even after the deformation we can still apply the powerful methods of integrability. 
In particular, given a $\sigma$-model that is integrable at the classical level, and for which we know the corresponding Lax connection, we can  construct also the Lax connection of the deformed model.
Constructing integrable deformations is interesting because it helps us understand how large is the family of integrable models. Moreover, in the presence of the deformation, we typically generate theories that are much more complicated than the original ones, and that in particular break (at least some of) the original symmetries. This scenario, therefore, offers the intriguing possibility of applying the integrability techniques in cases where other methods might fail.

\vspace{12pt}

The current-current deformations that we will construct in the following sections are related to 2-dimensional duality transformations that go under the name of \emph{T-dualities}. Because of this relation, in these notes we will review the concept of T-duality, which we will view as a canonical transformation, and we will use a ``double'' language that makes the duality transformations covariant. The deformations will be viewed as continuous transformations that are special elements of the $O(d,d)$ group.  This so-called duality group has  played the fundamental role in the formulation of Double Field Theory, see~\cite{Zwiebach:2011rg,Aldazabal:2013sca,Berman:2013eva,Hohm:2013bwa} for some reviews, which is a powerful language to recast the low energy limit of string theory.

In these notes we will focus only on certain aspects of the deformations, including when it comes to their relation to conformal symmetry, integrability and duality-covariant formulations.  In section~\ref{sec:concl} we give a list topics with references that the reader may want to consult to identify more connections and to get more details on the subjects.

\vspace{12pt}

The plan of the notes is the following. In section~\ref{sec:free} we give a pedagogical introduction to the current-current deformations  by deforming   theories of free scalar fields. We also give a basic review on T-duality to present the reinterpretation of the deformations as so-called TsT deformations. In section~\ref{sec:ChSch} we discuss the relation of the deformations to conformal invariance, when the seed theory is a CFT. In particular, we discuss the condition of integrable marginality for operators constructed as bilinears of chiral and antichiral currents of the CFT. These operators can be used to generate the infinitesimal deformations of the CFT. In section~\ref{sec:ODD} we consider general $\sigma$-models with commuting isometries, and we setup the ``double'' language that we use to describe canonical transformations of the field theory. According to our definition, these canonical transformations are maps that leave the Poisson brackets invariant, but may change the boundary conditions. In section~\ref{sec:finite}, then, we use this language to prove that the deformations  are indeed a special class of these canonical transformations. We also analyse the special case with chiral isometries that is relevant for deformations of CFTs. In section~\ref{sec:int} we discuss the relation to integrability and the reinterpretation of the deformed models as undeformed ones with twisted boundary conditions. We also give a discussion on the effect of the deformation at the level of the S-matrix of the quantum theory, and we end with the explicit example of a deformation of a Principal Chiral Model. In section~\ref{sec:concl} we provide a list of topics and references that have connections with the  previous sections, and that may be seen as motivations to go beyond the content of these notes.

\subsection{Preliminaries}
Let us start by setting some conventions and by making some generic considerations. We will consider 2-dimensional $\sigma$-models, and we will denote by $(\tau,\sigma)$ the coordinates of the 2-dimensional spacetime, which we will also call the ``worldsheet'' as in string theory. We will use $\alpha,\beta,\ldots$ to denote worldsheet indices. Depending on what is more useful for us, we may work with a Lorentzian or a Euclidean signature.

In the Lorentzian case, for the Minkowski metric we take $\eta=\text{diag}(-1,1)$. We will often prefer to work with light-cone coordinates \begin{equation}
    \sigma^\pm=\frac{\sigma\pm \tau}{\sqrt 2},
\end{equation} 
the inverse relations being $\sigma=(\sigma^++\sigma^-)/\sqrt 2$ and $\tau=(\sigma^+-\sigma^-)/\sqrt 2$. Notice that our definition of light-cone coordinates imply that the derivatives are given by $\partial_\pm=(\partial_\sigma\pm \partial_\tau)/\sqrt 2$ (where we denote $\partial_\pm=\frac{\partial}{\partial \sigma^\pm}$), whose inverse relations are $\partial_\sigma=(\partial_++\partial_-)/\sqrt 2, \partial_\tau=(\partial_+-\partial_-)/\sqrt 2$. In the light-cone coordinates the non-vanishing components of the metric and its inverse are just $\eta_{+-}=\eta_{-+}=\eta^{+-}=\eta^{-+}=1$. Later we will also use the antisymmetric tensor $\epsilon^{\alpha\beta}$ defined by $\epsilon^{\tau\sigma}=-1$. Then in the light-cone coordinates the non-vanishing components are $\epsilon^{+-}=-1,\epsilon^{-+}=1$. 
We will also use the projectors \begin{equation}\label{eq:Ppm}
    \Pi_{(\pm)}^{\alpha\beta}=\tfrac{1}{2}(\eta^{\alpha\beta}\pm\epsilon^{\alpha\beta}),
\end{equation}
so that in the light-cone coordinates the only non-vanishing components of these projectors are $\Pi^{-+}_{(+)}=1$ and $\Pi^{+-}_{(-)}=1$, respectively.
Taking into account that we lower and raise worldsheet indices with the metric $\eta_{\alpha\beta}$ and its inverse, we define the epsilon tensor with lower indices as $\epsilon_{\alpha\beta}=\eta_{\alpha\gamma}\eta_{\beta\delta}\epsilon^{\gamma\delta}$ so that $\epsilon_{+-}=1,\epsilon_{-+}=-1$.

When going to Euclidean signature, we may take the analytic continuation $\tau=iy, \sigma=x$ so that $x,y$ are the new Euclidean coordinates on the worldsheet. It is actually preferable to define the complex coordinates
\begin{equation}
    z=\frac{x+iy}{\sqrt 2}, \qquad \qquad \bar z=\frac{x-i y}{\sqrt 2},
\end{equation}
where the bar means complex conjugation.
Notice that the inverse relations are $x=(z+\bar z)/\sqrt 2, y=(z-\bar z)/(i\sqrt 2)$. Moreover, the relations between the derivatives are $\partial_x=(\partial+\bar \partial)/\sqrt 2, \partial_y=i/\sqrt 2(\partial-\bar\partial)$ (where we use the notation $\partial=\frac{\partial}{\partial z},\bar \partial=\frac{\partial}{\partial \bar z}$) and $\partial=(\partial_x-i\partial_y)/\sqrt 2,\bar\partial=(\partial_x+i\partial_y)/\sqrt 2$. The metric on the worldsheet is now proportional to the Kronecker delta, in the $x,y$ coordinates. In the complex coordinates $z,\bar z$  we have $\eta_{z\bar z}=\eta_{\bar z z}=\eta^{z\bar z}=\eta^{\bar z z}=1$. The Euclidean signature and the complex coordinates $z,\bar z$ are normally used when discussing 2-dimensional CFTs. However, the signature on the worldsheet is not very important for most of our considerations. Normally, we will prefer to use the ``Lorentzian'' notation with $\pm$ indices. As it is clear from the above formulae, going to the Euclidean notation only amounts to $\sigma^+\to z, \sigma^-\to \bar z$.

\vspace{12pt}

Conserved currents are important ingredients in our discussion. Given a current $J_\alpha$, its conservation reads
\begin{equation}
    \begin{aligned}
        0&=\partial_\alpha J^\alpha = \partial_+J^++\partial_-J^-=\partial_+J_-+\partial_-J_+\\
        0&=\partial_\alpha J^\alpha = \partial_zJ^z+\partial_{\bar z}J^{\bar z}=\partial  J_{\bar z}+\bar \partial J_z,
    \end{aligned}
\end{equation} 
respectively in the two conventions. From now on, we will only use the Lorentzian notation, the translation to the Euclidean one being trivial as explained above.
Let us now define the Hodge dual of the current as
\begin{equation}
    \tilde J^\alpha=\epsilon^{\alpha\beta}J_\beta.
\end{equation}
In components we may therefore write the following relations
\begin{equation}
    \tilde J_\pm=\pm J_\pm.
\end{equation}
\emph{If} not only $J_\alpha$ is conserved, but also $\tilde J_\alpha$ is conserved, then we have
\begin{equation}
    0=\partial_\alpha\tilde J^\alpha=-\partial_+J_-+\partial_-J_+.
\end{equation}
Together with the standard conservation equation for $J_\alpha$, then, we would have
\begin{equation}\label{eq:chir-comp}
    \partial_+J_-=0,\qquad\qquad \partial_-J_+=0,
\end{equation}
separately. Notice that the $-$ component does not depend on $\sigma^+$, and the $+$ component does not depend on $\sigma^-$. Under this assumption of conservation of both the current and its Hodge dual, let us define the ``left'' and ``right'' currents
\begin{equation}
    J^L_\alpha=\frac{J_\alpha+\tilde J_\alpha}{2},\qquad \qquad 
    J^R_\alpha=\frac{J_\alpha-\tilde J_\alpha}{2}.
\end{equation}
It is easy to see that, because of the relation between $J$ and $\tilde J$, their components are
\begin{equation}
    J^L_+=J_+\neq 0,\qquad J^L_-=0,\qquad
    J^R_+=0,\qquad J^R_-=J_-\neq 0.
\end{equation}
The above relations are valid even when $\tilde J$ is not conserved, but under the assumption that it is, then the left and right currents are also conserved currents. Moreover, because of~\eqref{eq:chir-comp}, we find that the left current only depends on $\sigma^+$ while the right current only depends on $\sigma^-$
\begin{equation}\label{eq:non0comp}
    J^L_+(\sigma^+),\qquad\qquad
    J^R_-(\sigma^-).
\end{equation}
They are then chiral and antichiral currents, respectively. In fact, $J^L$ only depends on chiral or left modes (at fixed $\sigma^+$, when $\tau$ increases then $\sigma$ must decrease and the mode moves to the left), while $J^R$ only depends on antichiral or right modes (at fixed $\sigma^-$, when $\tau$ increases then $\sigma$ must increase and the mode moves to the right). Notice that we may have started from the notion of left and right conserved currents, and then constructed a current and its Hodge dual, that would then be both conserved.

\vspace{12pt}

As anticipated, we will be interested in deforming actions of 2-dimensional $\sigma$-models by perturbations that are defined by current bilinears. Given that we want this combination to be a scalar (in order to be able to construct a deformation of the action), if for example  we have two different conserved currents $J^1_\alpha, J^2_\alpha$, then we have two options to construct the infinitesimal deformations, namely
\begin{equation}
    \eta^{\alpha\beta}J^1_\alpha J^2_
    \beta, \qquad\text{or}\qquad
\epsilon^{\alpha\beta}J^1_\alpha J^2_
    \beta.
\end{equation}
\emph{If} we are dealing with chiral and antichiral currents, then the two options are actually equivalent. In fact, going to the basis of left and right currents for convenience, it is easy to check that
\begin{equation}
    \eta^{\alpha\beta}J^{Li}_\alpha J^{Rj}_\beta=J^{Li}_+ J^{Rj}_-,\qquad
    \epsilon^{\alpha\beta}J^{Li}_\alpha J^{Rj}_\beta=-J^{Li}_+ J^{Rj}_-,
\end{equation}
where $i,j=1,2$.
We see that the two expressions are proportional to each other. Importantly, we must couple a left current with a right one, otherwise the bilinear would be trivially zero because of the non-vanishing components given in~\eqref{eq:non0comp}. In these notes we will construct deformations generated by current bilinears defined with the \emph{epsilon tensor}. In the presence of chiral and antichiral currents, as it is the case for CFTs, this does not imply a restriction of the scope of the construction.

\vspace{12pt}

In these lecture notes we will often employ a terminology that makes the target-space interpretation of the $\sigma$-models manifest. In particular, later we will be interested in the 2-dimensional action $S=-\int d^2\sigma \, \Pi_{(-)}^{\alpha\beta}\partial_\alpha x^m\partial_\beta x^nE_{mn}(x)$, where $x^m(\tau,\sigma)$ with $m=1,\ldots,D$ are scalar fields in two dimensions. These fields $x^m(\tau,\sigma)$ can be seen as maps from the 2-dimensional worldsheet to a $D$-dimensional target space. The metric of the target space is given by the symmetric combination $G_{mn}=\tfrac12(E_{mn}+E_{nm})$ of the generalised coupling $E_{mn}(x)$ appearing in the action $S$. As an example, consider the case in which the worldsheet coordinate $\sigma$ takes values in a circle. Then the time-slice at fixed $\tau$ of $x^m(\tau,\sigma)$ gives a snapshot of a closed string in target space. Moreover, the time evolution dictated by the $\tau$-dependence will describe the motion of the string in the target space. The antisymmetric combination $B_{mn}=\tfrac12(E_{mn}-E_{nm})$ has the target-space interpretation of a Kalb-Ramond field, which we will normally call just ``$B$-field'' following the usual terminology. To complete the target-space picture of the bosonic string, one may add also a scalar field called ``dilaton'', that will come back in section~\ref{sec:dilaton}. We will not go deeper into the target-space interpretation of the couplings that appear in the $\sigma$-model actions, because it will not be needed for these lecture notes. For an introduction to string $\sigma$-models and their target-space interpretation, we refer for example to the article~\cite{Callan:1985ia}, the lecture notes~\cite{Callan:1989nz} or the book~\cite{Zwiebach:2004tj}.

\section{Deformations and dualities of  free bosons}\label{sec:free}

Before going to the  case of current-current deformations of generic $\sigma$-models, we start with a simple discussion involving just free bosons. This will provide nice toy models to become familiar with the key ideas that we will generalise in the rest of the notes. 

\subsection{One boson}
Let us start with just one free massless scalar field $\phi$ in 2 dimensions, whose action we write as
\begin{equation}\label{eq:S1phi}
    S_0=-\frac{g}{2}\int d^2\sigma\, \partial_\alpha\phi\partial^\alpha \phi.
\end{equation}
We will not need to specify the topology of the 2-dimensional worldsheet. It may be the 2-dimensional Lorentzian plane, or it may have the topology of a cylinder (as in the case of closed strings). 
In the  case of the cylinder, the spatial coordinate is compactified on a circle of length $L$ as $\sigma\sim \sigma+L$, and to avoid multi-valuedness of $\phi$ we have to take it to be a compact boson. In other words,  we may think of $\phi$ as an angle: it takes values in the interval $[0,2\pi[$, and we identify these values with those outside the interval as $\phi\sim \phi+2\pi $. Strictly speaking, it is only when $\phi$ is compact that  the parameter $g$ in front of the action is physical. In fact, we may effectively set $g$ to 1 by redefining the scalar field as $\phi'=\sqrt g\phi$. However,  in the case of the compact boson, the rescaled scalar would satisfy a different periodicity condition, namely $\phi'\sim\phi'+2\pi R$ with $R=\sqrt g$, and  rather that an angle it would be a coordinate on a circle of radius $R$. For us the above action is just a toy model, and we will pretend that $g$ is a physical parameter also in the non-compact case. Later we will consider more complicated $\sigma$-models with various fields, and $g$ will be promoted to a generically field-dependent coupling. 

\vspace{12pt}

We now want to consider deformations of the above action. The simplest possible deformation consists of just modifying the parameter $g$. Guided by the symmetries of the model, can we construct a current-current deformation that results in $g+\delta g$? The answer is yes, as we will now see.

The first observation is that the action is invariant a $U(1)$ transformation implemented by shifting the scalar field, i.e.~$\phi\to \phi+c,\ c\in \mathbb R$. By Noether's theorem, we therefore know that there exists a current $J_\alpha$ which is conserved, $\partial_\alpha J^\alpha=0$. We write the current as
\begin{equation}
J_\alpha = g\, \partial_\alpha \phi,   
\end{equation}
where we decide to keep an explicit prefactor of $g$ because it comes directly from the action when writing the Noether's current.\footnote{As it is well known, one may use the trick of promoting the infinitesimal transformation to a local one to identify the Noether current $J$. Our convention is such that  the normalisation is fixed by taking $\delta S=-\int d^2\sigma\, \partial_\alpha(\delta\phi)J^\alpha$. } The conservation of the current is, as it should, ensured by the equations of motion $\partial_\alpha\partial^\alpha\phi=0$.

To construct a current-current deformation with two currents coupled by $\epsilon^{\alpha\beta}$, we need a second conserved current. The second current is identified by  observing that in this case the Hodge dual of the above Noether current is also conserved. In fact, if we define
\begin{equation}
    \tilde J^{\alpha}=\epsilon^{\alpha\beta}J_{\beta}=g\, \epsilon^{\alpha\beta}\partial_\beta\phi,
\end{equation}
it is obvious that $\tilde J_{\alpha}$ is conserved. It is conserved \emph{off-shell}, i.e.~without the need of imposing the equations of motion, because $\partial_\alpha\tilde J^\alpha=g\, \epsilon^{\alpha\beta}\partial_\alpha\partial_\beta\phi=0$ is just a consequence of the antisymmetry of $\epsilon^{\alpha\beta}$ and the fact that the partial derivatives commute. Therefore $\tilde J_\alpha$ is a \emph{topological} current. The charge that it measures gives us the difference of the boundary values of $\phi$ at the extremes $r_\pm$
\begin{equation}
    \tilde Q=\int_{r_-}^{r^+}d\sigma\, \tilde J_0=g\int_{r_-}^{r^+}d\sigma\, \partial_\sigma\phi=g(\phi(r_+)-\phi(r_-)).
\end{equation}
In the compact case, $\tilde Q$ is related to the winding number $n$ as
$\tilde Q=2\pi g\, n$, that is
 defined as the number of times the compact boson winds around when going from $x=0$ to $x=L$ (i.e.~$\phi(L)=\phi(0)+2\pi n$).

Being $\tilde J_\alpha$ conserved, from the  discussion of Section~\ref{sec:prel} we therefore know that we can define chiral and antichiral currents
\begin{equation}
    J^L_\alpha=\frac{J_\alpha+\tilde J_\alpha}{2},\qquad\qquad
    J^R_\alpha=\frac{J_\alpha-\tilde J_\alpha}{2},
\end{equation}
whose only non-vanishing components are \begin{equation}
    J^L_+(\sigma^+)=g\, \partial_+\phi,\qquad\qquad
    J^R_-(\sigma^-)=g\, \partial_-\phi.
\end{equation} 
In these formulas we indicated explicitly the dependence on the 2-dimensional coordinates $\sigma^\pm$. Once again,  the existence of chiral and antichiral currents is obvious from   the equations of motion that read as $\partial_\alpha\partial^\alpha\phi=2\partial_+\partial_-\phi=0$. 

\vspace{12pt}

Let us now construct a current-current deformation of the above action using $J_\alpha,\tilde J_\alpha$ (or equivalently $J^L,J^R$). We write 
\begin{equation}
    S_0\to S_0+\delta S_0,\qquad\text{where}\qquad \delta S_0=-\frac{\lambda}{2} \int d^2\sigma\, \epsilon^{\alpha\beta}J_\alpha\tilde J_\beta.
\end{equation}
Here $\lambda$ is a constant deformation parameter that, for the moment, we assume to be very small, $\lambda\ll 1$. Using the identity $\epsilon^{\alpha\beta}\eta_{\beta\gamma}\epsilon^{\gamma\delta}=\eta^{\alpha\delta}$ we find
\begin{equation}
    \delta S_0=-\frac{\lambda g^2}{2} \int d^2\sigma\, \partial_\alpha\phi\partial^\alpha \phi.
\end{equation}
In other words, we managed to construct a deformation of the original action that has the only effect of changing the overall parameter  as $g\to g+\lambda g^2$.

\vspace{12pt}

So far we have only discussed the \emph{infinitesimal} version of the deformation. We would like to ``integrate'' the above procedure and write a \emph{finite} version of the deformation. In other words, we want to promote the action to $S_\lambda$, depending on the deformation parameter $\lambda$ that now is not assumed to be small anymore. As ``initial-value condition'' we want to impose that $S_{\lambda=0}=S_0$. 

Consider the to-be-constructed deformed action $S_\lambda$ with $\lambda$ finite. This action is characterised by being multiplied by an overall  $g(\lambda)$, i.e.~now $g$ depends on the deformation parameter $\lambda$, with the obvious requirement that $g(0)=g$. An infinitesimal deformation of this action would be given by 
\begin{equation}
    S_\lambda\to S_{\lambda+\delta\lambda}= S_\lambda+\delta S_\lambda.
\end{equation}
Let us repeat the construction made when starting from $S_0$, and let us now deform $S_\lambda$ by
\begin{equation}
  \delta S_\lambda=-\frac{\lambda}{2} \int d^2\sigma\, \epsilon^{\alpha\beta}J^\lambda_\alpha\tilde J^\lambda_\beta.
\end{equation}
Importantly, now the currents $J^\lambda,\tilde J^\lambda$ are those of the \emph{deformed} model.
It is obvious that this leads to the infinitesimal deformation 
\begin{equation}
    g(\lambda)\to g(\lambda+\delta\lambda)= g(\lambda)+g(\lambda)^2\delta\lambda,
\end{equation}
so that we can identify the infinitesimal deformation of $g(\lambda)$ as $\delta g(\lambda)=g(\lambda)^2\delta\lambda$. We may  reinterpret this relation as a differential equation which is solved by
\begin{equation}
   \frac{dg(\lambda)}{d\lambda}=g(\lambda)^2\qquad\implies\qquad g(\lambda)=\frac{g}{1-\lambda g},
\end{equation}
where we fixed the integration constant by requiring $g(0)=g$.
Notice that we can extend the deformation only up to the value $\lambda=g^{-1}$, where we hit a singularity.\footnote{Interestingly, in~\cite{Rodriguez:2021tcz} it was shown that this limit of maximal deformation can actually be understood as an ultra-relativistic limit of the original undeformed model, yielding a theory with Carrollian symmetry. See also~\cite{Tempo:2022ndz,Parekh:2023xms}.}

The  example of the current-current deformation of one boson is nice and  instructive, but the deformation in this case is not very interesting: since it only changes the value of the parameter $g$ in front of the action, it amounts just to a redefinition (a rescaling) of the scalar field $\phi$. When we will consider $\sigma$-models on $D$-dimensional target spaces, this kind of redefinitions are just  coordinate transformations in target space.

\subsection{Two bosons}\label{sec:2bos}
 To have more interesting transformations emerging from the current-current deformations, we must consider the case of at least two fields. We now take the sum of two massless free-field actions
\begin{equation}
    S_0=-\frac{1}{2}\int d^2\sigma \left(g_1\partial_\alpha\phi_1\partial^\alpha \phi_1+g_2\partial_\alpha\phi_2\partial^\alpha \phi_2\right),
\end{equation}
for the scalar fields $\phi_i,\ i=1,2$, each with a parameter $g_i$. It is clear that in this case we can generate deformations using a total of 4 conserved currents, namely $J^\alpha_i,\tilde J^\alpha_i,\ i=1,2$. Working in the basis of chiral and antichiral currents $J^{iL}_\alpha, J^{iR}_\alpha,\ i=1,2$ is nice because we know that the current bilinears must be formed by a left and a right current. We therefore have 4 options of infinitesimal deformations. Two of them couple the flavour indices $i=1,2$ diagonally, while two off-diagonally
\begin{equation}
    J^{1L}J^{1R},\qquad
    J^{2L}J^{2R},\qquad
    J^{1L}J^{2R},\qquad
    J^{2L}J^{1R}.
\end{equation}
Obviously, the deformations generated by the diagonal bilinears $ J^{1L}J^{1R},
    J^{2L}J^{2R}$ will generate rescalings analogous to the one discussed in the case of one boson. The only difference here is that we can rescale $g_1$ and $g_2$ independently, because we can independently redefine $\phi_1$ or $\phi_2$. Actually, even the symmetric combination of the off-diagonal bilinears can be understood as a simple field redefinition (a coordinate transformation in target space). In fact, if we take the infinitesimal deformation of the action as
\begin{equation}
\begin{aligned}
     S_0+\delta S_0&=-\int d^2\sigma\left[\sum_{i=1}^2g_i\partial_+\phi_i\partial_-\phi_i+\lambda(J^{1L}_+J^{2R}_-+J^{2L}_+J^{1R}_-)\right]\\
    &=-\int d^2\sigma\left[\sum_{i=1}^2g_i\partial_+\phi_i\partial_-\phi_i+\lambda g_1g_2(\partial_+\phi_1\partial_-\phi_2+\partial_+\phi_2\partial_-\phi_1)\right],
    \end{aligned}
\end{equation}
we see that it is equivalent, for example, to the infinitesimal field redefinition $\phi_1\to\phi_1+\lambda g_2\phi_2$. A similar shift of $\phi_2$ would also be an option. The original target-space metric is diagonal, and it receives off-diagonal components just as a consequence of  the change of coordinates.

\vspace{12pt}

It is clear that the more interesting deformation is the one generated by the \emph{antisymmetric} bilinear in the currents. This will generate a new coupling that was not present in the original action, and that from the target-space point of view is understood as a $B$-field.

Let us therefore consider the deformation generated by $J^{1L}_+J^{2R}_--J^{2L}_+J^{1R}_-$. As done above in the case of one boson, we would like to understand how to promote the infinitesimal deformation to a finite one. To do that, we use the same strategy as before. We consider the action at a generic value of the deformation parameter $\lambda$, we construct its infinitesimal deformation by the  current bilinear, and then we study how the various couplings change with $\lambda$. In this case, we know that the deformation introduces a new coupling that is not present in the original action. Therefore, for the action at generic $\lambda$ we take
\begin{equation}
    S_\lambda=-\int d^2\sigma\left[\sum_{i=1}^2g_i(\lambda)\partial_+\phi_i\partial_-\phi_i+b(\lambda)(\partial_+\phi_1\partial_-\phi_2-\partial_+\phi_2\partial_-\phi_1)\right],
\end{equation}
where $b(\lambda)$ is the new coupling, and we impose the ``initial condition'' $g_i(0)=g_i, b(0)=0$. In principle, we could have included also a parameter for the symmetric off-diagonal coupling of $\phi_1$ and $\phi_2$ (in other words, the off-diagonal component of the target-space metric). But it turns out that this coupling is not generated by the antisymmetric bilinear that we are considering, so for simplicity we do not include it from the very beginning. We will see, instead, that it is important to allow  $g_i(\lambda)$ to depend on the deformation parameter because, although they are not affected by the infinitesimal deformation of $S_0$, these couplings do receive corrections at higher orders.

Let us pause for a moment to make a comment. Given that $b(\lambda)$ is constant (i.e.~it does not depend on the fields $\phi_1,\phi_2$) the term proportional to $b(\lambda)$ is in fact a total derivative off-shell (i.e. without the need to impose equations of motion), and it may be dropped from the action (under suitable boundary conditions, for example periodic boundary conditions). From the target-space point of view, this corresponds to a possible gauge transformation  of the $B$-field, that is shifted  by a constant. The fact that this term is trivial, however, is only a consequence of the fact that at this initial stage, for pedagogical reasons, we are working with the very simple toy-model of free bosons. Later we will consider more interesting $\sigma$-models, and in that case $\phi_1,\phi_2$ will be accompanied by other fields. In particular, $g_i$ and $b$ may depend on those additional fields (that now are set to zero), so that in general the antisymmetric term proportional to $b(\lambda)$ will not be a total derivative anymore.

Obviously, the above action is still invariant under the two $U(1)$ transformations $\phi_i\to \phi_i+c_i, c_i\in \mathbb R$. We can then  compute the corresponding Noether currents, that now will receive corrections proportional to $b(\lambda)$
\begin{equation}
    \begin{aligned}
        &J^1_\pm(\lambda)=g_1(\lambda)\partial_\pm\phi_1\mp b(\lambda)\partial_\pm \phi_2,\\
        &J^2_\pm(\lambda)=g_2(\lambda)\partial_\pm\phi_2\pm b(\lambda)\partial_\pm \phi_1.
    \end{aligned}
\end{equation}
Notice that in the last terms of each line one swaps the index $i$, and that there is a relative minus sign for the formula of $J^1$ and $J^2$. Not only these Noether currents are conserved, but also their Hodge duals. Once again, the conservation of $\tilde J^1, \tilde J^2$ is an identity that does not require imposing the equations of motion (because $g_i$ and $b$ are constants). We can therefore construct the left and right currents, that will have only the non-vanishing components $J^{iL}_+=J^i_+,J^{iR}_-=J^i_-$. If we construct the infinitesimal variation of the action $S_\lambda$ and we rearrange the terms
\begin{equation}
\begin{aligned}
    \delta S_\lambda&=-\delta\lambda \int d^2\sigma\left(J^{1L}_+J^{2R}_--J^{2L}_+J^{1R}_-\right)\\
    &=-\delta\lambda \int d^2\sigma\left(J^{1}_+J^{2}_--J^{2}_+J^{1}_-\right)\\
    &=-\delta\lambda\int d^2\sigma\, [-2b(\lambda)g_1(\lambda)\partial_+\phi_1\partial_-\phi_1-2b(\lambda)g_2(\lambda)\partial_+\phi_2\partial_-\phi_2\\
    &\qquad \qquad \qquad +(g_1(\lambda)g_2(\lambda)-b(\lambda)^2)(\partial_+\phi_1\partial_-\phi_2-\partial_+\phi_2\partial_-\phi_1)],
\end{aligned}
\end{equation}
we see that we can identify the infinitesimal variations of the couplings as
\begin{equation}
    \delta g_i=-2b(\lambda)g_i(\lambda)\delta \lambda,\qquad\qquad
    \delta b(\lambda)=(g_1(\lambda)g_2(\lambda)-b(\lambda)^2)\delta\lambda.
\end{equation}
We therefore get the following system of coupled differential equations
\begin{equation}
    \frac{dg_i(\lambda)}{d\lambda}=-2b(\lambda)g_i(\lambda),\qquad \qquad 
    \frac{d b(\lambda)}{d\lambda}=g_1(\lambda)g_2(\lambda)-b(\lambda)^2.
\end{equation}
We can separate variables if we take another derivative in the second equation (using the first equation to know what the derivative of $g_i$ is) and then taking linear combinations to remove the $g_i$-dependent terms. Doing this we find an equation for $b(\lambda)$ only
\begin{equation}
    \frac{d^2 b(\lambda)}{d\lambda^2}+6 b(\lambda)\frac{d b(\lambda)}{d\lambda}+4 b(\lambda)^3=0,
\end{equation}
which is solved by
\begin{equation}
    b(\lambda)=\frac{c_1(\lambda+c_2)}{1+c_1(\lambda+c_2)^2},
\end{equation}
where $c_1,c_2$ are integration constants. Demanding that $b(0)=0$ we see that we must set $c_2=0$.
The other constant of integration is fixed by demanding that we reproduce the infinitesimal deformation of the original action $S_0$. In that case, we see that we generate an infinitesimal coupling $b\sim \lambda g_1g_2$, which forces us to set $c_1=g_1g_2$ in the solution. We therefore obtain
\begin{equation}
    b(\lambda)=\frac{\lambda g_1g_2}{1+\lambda^2g_1g_2}.
\end{equation}
Having now a solution for $b(\lambda)$, we can use it to solve the differential equation for $g_i(\lambda)$. It is solved by
\begin{equation}
    g_i(\lambda)=\frac{g_i}{1+\lambda^2 g_1g_2},
\end{equation}
where we have already fixed the constant of integration demanding that $g_i(0)=g_i$.
Notice that we hit a singularity at a special value of $\lambda$ only if one and only one of the couplings is negative ($\lambda=(-g_1g_2)^{-1/2}$), otherwise there is no singularity and $\lambda$ can take any real value. From a target-space point of view, a negative $g_i$ corresponds to $\phi_i$ being a time coordinate.

\vspace{12pt} 

The deformation that we have just considered is very interesting: it generates a new coupling and it cannot be understood as a simple field redefinition (or coordinate transformation in target space). If we were considering a non-trivial example where the couplings depend on additional $\sigma$-model fields, it would generically be impossible to remove the term proportional to $b(\lambda)$ by a $B$-field gauge transformation. To understand better the above result, it is natural to ask whether there exists a more complicated transformation (neither a target space coordinate redefinition nor a $B$-field gauge transformation) that reproduces the above deformation. The answer turns out to be positive, but in order to understand it we need to briefly review the concept of T-duality.

\subsection{T-duality}\label{sec:T}
T-duality may be defined in different ways. In these notes we will focus on the interpretation of T-duality as a canonical transformation on the $\sigma$-model. T-duality may be implemented when there is a shift isometry on the $\sigma$-model action, and it is actually possible to implement factorised T-dualities when there is a collection of commuting isometries. Notice, however, that there are generalisations of this concept when the isometry group is non abelian (the so-called ``non-abelian T-duality''). As we will see later in Section~\ref{sec:actionOdd}, (the standard/abelian) T-duality may be understood as the swapping of the spatial derivatives of the fields with their conjugate momenta introduced in the Hamiltonian formalism. At the Lagrangian level, this transformation is a non-local map on the fields because, as we will soon review, the \emph{derivatives} of the original field are related to the \emph{derivatives} of the dual field. The relation is non-trivial, thus implying the non-locality of the transformation. Obviously, the brief review that we have here is not a complete presentation on the concept of T-duality, since we will focus only on the aspects that will be needed in the rest of the notes. We refer for example to the standard review~\cite{Giveon:1994fu} and to more literature listed in section~\ref{sec:concl}.

\vspace{12pt}

Let us start with the action for just one free scalar boson $\phi$ as in~\eqref{eq:S1phi}. We will take this as a toy model to review T-duality. Exercise~\ref{ex:Buscher} of section~\ref{sec:ex-T-TsT} suggests to repeat this derivation in the generic case of a $\sigma$-model with a shift isometry. The idea of T-duality is to rewrite the action for a new scalar field $\tilde \phi$ that is understood as the \emph{dual} of $\phi$. In order to see where the dual field comes from, we reinterpret the original action taking $A_\alpha=\partial_\alpha\phi$ as the fundamental field, rather than $\phi$ itself. This is equivalent to the original formulation if  we  are able to impose that locally $A_\alpha$ is just the derivative of a scalar. In other words, we demand that locally the Bianchi identity $\epsilon^{\alpha\beta}\partial_\alpha A_\beta=0$ is satisfied. We can achieve this if we write the following action
\begin{equation}\label{eq:1ordS}
    S=-\int d^2\sigma\left(\frac{g}{2}A_\alpha A^\alpha-\tilde \phi\epsilon^{\alpha\beta}\partial_\alpha A_\beta\right).
\end{equation}
On the one hand, we can easily integrate out $\tilde \phi$ from this action. In fact, its equations of motion just imply the desired Bianchi identity $\epsilon^{\alpha\beta}\partial_\alpha A_\beta=0$ which we locally solve as $A_\alpha=\partial_\alpha\phi$ for some scalar field $\phi$.\footnote{Imposing $\epsilon^{\alpha\beta}\partial_\alpha A_\beta=0$ globally and not just locally would forbid the possibility of having winding in the case of the compact boson. However, the equations of motion of $\tilde\phi$ imply a weaker constraint, namely that the \emph{integral} of $\epsilon^{\alpha\beta}\partial_\alpha A_\beta$ is proportional to $2\pi n$ with $n\in \mathbb Z$. In fact, even when $n\neq 0$, there is no contribution in the partition function because of the factor $e^{iS}$. For a 2-dimensional spacetime with the topology of a cylinder, using Stokes theorem  we can rewrite this requirement as $2\pi n=\int d^2\sigma\, \partial_\alpha(\epsilon^{\alpha\beta}A_\beta)=\oint d\sigma^\alpha A_\alpha=\oint d\sigma^\alpha \partial_\alpha\phi=\phi(L)-\phi(0)$, which is indeed compatible with a non-trivial winding for $\phi$. In the rest we will not be so careful with these global issues.} Substituting this solution in the action~\eqref{eq:1ordS}  we get back to the original action~\eqref{eq:S1phi}.

On the other hand, we have another option because we can integrate $A_\alpha$ out instead. To  first step is integrating by parts $\int d^2 \sigma\, \tilde \phi\epsilon^{\alpha\beta}\partial_\alpha A_\beta=-\int d^2 \sigma\, \epsilon^{\alpha\beta}(\partial_\alpha\tilde \phi) A_\beta$. Then we can compute the equations of motion of $A_\alpha$ and find
\begin{equation}\label{eq:solAT}
    gA^\alpha =\epsilon^{\alpha\beta}\partial_\beta\tilde \phi.
\end{equation}
Substituting this solution into the action~\eqref{eq:1ordS} we obtain
\begin{equation}
    S=-\int d^2\sigma\left[-\frac{g}{2} A_\alpha A^\alpha+A_\alpha\cancel{(gA^\alpha-\epsilon^{\alpha\beta}\partial_\beta\tilde \phi)}\right]=-\frac{g^{-1}}{2} \int d^2\sigma\, \partial_\alpha\tilde \phi\partial^\alpha\tilde \phi.
\end{equation}
The canceled expression vanishes because of the equations of motion of $A_\alpha$.
We see that the action for $\tilde\phi$ is analogous to the one of the original boson $\phi$, with the exception that now the parameter $g$ appears to be inverted.

Let us emphasize  that the relation between $\phi$ and the dual $\tilde \phi$ is quite non-trivial. From~\eqref{eq:solAT} and setting $A_\alpha=\partial_\alpha\phi$, we can relate the \emph{derivatives} of the fields as\footnote{For the sake of the explanation, here we are taking $A_\alpha=\partial_\alpha\phi$. However, the sign in this relation is ambiguous, and we could as well take $A_\alpha=-\partial_\alpha\phi$. In fact, the double formulation that we will use in Section~\ref{sec:actionOdd} will select the other sign, so that we will have the relations $g\partial_\pm\phi=\mp\partial_\pm\tilde\phi$.}
\begin{equation}
    g\partial_\alpha\phi=\epsilon_{\alpha\beta}\partial^\beta\tilde\phi.
\end{equation}
More explicitly, this is
\begin{equation}
g\partial_\pm\phi=\pm\partial_\pm\tilde\phi\qquad\implies\qquad 
\partial_\tau\tilde \phi=g\partial_\sigma\phi,\quad 
\partial_\sigma\tilde \phi=g\partial_\tau\phi,
\end{equation}
so that the time and space derivatives are exchanged. This observation makes it manifest that T-duality is a non-local transformation, because a relation between $\phi$ and $\tilde \phi$ would involve an integration that would introduce non-local terms.

\subsection{TsT deformations}\label{sec:TsT}
T-duality is a discrete map, in the sense that we do not have a continuous deformation parameter that brings us back to the original model when the parameter is sent to zero. T-duality is in fact a $\mathbb Z_2$ transformation, because acting twice with the T-duality map we go back to the original model. We can, however, generate a continuous transformation if we combine T-duality with something else. In particular, we will now perform the sequence of transformations T-duality -- shift -- T-duality, that we collectively call ``TsT''.  

To be able to implement a TsT transformation, it is enough to have two commuting isometries in the action. For pedagogical reasons, however, we will work once again with a toy model, where the original action is the sum of two free bosons. Exercise~\ref{ex:TsT} of section~\ref{sec:ex-T-TsT} suggests to compute the TsT deformation of a generic $\sigma$-model with 2 commuting isometries. Our starting point is therefore again
\begin{equation}\label{eq:S-2phi}
    S=-\int d^2\sigma\left(g_1\partial_+\phi_1\partial_-\phi_1+g_2\partial_+\phi_2\partial_-\phi_2\right).
\end{equation}
The first transformation that we want to apply is a T-duality on $\phi_1$
\begin{equation}
   \phi_1\to \tilde \phi_1. 
\end{equation}
From the previous discussion, we know that the only effect this has on the free-boson action is the inversion of the $g_1$ coupling. After this first T-duality transformation, we therefore obtain the new action
\begin{equation}
     S=-\int d^2\sigma \left(g_1^{-1}\partial_+\tilde\phi_1\partial_-\tilde\phi_1+g_2\partial_+\phi_2\partial_-\phi_2\right).
\end{equation}
At this point, we want to implement the shift transformation as
\begin{equation}
    \phi_2=\varphi_2+\lambda\tilde\phi_1.
\end{equation}
Here we are redefining the second scalar by a quantity that is proportional to the \emph{dual} of the first scalar. The parameter $\lambda$ controlling the shift will be our deformation parameter. Because of the shift, the action looks different, 
\begin{equation}
\begin{aligned}
    S=-\int d^2\sigma [&(g_1^{-1}+\lambda^2 g_2)\partial_+\tilde\phi_1\partial_-\tilde\phi_1+g_2\partial_+\varphi_2\partial_-\varphi_2\\
    &+\lambda g_2(\partial_+\tilde\phi_1\partial_-\varphi_2+\partial_+\varphi_2\partial_-\tilde\phi_1)],
\end{aligned}
\end{equation}
although we really did just a simple  \emph{local} field redefinition (i.e. a coordinate redefinition in target space). 
We now want to implement the last T-duality transformation
\begin{equation}
    \tilde\phi_1\to \tilde{\tilde\phi}_1\equiv\varphi_1.
\end{equation}
At this point we have to be careful. On this action, T-duality is \emph{not} implemented just as an inversion of the coupling in front of $\partial_+\tilde\phi_1\partial_-\tilde\phi_1$. There are in fact other terms in the action that contain $\tilde\phi_1$. To obtain the correct transformation we therefore repeat what we did in the previous example of T-duality: we replace $\partial_\pm\tilde\phi_1\to A_\pm$ in the action, together with a term, proportional to $\varphi_1$, ensuring that $A_\pm$ is a derivative when we integrate $\varphi_1$ out. The action is then
\begin{equation}
\begin{aligned}
    S=-\int d^2\sigma [&(g_1^{-1}+\lambda^2 g_2)A_+A_-+g_2\partial_+\varphi_2\partial_-\varphi_2\\
    &+\lambda g_2(A_+\partial_-\varphi_2+\partial_+\varphi_2A_-)+\varphi_1(\partial_+A_--\partial_-A_+)].
\end{aligned}
\end{equation}
To implement the T-duality transformation we integrate $A_\pm$ out. Now the equations of motion read
\begin{equation}
    (g_1^{-1}+\lambda^2 g_2)A_\pm+\lambda g_2 \partial_\pm\varphi_2\mp\partial_\pm\varphi_1=0.
\end{equation}
In order to derive them, it was again necessary to integrate by parts.
On the equations of motion, the previous action is
\begin{equation}
\begin{aligned}
    S&=-\int d^2\sigma[A_+\cancel{((g_1^{-1}+\lambda^2 g_2)A_-+\lambda g_2\partial_-\varphi_2+\partial_-\varphi_1)}
    +g_2\partial_+\varphi_2\partial_-\varphi_2\\
 &   \qquad\qquad+\lambda g_2\partial_+\varphi_2A_--\partial_+\varphi_1A_-]\\
 &=-\int d^2\sigma \left[\sum_{i=1}^2g_i(\lambda)\partial_+\varphi_i\partial_-\varphi_i+b(\lambda)(\partial_+\varphi_1\partial_-\varphi_2-\partial_+\varphi_2\partial_-\varphi_1)\right],
\end{aligned}
\end{equation}
where
\begin{equation}
   g_i(\lambda)=\frac{g_i}{1+\lambda^2 g_1g_2},\qquad
   b(\lambda)=\frac{\lambda g_1g_2}{1+\lambda^2g_1g_2}.
\end{equation}
We immediately recognise the finite version of the deformation of the action obtained in Section~\ref{sec:2bos} by the antisymmetric combination of current bilinears! We therefore conclude that such a current-current deformation is equivalent to a TsT deformation. While the deformations induced by \emph{symmetric} current bilinears are equivalent to \emph{local} field redefinitions of the fields, the deformation induced by the \emph{antisymmetric} current bilinear is equivalent to a \emph{non-trivial} transformation, which can be understood as a \emph{non-local} field redefinition because of the involvement of the T-duality map.

As a brief comment, let us remark that at each step of the implementation of the TsT deformation we do not spoil the  isometries of the action corresponding to the shifts of the two bosons. This is what allows us to carry out the chain of transformations to the very end. In fact, these isometries are present also in the final action.

\subsubsection{An alternative construction}\label{sec:TsT-alt}
Before concluding this section, let's see an alternative  construction that reproduces the above TsT deformation. At first, what we will do may seem odd, but it is justified by the construction of TsT in terms of $O(d,d)$ as in Section~\ref{sec:actionOdd}, and also by the non-abelian generalisation of TsT in terms of Yang-Baxter deformations, see section~\ref{sec:concl}.
We start again from the original action~\eqref{eq:S-2phi} but now we add to it a total derivative term proportional to a parameter that we call $\zeta$
\begin{equation}
    S=-\int d^2\sigma\left[g_1\partial_+\phi_1\partial_-\phi_1+g_2\partial_+\phi_2\partial_-\phi_2+\zeta(\partial_+\phi_1\partial_-\phi_2-\partial_+\phi_2\partial_- \phi_1)\right].
\end{equation}
Notice that the term proportional to $\zeta$ that we have added is in fact a total derivative, and we may drop it.\footnote{The comment made earlier about $b$, which later will be promoted to a field-dependent coupling when considering more interesting $\sigma$-models, does not apply to $\zeta$. The parameter $\zeta$ will always be taken to be a constant, and the corresponding term proportional to $\zeta$ will continue to be locally a total derivative also in more complicated $\sigma$-models.} However,  we now want to  apply T-duality transformations. The non-locality of the T-duality map will then promote $\zeta$ to a physical parameter that cannot be removed by a coordinate transformation or by a shift of the $B$-field.

In this case we want to apply T-duality transformations on $\phi_1$ and $\phi_2$ simultaneously. Of course, given that the corresponding isometries commute, we may as well implement them as a sequence, first dualising $\phi_1$ and then $\phi_2$ or vice versa. But we will save some time doing it simultaneously. We then rewrite the action as
\begin{equation}
    S=-\int d^2\sigma\, \left(\sum_{i=1,2}g_iA^i_+A^i_-+\zeta \epsilon_{ij}A^i_+A^j_- +\tilde\phi_i(\partial_+A^i_--\partial_-A^i_+)\right),
\end{equation}
where $\epsilon_{12}=-\epsilon_{21}=1$, $\epsilon_{11}=\epsilon_{22}=0$. For convenience, we will now start using a matrix notation that we will employ also later in the lecture notes. We do so by combining the fields into column vectors 
\begin{equation}
A=\left(\begin{array}{c}
    A^1  \\
     A^2
\end{array}\right),\qquad \qquad \tilde \phi=
\left(\begin{array}{c}
    \tilde\phi^1  \\
     \tilde\phi^2
\end{array}\right).
\end{equation} 
Defining the matrices
\begin{equation}
    G=\left(\begin{array}{cc}
      g_1   &  0\\
        0 & g_2
    \end{array}\right),\qquad\qquad
    \epsilon=\left(\begin{array}{cc}
      0   &  1\\
        -1 & 0
    \end{array}\right)
\end{equation}
we can then rewrite the action as
\begin{equation}
    S=-\int d^2\sigma\, \left[A_+^t(G+\zeta \epsilon)A_-+\tilde\phi^t(\partial_+A_--\partial_-A_+)\right],
\end{equation}
with $t$ denoting transposition.
To implement T-duality on $\phi_1,\phi_2$ we compute the equations of motion for the vector $A_\pm$. We find
\begin{equation}
    A_\pm=\pm(G\mp \zeta\epsilon)^{-1}\partial_\pm\tilde\phi,
\end{equation}
where the matrix indices are omitted.
The different sign for $A_\pm$ in front of $\zeta\epsilon$ is due to the fact that when computing the equations of motion for $A_-$ we have to take the transposition of the matrix that multiplies it. Notice that the matrix $G$ is invertible, and therefore $G\mp \zeta\epsilon$ is invertible at least in a neighbourhood of $\zeta=0$.
Substituting these solutions in the action we find
\begin{equation}
    \begin{aligned}
    S&=-\int d^2\sigma\, \left[\cancel{A_+^t\left((G+\zeta \epsilon)A_-+\partial_-\tilde\phi^t\right)}-\partial_+\tilde\phi^tA_-)\right] \\
    &=-\int d^2\sigma\, \left[\partial_+\tilde\phi^t(G+\zeta \epsilon)^{-1}\partial_-\tilde\phi)\right].
    \end{aligned}
\end{equation}
Let's call $\tilde E=(G+\zeta \epsilon)^{-1}=G^{-1}(1+\zeta\epsilon G^{-1})^{-1}$.
It is just a $2\times 2$ matrix which we can easily compute, and we find
\begin{equation}
    \tilde E=\frac{1}{1+\frac{\zeta^2}{g_1g_2}}\left(\begin{array}{cc}
      \frac{1}{g_1}   &  -\frac{\zeta}{g_1g_2}\\
        \frac{\zeta}{g_1g_2} & \frac{1}{g_2}
    \end{array}\right).
\end{equation}
The symmetric and antisymmetric parts of $\tilde E$ have the interpretation of metric $\tilde G=(\tilde E+\tilde E^t)/2$ and $B$-field $\tilde B=(\tilde E-\tilde E^t)/2$ in target space, and they are
\begin{equation}
    \tilde G=\left(\begin{array}{cc}
      \frac{g_2}{g_1g_2+\zeta^2}   &  0\\
        0 & \frac{g_1}{g_1g_2+\zeta^2}
    \end{array}\right),\qquad\qquad
    \tilde B=\left(\begin{array}{cc}
      0& -\frac{\zeta}{g_1g_2+\zeta^2}   \\
        \frac{\zeta}{g_1g_2+\zeta^2} & 0
    \end{array}\right).
\end{equation}
We can interpret this model as a deformation of the T-dual model, that we recover when sending $\zeta\to 0$. As anticipated, the parameter $\zeta$ is now physical.

We now want to implement the following field redefinition $\tilde\phi\to U\tilde \phi$ with 
\begin{equation}
    U=\left(\begin{array}{cc}
      0  &  \zeta\\
        -\zeta & 0
    \end{array}\right).
\end{equation}
Explicitly, it would be $\phi_1\to \zeta\phi_2$ and $\phi_2\to - \zeta\phi_1$. On the one hand, it is just a field redefinition, or equivalently a coordinate transformation in target space. On the other hand, we can implement it only as long as $\zeta\neq 0$, otherwise it is singular.
The transformed metric is
\begin{equation}
    U^t\tilde GU=\left(\begin{array}{cc}
      \frac{\zeta^2g_1}{g_1g_2+\zeta^2}   &  0\\
        0 & \frac{\zeta^2g_2}{g_1g_2+\zeta^2}
    \end{array}\right),
\end{equation}
and if we set $\zeta=1/\lambda$  we find
\begin{equation}
    U^t\tilde GU=\left(\begin{array}{cc}
      \frac{g_1}{1+\lambda^2g_1g_2}   &  0\\
        0 & \frac{g_2}{1+\lambda^2g_1g_2}
    \end{array}\right),\qquad \qquad \text{for }\zeta=1/\lambda.
\end{equation}
This is precisely the target-space metric  that appeared when doing the TsT deformation!
The $B$-field is not exactly the one of the TsT deformation, but it is related to that simply by a constant shift
\begin{equation}
\begin{aligned}
    U^t\tilde BU&=\left(\begin{array}{cc}
      0 & -\frac{1}{\lambda(1+\lambda^2g_1g_2)}\\
        \frac{1}{\lambda(1+\lambda^2g_1g_2)} & 0
    \end{array}\right)\\
  &  =\left(\begin{array}{cc}
      0 & -\frac{\lambda g_1g_2}{1+\lambda^2g_1g_2}\\
        \frac{\lambda g_1g_2}{1+\lambda^2g_1g_2} & 0
    \end{array}\right)+\left(\begin{array}{cc}
      0 & -\frac{1}{\lambda}\\
        \frac{1}{\lambda} & 0
    \end{array}\right).
\end{aligned}
\end{equation}
In this particular example we have shown that a TsT deformation is equivalent to the combination of these two operations: first adding a total derivative to the original $\sigma$-model and then applying T-duality on both directions. While the TsT is normally understood as a deformation of the original $\sigma$-model (which is recovered sending $\lambda\to 0$), this second construction is naturally understood as a deformation of the \emph{T-dual} $\sigma$-model (which is recovered sending $\zeta\to 0$). We can therefore think of the TsT deformation as an interpolation between the original and the T-dual model. Importantly, this reinterpretation of the TsT deformation involves coordinate redefinitions and shifts of the $B$-field that are singular in the $\lambda\to 0(\zeta\to\infty)$ and $\lambda\to \infty(\zeta\to 0)$ limits. It is only at \emph{finite} values of the deformation parameter that we can safely translate from one language to the other.

\subsection{Exercises}\label{sec:ex-T-TsT}
\begin{enumerate}
    \item \label{ex:Buscher} \textbf{Buscher rules ---} Consider the action of a generic $\sigma$-model \begin{equation}
        S=-\int d^2\sigma \, \Pi_{(-)}^{\alpha\beta}\partial_\alpha x^m\partial_\beta x^nE_{mn},
    \end{equation}
    where $x^m$ with $m=1,\ldots,D$ are the $\sigma$-model scalar fields, interpreted as coordinates in target-space. The coupling $E_{mn}$ can be written in terms of its symmetric and antisymmetric parts $E_{mn}=G_{mn}+B_{mn}$ that are understood as a metric and a $B$-field in target space. Assume that the action is invariant under the shift $x^1\to x^1+c, c\in\mathbb R$. That means that $E_{mn}$ does not depend on $x^1$ (although it may still depend on the remaining fields). Derive the so-called ``Buscher rules'' by implementing the T-duality map as explained in Section~\ref{sec:T}. In other words, replace $\partial_\alpha x^1\to A_\alpha$, add a term imposing the relevant Bianchi identity, and integrate $A_\alpha$ out. If we denote the coordinates as $x^m=\{x^1,x^\mu\}$ with $\mu=2,\ldots,D$, then you should find
    \begin{equation}\label{eq:Buscher}
    \begin{aligned}
        &\tilde E_{11}=\frac{1}{E_{11}},\qquad\qquad &&\tilde E_{1\nu}=\frac{E_{1\nu}}{E_{11}},\\ 
        &\tilde E_{\mu 1}=-\frac{E_{\mu 1}}{E_{11}},\qquad\qquad &&\tilde E_{\mu\nu}=E_{\mu\nu}-\frac{E_{\mu 1}E_{1\nu}}{E_{11}}.
    \end{aligned}
    \end{equation}
    You may translate these formulas as transformation rules for $G_{mn}, B_{mn}$.

    \item \label{ex:TsT} \textbf{TsT deformations ---} Assume that the sigma-model action has a second isometry realised by shifts of $x^2$. Notice that the two isometries commute. Implement a TsT transformation of the action generalising what done in Section~\ref{sec:TsT}. Taking $x^m=\{x^1, x^2,x^\mu\}$ with $\mu=3,\ldots,D$, you should find
    \begin{equation}\label{eq:TsT}
        \begin{aligned}
        &E'_{ij}=\left(E_{ij}+\lambda \epsilon^{kl}E_{ki}E_{lj}\right)\Delta^{-1},\\
        &E'_{i\nu}=\left(E_{i\nu}+\lambda\epsilon^{kl}E_{ki}E_{l\nu}\right)\Delta^{-1},\\
            &E'_{\mu i}=\left(E_{\mu i}-\lambda\epsilon^{kl}E_{ik}E_{\mu l}\right)\Delta^{-1},\\
            &E'_{\mu\nu}=E_{\mu\nu}+\left(\lambda\epsilon^{kl}E_{\mu k}E_{l\nu}+\lambda^2\epsilon^{ij}\epsilon^{kl} E_{il}E_{\mu k}E_{j\nu}\right)\Delta^{-1}\\
            &\Delta=1+\lambda\epsilon^{ij}E_{ij}+\frac{\lambda^2}{2}\epsilon^{ij}\epsilon^{kl}E_{ik}E_{jl},
        \end{aligned}
    \end{equation}
    where $i,j,k,l=1,2$ and $\epsilon^{12}=1=-\epsilon^{21}$.    
\end{enumerate}

\section{Integrably-marginal current-current deformations of CFTs}\label{sec:ChSch}
So far we have only looked at current-current deformations in a toy model, and we have not discussed the relation to the (possible) conformal symmetry of the $\sigma$-model, nor the (possible) integrability. 
In this section we consider current-current deformations and analyse the conditions that arise by demanding that they are \emph{marginal} deformations of CFTs.  The starting point is  a CFT that we deform  by a marginal operator $\mathcal O(z,\bar z)$ that is built from the product of chiral and antichiral currents. Soon we will add structure to this. 

An important point is that we want to distinguish between operators $\mathcal O(z,\bar z)$ that are ``potentially'' marginal and operators that are \emph{integrably} marginals. In the latter case, the infinitesimal deformation generated by $\mathcal O(z,\bar z)$ can be iterated and ``integrated'' to give rise to a \emph{finite} deformation, that remains marginal to all orders. In other words, if we start with an infinitesimal deformation generated by $\lambda \mathcal O(z,\bar z)$ with $\lambda\ll 1$, we demand that it is possible to promote the deformation to finite values of $\lambda$, in such a way that $\lambda$ continues to be a marginal coupling.

In this section we will adapt our notation to the standard one used in the context of CFTs. In particular, we will take the worldsheet to be euclidean, and we will use complex coordinates $z,\bar z$ there. See Section~\ref{sec:prel}.

\subsection{Initial setup and results}
In~\cite{Chaudhuri:1988qb} Chaudhuri and Schwartz identified a necessary and sufficient condition to characterise integrably marginal operators. Obviously, in order to do so, they do not consider the most generic setup. Let's see now what assumptions they used for their construction.

 First of all,  they consider CFTs with chiral and antichiral currents $J_a(z), \bar J_{\bar a}(\bar z)$. 
 The indices $a,\bar a$ label the currents in the set. The currents $J_a(z), \bar J_{\bar a}(\bar z)$ have conformal dimensions $(1,0)$ and $(0,1)$ respectively. Moreover, we assume that they satisfy current algebra relations
\begin{equation}\label{eq:OPE-JJ}
    \begin{aligned}
        &J_a(z)J_b(w)\sim \frac{K_{ab}\mathbf 1}{(z-w)^2}+\frac{i\, f_{ab}{}^cJ_c(w)}{z-w}+\text{regular},\\
        &\bar J_a(\bar z)\bar J_b(\bar w)\sim \frac{\bar K_{ab}\mathbf 1}{(\bar z-\bar w)^2}+\frac{i\, \bar f_{ab}{}^c\bar J_c(\bar w)}{\bar z-\bar w}+\text{regular}.
    \end{aligned}
\end{equation}
Here ``regular'' stands for terms that do not diverge in the limit $z\to w$. Above we have written explicitly the identity operator $\mathbf 1$ appearing with the double pole. Later, however, we will always omit it.
Moreover, $f_{ab}{}^c,\bar f_{ab}{}^c, K_{ab},\bar K_{ab}$ are constant real coefficients. In particular, $f_{ab}{}^c$ are antisymmetric in the indices $a,b$ and satisfy the Jacobi identity. They are therefore  interpreted as the structure constants of a Lie algebra, that we will denote by $\mathfrak g$. The coefficients $K_{ab}$, on the other hand, can be identified with a non-degenerate  ad-invariant bilinear form on the Lie algebra $\mathfrak g$. We will denote the inverse metric by $K^{ab}$, that we can use to raise the algebra  indices. Then $f^{abc}$ is totally antisymmetric. Importantly, later we will restrict ourselves to the case of \emph{compact} Lie algebras. In that case we can take $K_{ab}$ to be diagonal with \emph{positive} entries. Similar comments hold for the barred coefficients, keeping in mind that we don't need to require that the Lie algebras identified by $f_{ab}{}^c$ and $\bar f_{ab}{}^c$ are the same, so that in general we have two different Lie algebras $\mathfrak g$ and $\bar{\mathfrak g}$.
Examples of CFTs with current algebras are Wess-Zumino-Witten (WZW) models, see for example the lecture notes~\cite{Blumenhagen:2009zz}.

\vspace{12pt}

We now consider marginal deformations generated by
\begin{equation}
    O(z,\bar z)=c^{a\bar a}J_a(z)\bar J_{\bar a}(\bar z),
\end{equation}
where the repeated indices imply summation. The coefficients $c^{a\bar a}$ are taken to be real, and at the infinitesimal level the Lagrangian is deformed as $\mathcal L(z,\bar z)\to \mathcal L(z,\bar z)+\lambda O(z,\bar z)$. 

In general, not all coefficients $c^{a\bar  a}$ need to be non-vanishing. Let us introduce a new notation with indices $i,j$ corresponding  only to the non-vanishing entries $c^{i\bar \imath}\neq 0$. What we will see is that, under the condition of dealing with $\mathfrak g$ and $\bar{\mathfrak g}$ compact, the \emph{necessary and sufficient} condition for $O(z,\bar z)$ to be \emph{integrably} marginal is that we can write it again as $O(z,\bar z)=c^{i\bar \imath}J_i(z)\bar J_{\bar \imath}(\bar z)$ for currents satisfying
\begin{equation}
    \begin{aligned}
        &J_i(z)J_j(w)\sim \frac{K_{ij}}{(z-w)^2}+\text{regular},\\
        &\bar J_{\bar \imath}(\bar z)\bar J_{\bar \jmath}(\bar w)\sim \frac{\bar K_{\bar \imath\bar \jmath}}{(\bar z-\bar w)^2}+\text{regular}.
    \end{aligned}
\end{equation}
Notice that only the double pole contribution is present now, while the single pole contribution with the structure constants $f$ and $\bar f$ has disappeared.
In other words, the \emph{subset} of chiral and antichiral currents entering the definition of $O(z,\bar z)$ identify  \emph{abelian} subalgebras $\mathfrak h\subset \mathfrak g,\bar{\mathfrak h}\subset \bar{\mathfrak g}$, because $f_{ij}{}^k=\bar f_{\bar \imath\bar \jmath}{}^{\bar k}=0$.
In general, as we will see, a redefinition of the basis for the currents is necessary in order to achieve that.

\subsection{The marginality condition}
The operator $O(z,\bar z)$ has conformal weights $(h,\bar h)=(1,1)$. Therefore, it has classical conformal dimension $\Delta=h+\bar h=2$ and it is classically marginal. To understand whether this is still true at the quantum level when we deform the Lagrangian by $\lambda O(z,\bar z)$, the strategy is to look at the conformal dimension of this operator \emph{in the presence of the deformation induced by itself}. We will therefore  demand that even in the presence of the deformation the conformal dimension stays $\Delta=2$. If $\Delta$ received quantum corrections, perhaps at higher orders in the $\lambda$-expansion, then the operator would cease to be marginal and the deformation would break the conformal invariance of the deformed model. 

We know that in a CFT we can identify the conformal dimension of operators if we compute their 2-point functions, because conformal invariance fixes them to be
\begin{equation}
    \braket{O(z,\bar z)O(w,\bar w)}_\lambda \propto \frac{1}{|z-w|^{2\Delta}},
\end{equation}
where we do not care about the overall proportionality coefficient. The notation $\braket{\ldots}_\lambda$ indicates that the correlator is computed \emph{in the deformed theory}. For a marginal operator, the exponent of $|z-w|$ is $-4$, and we demand that this remains unchanged to all orders in $\lambda$. 

As soon as the conformal invariance is lost (for example because $O(z,\bar z)$ ceases to be marginal), the 2-point function is not necessarily of the above form. But let's suppose that the deformed theory remains conformal until a certain order in the expansion of the parameter $\lambda$, and that it ceases to be conformal at the next order. That means that there is a moment in which the 2-point function is still constrained to be of the above form as dictated by the conformal symmetry, but the operator $O(z,\bar z)$ acquires an anomalous dimension $\delta\Delta$ (i.e.~a quantum correction to its classical conformal dimension). If we write $\Delta'=\Delta +   
\delta\Delta+\ldots$ where we ignore higher orders, then we  have 
\begin{equation}
\begin{aligned}
    \frac{1}{|z-w|^{2(\Delta+ \delta\Delta)}}&=\exp(-2(\Delta+ \delta\Delta)\log |z-w|)\\
    &=\frac{1}{|z-w|^{2\Delta}}\left(1-2 \delta\Delta \log|z-w|+\ldots\right).
    \end{aligned}
\end{equation}
In other words, in order to identify a possible anomalous dimension for $O(z,\bar z)$, we must consider the 2-point function of $O(z,\bar z)$ and look for terms proportional to
\begin{equation}
    |z-w|^{-4}\log|z-w|.
\end{equation} 
Demanding that there are no such log-terms ensures the marginality of the operator at each order in $\lambda$.

\vspace{12pt}

The 2-point function in the deformed theory can be written in terms of the correlators of the original \emph{undeformed} CFT. In fact, if the latter are denoted simply as $\braket{\ldots}$ (so that $\braket{\ldots}=\braket{\ldots}_{\lambda=0}$) and if we consider that in the presence of the deformation the weight in the path integral is given by the exponential of the deformed action
\begin{equation}
    \exp(-S)=\exp\left(-S_0-\lambda\int d^2z\, O(z,\bar z)\right),
\end{equation}
then we have{\small 
\begin{equation}\label{eq:2pf-l}
   \braket{O(z,\bar z)O(w,\bar w)}_\lambda = \frac{\sum_{n=0}^\infty\frac{(-\lambda)^n}{n!}\int d^{2n}z\braket{O(z,\bar z)O(w,\bar w)O(z_1,\bar z_1)\cdots O(z_n,\bar z_n)}}{\sum_{n=0}^\infty\frac{(-\lambda)^n}{n!}\int d^{2n}z\braket{O(z_1,\bar z_1)\cdots O(z_n,\bar z_n)}},
\end{equation}
}
where $d^{2n}z\equiv d^2z_1\ldots d^2z_n$.
Here we should divide the action by $\hbar$, but for simplicity we are setting $\hbar=1$. The expansion in~\eqref{eq:2pf-l} is consequence of the expansion of the exponential map, and formally we may think of it as an expansion in powers of  $\hbar^{-1}$. 
Importantly, here we are only looking at the \emph{infinitesimal} deformation of the classical action, which is why $\lambda$ appears linearly in $S$.  However, as anticipated in section~\ref{sec:2bos} and as seen later for the case of generic $\sigma$-models,  the finite form of the deformation includes corrections to the classical action also at higher powers of $\lambda$. When computing correlation functions, then, their corrections at higher powers of $\lambda$ are due to  competing contributions, coming both from the $\lambda$-dependence of the classical action and from the quantum corrections that result from the expansion of the exponential map. We may formally consider this as a double-scaling expansion in powers of $\lambda$ and of $\hbar^{-1}$. Having truncated the $\lambda$-expansion of the classical action at linear order, though, the $\hbar$-expansion may be formally organised in terms of powers of $\lambda$, and to agree with the terminology commonly used in this case in the literature, we will simply talk about a ``$\lambda$-expansion''. Nevertheless, the reader should keep in mind that the terms at order $\lambda^2$ that we will now discuss, correspond to the linear truncation of the $\lambda$-expansion in the classical action and to the second order of the expansion of the exponential map. The kind of divergences that arise from these terms should not mix with those related to the $\lambda^2$-corrections  to the classical action.

\subsection{Marginality condition at order $\mathcal O(\lambda^2)$}
In principle one should consider each order of the $\lambda$-expansion of the above 2-point function and demand that no anomalous dimension appears. As shown in~\cite{Chaudhuri:1988qb}, it is actually enough to focus on the order $\mathcal O(\lambda^2)$ of the expansion of the correlator: in fact, it turns out that the necessary requirement that ensures that the operator is marginal at that order is sufficient to make it marginal to all orders. 

Let's therefore focus just on the  $\mathcal O(\lambda^2)$-contribution to the above 2-point function. It is clearly given by
\begin{equation}
\begin{aligned}
    &\frac{\lambda^2}{2}\int d^2z_1d^2z_2 \braket{O(z,\bar z)O(w,\bar w)O(z_1,\bar z_1) O(z_2,\bar z_2)}\\
    &-\frac{\lambda^2}{2}\braket{O(z,\bar z)O(w,\bar w)}\int d^2z_1d^2z_2 \braket{O(z_1,\bar z_1) O(z_2,\bar z_2)}\\
    &+\lambda^2\braket{O(z,\bar z)O(w,\bar w)}\left(\int d^2z_1\braket{O(z_1,\bar z_1) }\right)^2\\
    &-\lambda^2\left(\int d^2z_1\braket{O(z,\bar z)O(w,\bar w)O(z_1,\bar z_1}\right)\left(\int d^2z_1\braket{O(z_1,\bar z_1) }\right),
    \end{aligned}
\end{equation}
where the first term comes from the numerator of~\eqref{eq:2pf-l},  the second and third terms from the denominator and the fourth term from both. The last two terms vanish because  the 1-point function vanishes, $\braket{O(z,\bar z)}=0$. The second term does not contribute to the anomalous dimension, because it is obvious that it cannot give rise to an expression with a term $|z-w|^{-4}\log|z-w|$. Certainly, the integral $\int d^2z_1d^2z_2 \braket{O(z_1,\bar z_1) O(z_2,\bar z_2)}$ will give rise to divergences (when $z_1$ and $z_2$ are close to each other) and it will therefore need to be renormalised, but eventually these infinities will be taken care of, and they will not be responsible for generating logs of the type $\log|z-w|$.   In general, terms with $|z-w|^{-4}\log|z-w|$ that therefore can give rise to an anomalous dimension  can potentially appear only from the first term above. Using the definition of $O(z,\bar z)$ we can rewrite it as 
\begin{equation}\label{eq:2pf-l2}
\begin{aligned}
    &\frac{\lambda^2}{2}\int d^2z_1d^2z_2 \braket{O(z,\bar z)O(w,\bar w)O(z_1,\bar z_1) O(z_2,\bar z_2)}\\
    &=\frac{\lambda^2}{2}\int d^2z_1d^2z_2 \, c^{a\bar a}c^{b\bar b}c^{c\bar c}c^{d\bar d}\times\\
    &\qquad \braket{J_a(z)J_b(w)J_c(z_1)J_d(z_2)}\braket{\bar J_{\bar a}(\bar z)\bar J_{\bar b}(\bar w)\bar J_{\bar c}(\bar z_1)\bar J_{\bar d}(\bar z_2)},
    \end{aligned}
\end{equation}
which has a nice factorised form in terms of correlators of chiral/antichiral currents. Thanks to this factorisation, for the time being we can focus  on  calculations involving the chiral currents only, for example, since the other ones will be analogous. 

In general, the exact computation of the above integral is quite difficult, especially because we would need to regularise and renormalise our integral. We expect, in fact, to have divergences  corresponding to the limit of different points coming close to each other. The renormalisation of these divergences can then produce the log corrections to the 2-point function that are responsible for the anomalous dimension. Luckily, we won't need to do the exact computation: using the OPE we can focus on the regions where the points collide, and identify the potentially divergent contributions.

\subsection{Marginality condition from the OPE}
Let's now discuss what happens when the points $z,w,z_1,z_2$ come close to each other. In this limit we can rewrite the correlators in~\eqref{eq:2pf-l2}  because, thanks to the OPE~\eqref{eq:OPE-JJ}, we know how products of two currents behave in this situation. Given that we have 4 different points, in general one will get different contributions \emph{depending on the order} in which we bring the different points close to each other. For example, taking $z_1\sim z$ and then $z\sim w$ is different from taking first $z\sim w$ and then $z_1\sim z$. Taking into account that we are integrating over $z_1,z_2$, however, we know that it will be enough to consider only half of all the possible orders for the OPEs, because the other half obtained by swapping $z_1\leftrightarrow z_2$ will be automatically taken care of under the integration.

Given that we are interested in $|z-w|^{-4}\log|z-w|$ and that $|z-w|^{-4}=(z-w)^{-2}(\bar z-\bar w)^{-2}$, if we focus on the correlator with chiral currents only, we must see, in particular, how to generate a factor of $(z-w)^{-2}$ from the OPE. Knowing the current algebra relations, we know that this factor will appear from the OPE of $J_a(z)J_b(w)$ with coefficient $K_{ab}$. We therefore conclude that one contribution that we must take into account is the one coming from the OPEs

  \begin{center}
     \begin{tikzpicture}[scale=0.8, every node/.style={scale=1}]
\node (a) at (0,0) {$z$};
\node (b) at (2,0) {$w$};
\node (c) at (1,-1) {$\mathbf 1$};
\node (e) at (5,0) {$z_1$};
\node (f) at (7,0) {$z_2$};
\node (g) at (6,-1) {$\mathbf 1$};
    \draw   (a) -- (c) -- (b);
    \draw   (e) -- (g) -- (f);
\end{tikzpicture}
\end{center}

With this graphical notation we mean that we must consider the OPEs where $z\sim w$ and $z_1\sim z_2$, and keep only the terms that in each case give rise to the identity. We therefore have
\begin{equation}\label{eq:contr1}
   \braket{ \frac{K_{ab}\mathbf 1}{(z-w)^2}\frac{K_{cd}\mathbf 1}{(z_1-z_2)^2}}=\frac{K_{ab}K_{cd}}{(z-w)^2(z_1-z_2)^2},
\end{equation}
where we used that $\braket{\mathbf 1}=1$.
Among all the contributions that do not have the $f$ terms, this is the only one that is interesting for us, because the other ones do not have the right $z,w$-dependence.

Let us now analyse the contributions with $f$ terms that we should include in our computation.
It is clear that among all the possible OPEs (that are obtained by performing the collisions in different orders) we must consider only those that leave the collision of $z$ and $w$ \emph{to the very end}. This is in fact the only way we would get the desired factor $(z-w)^{-2}$ and $f$ terms. 
As an example of something that is \emph{not} interesting for us, suppose that we consider a chain of OPE such that the first collision is $z\sim w$. Apart from the double pole contribution already taken into account in~\eqref{eq:contr1}, this OPE will generate a term proportional to $(z-w)^{-1}$ and $J_e(w)$ which, together with the remaining $J_c(z_1)J_d(z_2)$ currents in the correlator, does not have the chance to help generate the overall $(z-w)^{-2}$ factor.

The possible chains of OPEs that we need to take into account are summarised in  Table~\ref{tab:OPE}. With the graphic notation we try to indicate the order of the OPEs in the chain. Notice that every time we identify an entry in the table, we can immediately write another one by exchanging $z$ and $w$. It is clear that in each case we must include 3 collisions of points, until the last one gives us the desired factor proportional to $(z-w)^{-2}$ and the identity operator, so that we can use $\braket{\mathbf 1}=1$.

\begin{table}[t]
    \centering
    \begin{tabular}{|c|c|c|}
    \hline
         I & 
         \parbox[c]{95pt}{\figI}
          & 
           {\Large    $
             \frac{if_{cd}{}^e}{z_1-z_2}\, \frac{if_{ae}{}^f}{z-z_1}\, \frac{K_{fb}}{(z-w)^2}
         $} 
          \\
          \hline
         II & 
         \parbox[c]{95pt}{\figII}
         & {\Large $\frac{if_{cd}{}^e}{z_1-z_2}\, \frac{if_{be}{}^f}{w-z_1}\, \frac{K_{fa}}{(z-w)^2}$ }\\
         \hline
         III & 
         \parbox[c]{95pt}{\figIII}
          & {\Large $  \frac{if_{bd}{}^e}{w-z_2}\, \frac{if_{ec}{}^f}{w-z_1}\, \frac{K_{fa}}{(z-w)^2}$}\\
          \hline
         IV &   \parbox[c]{95pt}{\figIV}& {\Large $  \frac{if_{ad}{}^e}{z-z_2}\, \frac{if_{ec}{}^f}{z-z_1}\, \frac{K_{fb}}{(z-w)^2}$}\\
\hline
         V & 
         \parbox[c]{95pt}{\figV}
& {\Large $  \frac{if_{bd}{}^e}{w-z_2}\, \frac{if_{ac}{}^f}{z-z_1}\, \frac{K_{ef}}{(z-w)^2}$}\\
\hline
         VI & 
          \parbox[c]{95pt}{\figVI}
&{\Large $  \frac{if_{ad}{}^e}{z-z_2}\, \frac{if_{bc}{}^f}{w-z_1}\, \frac{K_{ef}}{(z-w)^2}$}\\
\hline
    \end{tabular}
    \caption{All possible OPEs of chiral currents that, at least in principle, we need to consider to identify the anomalous dimension.}
    \label{tab:OPE}
\end{table}
Putting these results together, the correlators that appear inside the integral are
\begin{equation}
\begin{aligned}
    \braket{J_a(z)J_b(w)J_c(z_1)J_d(z_2)}&=\frac{K_{ab}K_{cd}}{(z-w)^2(z_1-z_2)^2}\\
    & -\frac{1}{(z-w)^2}\Bigg[\frac{f_{cd}{}^ef_{aeb}}{(z_1-z_2)(z-z_1)}+\frac{f_{cd}{}^ef_{bea}}{(z_1-z_2)(w-z_1)}\\
    &+\frac{f_{bd}{}^ef_{eca}}{(w-z_2)(w-z_1)}+\frac{f_{ad}{}^ef_{ecb}}{(z-z_2)(z-z_1)}\\
    &+\frac{f_{bd}{}^ef_{ace}}{(w-z_2)(z-z_1)}+\frac{f_{ad}{}^ef_{bce}}{(z-z_2)(w-z_1)}\Bigg]+\ldots\\
    \braket{\bar J_{\bar a}(\bar z)\bar J_{\bar b}(\bar w)J_{\bar c}(\bar z_1)\bar J_{\bar d}(\bar z_2)}&=\frac{\bar K_{\bar a\bar b}\bar K_{\bar c\bar d}}{(\bar z-\bar w)^2(\bar z_1-\bar z_2)^2}\\
    & -\frac{1}{(\bar z-\bar w)^2}\Bigg[\frac{\bar f_{\bar c\bar d}{}^{\bar e}\bar f_{\bar a\bar e\bar b}}{(\bar z_1-\bar z_2)(\bar z-\bar z_1)}+\frac{\bar f_{\bar c\bar d}{}^{\bar e}\bar f_{\bar b\bar e\bar a}}{(\bar z_1-\bar z_2)(\bar w-\bar z_1)}\\
    &+\frac{\bar f_{\bar b\bar d}{}^{\bar e}\bar f_{\bar e\bar c\bar a}}{(\bar w-\bar z_2)(\bar w-\bar z_1)}+\frac{\bar f_{\bar a\bar d}{}^{\bar e}\bar f_{\bar e\bar c\bar b}}{(\bar z-\bar z_2)(\bar z-\bar z_1)}\\
    &+\frac{\bar f_{\bar b\bar d}{}^{\bar e}\bar f_{\bar a\bar c\bar e}}{(\bar w-\bar z_2)(\bar z-\bar z_1)}+\frac{\bar f_{\bar a\bar d}{}^{\bar e}\bar f_{\bar b\bar c\bar e}}{(\bar z-\bar z_2)(\bar w-\bar z_1)}\Bigg]+\ldots
\end{aligned}
\end{equation}
where we wrote explicitly also the correlator of antichiral currents that is obtained simply by adding bars everywhere. In both cases the dots stand for contributions that are not interesting for us because they cannot contribute to the term in the 2-point function that would yield an anomalous dimension.

Multiplying the two correlators we will obtain terms of 4 different types: one term with no $f$ or $\bar f$, various terms with $ff$, as many terms with $\bar f\bar f$, and finally terms with $ff\bar f\bar f$. Notice that after this multiplication all these contributions come with an overall factor of $|z-w|^{-4}$, as desired. The term with no $f$ or $\bar f$ does not give rise to the  $\log|z-w|$ that we want to identify. Of the terms  with $ff$ or $\bar f\bar f$, we should keep only those arising from the contributions V and VI in the table, because they are the only ones that have both $z$ and $w$ inside the integral over $z_1,z_2$ and have the chance of giving rise to the $\log|z-w|$. In particular, once we reintroduce the $c^{a\bar a}$ coefficients entering the definition of $O(z,\bar z)$, we have that these terms are proportional to 
\begin{equation}
\begin{aligned}
    &\frac{\lambda^2}{2}\int d^2z_1d^2z_2 \braket{O(z,\bar z)O(w,\bar w)O(z_1,\bar z_1) O(z_2,\bar z_2)}\\
    &=\frac{\lambda^2}{2}\frac{\alpha\log\Lambda}{|z-w|^4} c^{a\bar a}c^{b\bar b}c^{c\bar c}c^{d\bar d}(\bar K_{\bar a\bar b}\bar K_{\bar c\bar d}f_{bd}{}^ef_{ace}+ K_{ab}K_{cd}\bar f_{\bar b\bar d}{}^{\bar e}\bar f_{\bar a\bar c\bar e})+\ldots
    \end{aligned}
\end{equation}
Notice that we were able to group together the contributions from V and VI of Table~\ref{tab:OPE} thanks to the contraction with the Lie algebra metrics, which allows us to permute indices and relate the two terms. The numerical coefficient $\alpha$ can be computed explicitly when doing the complete calculation, and $\Lambda$ is the UV cutoff that is introduced to regularise the divergences that will give rise to the log corrections that we are after. 
The dots in the equation stand both for terms that do not contribute to the log corrections that we want to analyse, and for terms of the type $ff\bar f\bar f$. These last terms also contribute to the anomalous dimension, with a different dependence on UV cutoffs, therefore we should take them into account and demand that they vanish. But we will see that the condition arising from the terms that we have written explicitly will be enough to cancel also the others.

\subsection{Analysis of the necessary condition}
Obviously, given the above results, a necessary condition to have no anomalous dimension is to require that 
\begin{equation}
    c^{a\bar a}c^{b\bar b}c^{c\bar c}c^{d\bar d}(\bar K_{\bar a\bar b}\bar K_{\bar c\bar d}f_{bd}{}^ef_{ace}+ K_{ab}K_{cd}\bar f_{\bar b\bar d}{}^{\bar e}\bar f_{\bar a\bar c\bar e})=0.
\end{equation}
Let's try to rewrite this equation in a more readable way. Let's define
\begin{equation}
             \bar C^{\bar a\bar bc}=c^{a\bar a}c^{b\bar b}f_{ab}{}^c,
             \qquad 
              C^{ab\bar c}=c^{a\bar a}c^{b\bar b}\bar f_{\bar a\bar b}{}^{\bar c},
\end{equation}
then the above condition is just
\begin{equation}
    \bar C^{\bar b\bar d e}\bar C^{\bar a\bar c f}\bar K_{\bar a\bar b}\bar K_{\bar c\bar d}K_{ef}+
     C^{bd\bar e} C^{ac\bar f}K_{ab}K_{cd}\bar K_{\bar e\bar f}=0.
\end{equation}
At this point we make the crucial assumption that both $\mathfrak g$ and $\bar{\mathfrak g}$ are \emph{compact}. Then the metrics on the Lie algebras can be taken to be just the identity, $K_{ab}=\delta_{ab}, \bar K_{\bar a\bar b}=\delta_{\bar a\bar b}$. The above condition then reduces to
\begin{equation}
    \sum_{\bar b,\bar d, e}(\bar C^{\bar b\bar d e})^2+\sum_{bd\bar e} (C^{bd\bar e} )^2=0.
\end{equation}
Being a sum of squares, this is satisfied if and only if
\begin{equation}
    \bar C^{\bar b\bar d e}=0,\qquad \qquad 
    C^{bd\bar e} =0.
\end{equation}
It's important to stress that this simplified condition always implies the vanishing of the anomalous dimension at order $\mathcal O(\lambda^2)$, but it is \emph{equivalent} to the requirement of having no anomalous dimension only in the \emph{compact} case. In the non-compact case there may be weaker conditions for the marginality of $O(z,\bar z)$.

Clearly, the above conditions are satisfied if we restrict our coefficients $c^{a\bar a}$ to take values only in abelian subalgebras of $\mathfrak g,\bar{\mathfrak g}$, because then the structure constants contributing to $\bar C^{\bar b\bar d e}, C^{bd\bar e}$ are zero. We can actually show that, if we allow ourselves to redefine the basis for the currents, then this is actually the most generic situation. In fact, let's define for example
\begin{equation}
    V_i(z)=\delta_{i\bar a}c^{a\bar a}J_a(z).
\end{equation}
In terms of these $V_i(z)$, the operator $O(z,\bar z)$ now reads as $O(z,\bar z)=c^{a\bar a}J_a(z)\bar J_{\bar a}(\bar z)=\delta^{i\bar a}V_i(z)\bar J_{\bar a}(\bar z)$. We have, in essence, replaced $c$ by the Kroenecker $\delta$ thanks to a proper choice of the basis for the currents. Notice that in general the range of the index $i$ may be more limited than that of the index $a$, because some entries of the coefficients $c^{a\bar a}$ may be 0. Let us now look at the OPE for these new objects
\begin{equation}
\begin{aligned}
    V_i(z)V_j(w)&=\delta_{i\bar a}c^{a\bar a}\delta_{j\bar b}c^{b\bar b}J_a(z)J_b(w)\\
    &=\delta_{i\bar a}c^{a\bar a}\delta_{j\bar b}c^{b\bar b}\left(\frac{\delta_{ab}}{(z-w)^2}+\frac{i\, f_{ab}{}^cJ_c(z)}{z-w}\right)+\text{regular}\\
    &=\frac{K'_{ij}}{(z-w)^2}+\delta_{i\bar a}\delta_{j\bar b}\frac{i\, C^{\bar a\bar b c}J_c(z)}{z-w}+\text{regular},\qquad 
    K'_{ij}=\delta_{i\bar a}c^{a\bar a}\delta_{j\bar b}c^{b\bar b}\delta_{ab}.
    \end{aligned}
\end{equation}
We see that thanks to the condition $C^{\bar a\bar b c}=0$ the term with the single pole drops, and we just have
\begin{equation}
    V_i(z)V_j(w)=\frac{K'_{ij}}{(z-w)^2}+\text{regular}.
\end{equation}
The matrix $K'=c^tc$ is non-degenerate if we restrict the indices to the appropriate range. Moreover it is real and symmetric, and therefore it can be diagonalised. After this is done, given that the basis for the $V_i$ has changed, the operator will now read as $O(z,\bar z)=c'{}^{i\bar a}V_i(z)\bar J_{\bar a}(\bar z)$ for some new coefficients $c'{}^{i\bar a}$. At this point one can repeat the same reasoning for the antichiral currents and define $\bar V_{\bar \imath}(\bar z)$, which will also have no single pole in their OPE. The operator will finally read $O(z,\bar z)=c''{}^{i\bar \imath}V_i(z)\bar V_{\bar \imath}(\bar z)$. 

\subsection{Sufficiency of the constraints from the marginality condition}
Importantly, so far we have only discussed a \emph{necessary} condition that comes from demanding that a particular contribution to the $\mathcal O(\lambda^2)$ anomalous dimension cancels. Remember that this was obtained by looking at the terms with $ff$ and $\bar f\bar f$. It turns out, however, that this condition is sufficient to cancel also the terms with $ff\bar f\bar f$, that potentially contribute to the anomalous dimension as well, and that we have previously omitted. In fact, it is immediate to see this: all expressions have the structure constants contracted with the $c$-coefficients, and therefore vanish thanks to  the above marginality condition.

Even more, this condition turns out to be enough to ensure the marginality of the operator $O(z,\bar z)$ to \emph{all} orders in the $\lambda$-expansion. In fact, when working in the basis of the $V_i,\bar V_{\bar \imath}$, we can formally map the computations for the OPEs to those of free bosons if we take
\begin{equation}
    V_i(z)\to i\, \partial_z\phi_i(z),
    \qquad
    \bar V_{\bar \imath}(\bar z)\to i\, \partial_{\bar z}\phi_{\bar \imath}(\bar z),
\end{equation}
because they satisfy
\begin{equation}
     \partial_z\phi_i(z) \partial_w\phi_i(w)=-\frac{\delta_{ij}}{(z-w)^2}+\ldots,\qquad 
     \partial_{\bar z}\phi_{\bar \imath}(\bar z) \partial_{\bar w}\phi_{\bar \jmath}(\bar w)=-\frac{\delta_{\bar \imath\bar \jmath}}{(\bar z-\bar w)^2}+\ldots
\end{equation}
Therefore, when computing the 2-point function of $O(z,\bar z)=c''^{i\bar \imath}V_i(z)\bar V_{\bar \imath}(\bar z)$ in the deformed theory,  we can pretend that we are  morally computing the  2-point function of $O(z,\bar z)=-c''^{i\bar \imath}\partial_z\phi_i(z)\partial_{\bar \imath}\phi(\bar z)$ in the presence of the deformation $\lambda O(z,\bar z)$ of the  Lagrangian for the free bosons. No anomalous dimension arises from this computation. 

\vspace{12pt}

We have then reached the result  anticipated at the beginning of the section: in the case of current algebras on compact Lie algebras,  integrably marginal current-current deformations are identified by \emph{abelian} subalgebras of $\mathfrak g$ and $\bar{\mathfrak g}$. The most generic deformations, then, are generated by operators
\begin{equation}
    O(z,\bar z)=c^{i\bar \imath}J_i(z)\bar J_{\bar \imath}(\bar z),
\end{equation}
where the indices $i,\bar \imath$ span the Cartan subalgebras $\mathfrak h,\bar{\mathfrak h}$ of $\mathfrak g,\bar{\mathfrak g}$.

This result is very nice, but a natural question arises: although we have identified the condition for \emph{integrable} marginality, what is the actual \emph{finite} expression for the deformation of the Lagrangian that a lowest order is given by $\mathcal L(z,\bar z)+\lambda O(z,\bar z)$? It turns out that the answer is given by a certain subgroup of the $O(d,d)$ duality group. In the following section we will introduce this duality group and we will show that it contains elements that give rise to the finite form of the marginal deformations that we have identified. We will then show that these are actually just TsT deformations.

\section{The $O(D,D)$ duality group and the $\sigma$-model}\label{sec:ODD}
Useful reviews for the $O(D,D)$ duality group are for example those belonging to the literature on Double Field Theory~\cite{Zwiebach:2011rg,Aldazabal:2013sca,Berman:2013eva,Hohm:2013bwa}. The presentation here will follow that of~\cite{Osten:2019ayq} and~\cite{Borsato:2021gma}.

The reason  to consider the $O(D,D)$ duality group here is related to the fact that we will be led to consider deformations of sigma-models with an abelian group of $D$ symmetries. This is analogous to the restriction to  abelian subalgebras of chiral/antichiral currents, that was found in the previous section when looking at the integrably marginals operators. We will use $O(D,D)$ to construct the deformations in section~\ref{sec:finite}, and the connection to the marginal deformations of CFTs will become clear in section~\ref{sec:chiral} when discussing the case with chiral isometries.

\subsection{A ``double'' language for the Hamiltonian of the $\sigma$-model}
Let's start by considering the action of a generic $\sigma$-model
\begin{equation}\label{eq:sigma-model}
        S=-\int d^2\sigma \, \Pi_{(-)}^{\alpha\beta}\partial_\alpha x^m\partial_\beta x^n(G_{mn}+B_{mn}),
    \end{equation}
    where $x^m$ with $m=1,\ldots ,D$ are the fields and $G_{mn},B_{mn}$ generalised couplings (respectively symmetric and antisymmetric in the $m,n$ indices) that in general may be $x$-dependent. From the target-space point of view, $x^m$ are coordinates, while $G_{mn},B_{mn}$ are the metric and the Kalb-Ramond field. We will often switch from one terminology to the other, so that $x^m$ may be called fields or coordinates.
We can write the above action as
$
S=\int d^2 \sigma\, \mathcal L$
for the Lagrangian
\begin{equation}
\mathcal L=\tfrac12 G_{mn}\dot x^m\dot x^n-\tfrac12 G_{mn}x'{}^mx'{}^n-B_{mn}\dot x^mx'{}^n,
\end{equation} 
where we use the notation $\dot x^m=\partial_\tau x^m, x'{}^m=\partial_\sigma x^m$.
To go to Hamiltonian formalism we define momenta in the standard way
\begin{equation}
p_m\equiv \frac{\delta \mathcal S}{\delta \dot x^m}
=G_{mn}\dot x^n-B_{mn}x'{}^n,\quad\implies\quad
\dot x^m=G^{mn}(p_n+B_{np}x'{}^p).
\end{equation}
The Hamiltonian density is identified by writing the Lagrangian as $\mathcal L=p_m\dot x^m-\mathcal H$.
It is possible to check by direct computation that the  Hamiltonian density $\mathcal H$ can be written in a nice way if we collect all the $p_m$ and $x'{}^m$ in a ``double vector'' $\Psi_M$ with a ``double index'' $M$ running from 1 to $2D$. If we define it as
\begin{equation}
\Psi_M\equiv\left(\begin{array}{c} x'{}^m\\ p_m\end{array}\right),
\end{equation}
then the Hamiltonian density is
\begin{equation}
    \mathcal H=\tfrac12 \Psi_M\mathcal H^{MN}\Psi_N,
\end{equation}
with 
\begin{equation}\label{eq:par-HMN}
    \mathcal H^{MN}=\left(\begin{array}{cc}
 G_{mn}-B_{mp}G^{pq}B_{qn} & B_{mp}G^{pn}\\
-G^{mp}B_{pn}&G^{mn}
\end{array}
\right).
\end{equation}
The object $  \mathcal H^{MN}$ is called the ``generalised metric'' and it is a fundamental object in generalised geometry and double field theory.
Often, we will omit the indices in the blocks of ``double matrices''. The convention is always that the position of indices is
\begin{equation}
    \mathcal M^{MN}=\left(\begin{array}{cc}
m_{mn}& n_m{}^n\\
p^m{}_n&\quad q^{mn}
\end{array}
\right).
\end{equation}
Conversely, for a double matrix with lower indices we have
\begin{equation}
    \mathcal M_{MN}=\left(\begin{array}{cc}
r^{mn}& s^m{}_n\\
t_m{}^n&\quad u_{mn}
\end{array}
\right).
\end{equation}
The rule is that the position of the indices of the lower-right block matches the position of the indices of the double matrix. The position of the indices of other blocks are found from that starting point.
We will see later how to raise and lower double indices and how the respective  matrices and their entries are related.
Following the above rules, we can omit the indices in the generalised metric and write
\begin{equation}\label{eq:Hcan}
    \mathcal H\equiv\left(\begin{array}{cc}
 G-BG^{-1}B& 
BG^{-1}\\-G^{-1}B&G^{-1}
\end{array}
\right).
\end{equation}
We trust that despite the omitted indices there will be no confusion with the Hamiltonian density $\mathcal H$.

\vspace{12pt}

In the Hamiltonian formalism the dynamics is controlled by the Poisson brackets.
If we use the standard Poisson brackets
\begin{equation}
\begin{aligned}
&\{x^m(\sigma_1),x^n(\sigma_2)\}=0,\qquad
\{p_m(\sigma_1),p_n(\sigma_2)\}=0,\\
&\{x^m(\sigma_1),p_n(\sigma_2)\}=\delta^m_n\delta(\sigma_1-\sigma_2),
\end{aligned}
\end{equation}
we can check that in terms of $\Psi_M$ they read
\begin{equation}
\{\Psi_M(\sigma_1),\Psi_N(\sigma_2)\}=\tfrac12\eta_{MN}(\partial_1-\partial_2)\delta_{12}-\tfrac12 \Omega_{MN}(\partial_1+\partial_2)\delta_{12},
\end{equation}
where
\begin{equation}\label{eq:eta-Omega}
\eta_{MN}\equiv\left(\begin{array}{cc}
0& \mathbf 1\\
\mathbf 1& 0
\end{array}
\right),\qquad\qquad
\Omega_{MN}\equiv\left(\begin{array}{cc}
0& \mathbf 1\\
-\mathbf 1& 0
\end{array}
\right),
\end{equation}
where $\mathbf 1$ denotes the $D\times D$ identity matrix and we use the shorthand notation $\partial_{1}=\partial_{\sigma_{1}}$, $\partial_{2}=\partial_{\sigma_{2}}$ and $\delta_{12}=\delta(\sigma_1-\sigma_2)$. The above expressions must be understood in the sense of distributions, and because we have both derivatives $\partial_1$ and $\partial_2$ on the delta function, we think of the above expressions as being applied to test functions as
\begin{equation}
\int d\sigma_1 d\sigma_2 \ldots \varphi_1\varphi_2,
\end{equation}
where it is assumed that $\varphi_i$ depends only on $\sigma_i$.
If we do this, and if we assume reasonable boundary conditions for the test functions, then the term with $\Omega_{MN}$ in the Poisson brackets can be dropped. In fact, it is easy to see that it would never contribute when integrated, because
\begin{equation}
\begin{aligned}
\int d\sigma_1 d\sigma_2 [(\partial_1+\partial_2)\delta_{12}] \varphi_1\varphi_2
&=-\int d\sigma_1 d\sigma_2 \delta_{12}(\varphi'_1\varphi_2+\varphi_1\varphi'_2)\\
&=-\int d\sigma \partial_\sigma(\varphi_1(\sigma)\varphi_2(\sigma))=0.
\end{aligned}
\end{equation}
The last step relies on the boundary conditions used for the test functions: on the line we require that $\varphi_i$ go to a constant (typically 0) at the boundaries, and on the circle that the test functions are periodic.\footnote{See~\cite{Osten:2019ayq} for a discussion of what happens when relaxing these boundary conditions for test functions.} In short, we may assume that $\partial_2\delta_{12}=-\partial_1\delta_{12}$, and we may use this to rewrite also the term with $\eta_{MN}$, so that the Poisson brackets become
\begin{equation}\label{eq:Poisson-bracket-double}
\{\Psi_M(\sigma_1),\Psi_N(\sigma_2)\}=\eta_{MN}\partial_1\delta_{12}.
\end{equation}
Notice that in our rewriting of the Poisson brackets only $x'{}^m$ and not $x^m$ participate. This means that the zero-modes of $x^m$ are not contemplated in the discussion. In particular, suppose that we expand $x^m$ in Fourier modes $x^m=\sum_\alpha x^m_\alpha(\tau)e^{i\alpha\sigma}$, so that $x'{}^m=\sum_\alpha i\alpha x^m_\alpha(\tau)e^{i\alpha\sigma}$. Then the zero-mode $x^m_0(\tau)$ does not appear in the expansion of $x'{}^m$, and therefore the Poisson brackets written in terms of $\Psi_M$ do not capture it.

\subsection{Some canonical transformations in the double formalism}
At this point we want to consider non-trivial canonical transformations that give rise to interesting new $\sigma$-models. We want to work with canonical transformations because we know that they preserve the integrability of the original $\sigma$-model (if it is there in the first place). In fact, given an integrable Hamiltonian, the integrability cannot be broken by a canonical transformation because that is just a different way to parameterise the phase space. At the same time, some canonical transformations are not interesting because they do not generate new $\sigma$-models. For example, one may redefine the fields $x^m\to \tilde x^m(x)$ and consequently promote this to a canonical transformation by saying how the conjugate momenta transform. However, this will not give rise to a new $\sigma$-model, since we are simply redefining the Lagrangian fields by a local transformation (equivalently, we are doing a coordinate transformation in target space). We learn that interesting canonical transformations must necessarily mix the fields $x^m$ and their conjugate momenta $p_m$. Then, from a Lagrangian point of view, the map will be understood as non-local: \emph{derivatives} of fields are mapped to \emph{derivatives} of fields, in the spirit of the transformation induced by T-duality. 

We will see that the double notation becomes a very useful language in order to do that. Suppose that we implement a transformation of the phase-space variables that in the double language we write as

\begin{equation}
    \Psi_M=\tilde\Psi_N\ \mathcal O^N{}_M.
\end{equation}
If we specify the entries of the matrix $\mathcal  O^N{}_M$, this operation is equivalent to defining a map $(p_m,x^m)\to(\tilde p_m,\tilde x^m)$.\footnote{Again, up to the zero modes. Recall, in fact, the rewriting of the Poisson brackets as in~\eqref{eq:Poisson-bracket-double}, where $x'{}^m$ but not $x^m$ participate.}
This transformation is canonical if  it preserves the Poisson brackets. In the double language that is equivalent to requiring
\begin{equation}
\begin{aligned}
\eta_{MN}\partial_1\delta_{12}&=\{\Psi_M(\sigma_1),\Psi_N(\sigma_2)\}=\{\tilde\Psi_M(\sigma_1),\tilde\Psi_N(\sigma_2)\}\\
&=\mathcal O^P{}_M\mathcal O^Q{}_N\{\Psi_P(\sigma_1),\Psi_Q(\sigma_2)\}=\mathcal O^P{}_M\mathcal O^Q{}_N\eta_{PQ}\partial_1\delta_{12},
\end{aligned}
\end{equation}
where we have already made an important assumption by assuming that $\mathcal O^N{}_M$ is \emph{constant}. Only thanks to the  assumption that $\mathcal O^N{}_M$ is constant we can bring it outside the Poisson brackets as we did above. Then the requirement that the transformation is canonical is equivalent to the following very simple condition on the matrix $\mathcal O^N{}_M$
\begin{equation}
    \mathcal O^P{}_M\ \eta_{PQ}\ \mathcal O^Q{}_N=\eta_{MN}.
\end{equation}
Getting rid of the explicit indices, we can equivalently write it as
\begin{equation}
    \mathcal O^t\eta\mathcal O=\eta,
\end{equation}
where $t$ denotes transposition.
Because of the definition of $\eta$, therefore, $\mathcal O$ is an element of the $O(D,D)$ group. It is a variant of the group of orthogonal matrices (like the Lorentz group) where now half of the entries of the metric are positive and the other half are negative.\footnote{ Normally, the $O(D,D)$ group is defined as the group of matrices $\mathcal O$ that leave invariant the $O(D,D)$ metric $\eta$ defined as \begin{equation}
\eta=\left(\begin{array}{cc}
\mathbf 1& 0\\
0& -\mathbf 1
\end{array}
\right),
\end{equation} so that $\mathcal O^t\eta\mathcal O=\eta$. Although our $\eta$ in~\eqref{eq:eta-Omega} takes a different form compared to the standard $O(D,D)$ metric, the two are related by a similarity transformation. It turns out that in our case the natural expression to work with is the one written in the main text.}

Demanding that the Hamiltonian density $\mathcal H$ remains invariant under the above transformation implies the following transformation of the generalised metric
\begin{equation}
\tilde {\mathcal H}^{MN}=\mathcal O^M{}_P\mathcal H^{PQ}\mathcal O^N{}_Q.
\end{equation}
In fact, with these rules we have
\begin{equation}
    \mathcal H=\tfrac12 \Psi_M\mathcal H^{MN}\Psi_N=\tfrac12 \tilde \Psi_M\tilde {\mathcal H}^{MN}\tilde \Psi_N.
\end{equation}
Given the transformation of the generalised metric, we are tempted to reinterpret the transformation as a map at the level of the background fields $G_{mn},B_{mn}\to \tilde G_{mn},\tilde B_{mn}$ where, obviously, the transformed background fields are identified by the components of $\tilde {\mathcal H}^{MN}$, if we parameterise this as in~\eqref{eq:par-HMN} with tildes everywhere. However, in general we expect to run into trouble. The reason is that the background fields $G_{mn},B_{mn}$ can in general depend on the coordinates $x^m$. After the map $(p_m,x^m)\to (\tilde p_m,\tilde x^m)$ we may introduce a dependence of the background fields on the transformed momenta $\tilde p_m$. When that happens, we cannot reinterpret the Hamiltonian as coming from a $\sigma$-model action as the one written in~\eqref{eq:sigma-model}.

\vspace{12pt}

To avoid the above problem, we further restrict the scope of our construction. We do not consider \emph{generic} $\sigma$-models, rather only those that satisfy a certain condition that we are about to impose. In any case, we will see that this restricted class of models is still large enough and certainly very interesting. First, we separate the original coordinates as $x^m=\{\phi^i,\hat x^{ \mu}\}$, where the index $i=1,\ldots,d$ and the index $ \mu=d+1,\ldots,D$. In the double language we will have a similar splitting, using now the capital indices  $M=(I,\hat M)$. At this point we assume that the original background has a set of $d$ commuting isometries implemented as shifts of the coordinates $\phi^i$ as $\phi^i\to\phi^i+c^i, c^i\in \mathbb R$. When this is the case, the action depends only on the \emph{derivatives} of $\phi^i$, and  the background fields do not depend on $\phi^i$. Background fields can still depend on the $\hat x^{ \mu}$ coordinates. We can therefore repeat the above reasoning but assuming that $\mathcal O^M{}_N$ acts as the identity on the hatted indices $\hat M$, and it is non-trivial only on the $I$ indices. We are sure, then, that the corresponding transformation will generate an Hamiltonian that comes from a Lagrangian of the form~\eqref{eq:sigma-model}. In fact, the map acts as the identity (i.e.~nothing changes) on the ``problematic coordinates'' $\hat x^{ \mu},\hat p_{ \mu}$. The map acts non-trivially on the $\phi^i$ coordinates and the corresponding conjugate momenta, but by assumption the Hamiltonian density will depend on them only through the vector $\Psi_I$, while the generalised metric does not depend on them.

To see this more explicitly, let us rearrange the basis for $\Psi_M$ in a convenient way, in particular
\begin{equation}
\Psi_M=\left(\begin{array}{c}  x'{}^m \\ p_m\end{array}\right)\qquad\to\qquad
\Psi_M=\left(\begin{array}{c} \Psi_I\\ \Psi_{\hat M}\end{array}\right)=\left(\begin{array}{c} \phi'{}^i\\ p_i\\ \\ \hat x'{}^{ \mu}\\ \hat p_{ \mu}\end{array}\right),
\end{equation}
so that we can write a simple expression for $\mathcal O^M{}_N$ as
\begin{equation}
\mathcal O^M{}_N=\left(\begin{array}{cc}
\mathcal O^I{}_J& \mathsf 0\\
\mathsf 0& \delta^{\hat M}{}_{\hat N}
\end{array}
\right)\qquad\implies \qquad
\mathcal O=\left(\begin{array}{cc}
\mathsf O& \mathbf 0\\
\mathbf 0& \mathbf 1
\end{array}
\right). 
\end{equation}
The second equation is the index-free version of the first one, where $\mathsf O$ is the only non-trivial block. Notice that, therefore, we are not interested in the most generic elements of $O(D,D)$, rather only in an $O(d,d)$ subgroup.
In the same basis, we may decompose the generalised metric as
\begin{equation}
    \mathcal H^{MN}=\left(\begin{array}{cc}
\mathcal H^{IJ}& \mathcal H^{I\hat N}\\
\mathcal H^{\hat MJ}& \mathcal H^{\hat M\hat N}
\end{array}
\right)\qquad \implies\qquad
\mathcal H=\left(\begin{array}{cc}
\mathsf H& \mathsf h\\
\mathsf h^t& \hat{\mathsf H}
\end{array}
\right) ,
\end{equation}
where we have chosen some new names for the various blocks in the index-free notation.
Then the transformation of the generalised metric is
\begin{equation}
\tilde{\mathcal H}=\mathcal O\mathcal H\mathcal O^t=\left(\begin{array}{cc}
\mathsf O\mathsf H\mathsf O^t& \mathsf O\mathsf h\\
\mathsf h^t\mathsf O^t& \hat{\mathsf H}
\end{array}
\right)
 .
\end{equation}
By assumption, the original blocks $\mathsf H,\mathsf h,\hat{\mathsf H}$ do not depend on $\phi^i$, they can only depend on $\hat x^\mu$. After applying the map, some blocks are transformed by the multiplication by the matrix $\mathsf O$, which  is constant.

\subsection{A brief introduction to the $O(D,D)$ group}
Before continuing, we give a brief introduction to the $O(D,D)$ group.
Let us parameterise a generic element of $O(D,D)$ as
\begin{equation}\label{eq:Ocan}
   \mathcal O^M{}_{N}=\left(\begin{array}{cc}
         a_m{}^n & b_{mn} 
    \\
         c^{mn} &  d^m{}_n\end{array}\right),
\end{equation}
in terms of the $D\times D$ blocks $a,b,c,d$.
Given that $\eta_{MN}$ is the metric of $O(D,D)$, we will use it to lower and raise the double indices. In particular, notice that multiplication by $\eta$ from the left swaps the blocks horizontally as if they were two rows, while multiplication by $\eta$ from the right swaps the blocks vertically as if they were two columns. More explicitly, we have
\begin{equation}
\begin{aligned}
 &     \mathcal O^{MN}=\left(\begin{array}{cc}
         b_{mn} & a_m{}^n
    \\
        d^m{}_n & c^{mn}\end{array}\right),\qquad
        \mathcal O_{MN}=\left(\begin{array}{cc}
         c^{mn} &  d^m{}_n
    \\
         a_m{}^n & b_{mn} \end{array}\right),\\
&         \mathcal O_M{}^N=\left(\begin{array}{cc}
        d^m{}_n & c^{mn} \\
        b_{mn} & a_m{}^n
    \end{array}\right).
\end{aligned}
\end{equation}
Given that raising or lowering the indices changes the position of the blocks, if we want to have an index-free notation it is important to specify what is the ``canonical'' position of the indices that are being omitted. For $\mathcal O$, we will always take the convention that the indices are in the canonical position $\mathcal O^M{}_{N}$ as in~\eqref{eq:Ocan}. For the generalised metric $\mathcal H$ the canonical position is the one with both indices upper $\mathcal H^{MN}$ as in~\eqref{eq:Hcan}. Notice that $\eta$ agrees with its inverse, so that $\eta^{MN}=\eta_{MN}$, but $\eta_M{}^N=\delta_M{}^N$ and $\eta^M{}_N=\delta^M{}_N$.

It is easy to check that the condition $\mathcal O^t\eta \mathcal O=\eta$ implies the following quadratic conditions on the $D\times D$ blocks $a,b,c,d$
\begin{equation}\label{eq:ODD-const-bl1}
    c^ta+a^tc=\mathbf 0,\qquad d^tb+b^td=\mathbf 0,\qquad a^td+c^tb=\mathbf 1.
\end{equation}
Moreover, the $O(D,D)$ condition implies that the inverse of $\mathcal O$ is
\begin{equation}
    \mathcal O^{-1}=\eta^{-1}\mathcal O^{t}\eta=\left(\begin{array}{cc}
        d^t & c^t \\
        b^t & a^t
    \end{array}\right).
\end{equation}
Given that $O(D,D)$ is a group, the inverse of $\mathcal O$ also satisfies the $O(D,D)$ condition $(\mathcal O^{-1})^t\eta\mathcal O^{-1}=\eta$ which (taking into account that $\eta^{-1}=\eta$) implies $\mathcal O\eta\mathcal O^t=\eta$. Therefore, we also have the conditions
\begin{equation}\label{eq:ODD-const-bl2}
    ab^t+ba^t=\mathbf 0,\qquad cd^t+dc^t=\mathbf 0,\qquad ad^t+bc^t=\mathbf 1.
\end{equation}
These latter conditions are not independent from the previous ones, but it turns out that sometimes it is useful to have them written explicitly.

\vspace{12pt}

The $O(D,D)$ group is generated by the elements
\begin{equation}\label{eq:RST}
    \mathcal R=\left(\begin{array}{cc}
       \rho & \mathbf 0 \\
        \mathbf 0 & \rho^{-t}
    \end{array}\right),\qquad
        \mathcal S=\left(\begin{array}{cc}
       \mathbf 1 & s \\
        \mathbf 0 & \mathbf 1
    \end{array}\right),\qquad
    \mathcal T_{(p)}=\left(\begin{array}{cc}
       \mathbf 1-u_{(p)} & -u_{(p)} \\
       - u_{(p)}& \mathbf 1-u_{(p)}
    \end{array}\right),
\end{equation}
where $\rho\in GL(D)$ (so that we can define its inverse), $s$ is antisymmetric $(s^t=-s)$, and $u_{(p)}=\text{diag}(0,\ldots,0,1,0,\ldots,0)$ where the $1$ is at position $p$. In other words, the statement is that a generic element of $O(D,D)$ can be written as a chain of products of the above elements.
Notice that matrices of the type $\mathcal R$ form a subgroup of $O(D,D)$ containing the identity. Because of their upper-triangular form, matrices of $\mathcal S$-type form another subgroup, that also contains the identity. Finally, one can check that $\mathcal T_{(p)}^2=\mathbf 1$. See also exercise~\ref{ex:simple} of section~\ref{sec:ex-Odd}.

\subsection{Action of $O(D,D)$ on the background fields}\label{sec:actionOdd}
Let us now see what is the action of an $O(D,D)$ transformation on the background fields. In other words, we want to know how the metric and $B$-field are transformed under an $O(D,D)$ transformation of the generalised metric. In general, the formulas for a generic $O(D,D)$ transformation of $(G,B)\to (\tilde G,\tilde B)$ may be quite complicated, as it can be expected from the parameterisation of the generalised metric in terms of the background fields and by its transformation rule $\tilde {\mathcal H}=\mathcal O\mathcal H\mathcal O^t$. However, it turns out that a nice formula can be written for the transformation of $E_{mn}=G_{mn}+B_{mn}$, just the sum of the metric and the $B$-field. In fact, if we parameterise the $O(D,D)$ transformation in terms of the  $D\times D$ blocks $a,b,c,d$ as done in~\eqref{eq:Ocan}, then the transformation rules are\footnote{Obviously, here we are assuming that the matrix $cE+d$ is invertible. All the $O(D,D)$ transformations that we will consider satisfy this property.}
\begin{equation}\label{eq:tr-E}
    \tilde E=(aE+b)(cE+d)^{-1}.
\end{equation}
The verification of this statement is proposed as the guided exercise~\ref{ex:transf-E} of section~\ref{sec:ex-Odd}.

\vspace{12pt}

It is particularly interesting to look at the transformation rules induced by the generators $\mathcal R,\mathcal S,\mathcal T_{(p)}$. One may just apply the above formula, but it is nice to do this also in the following alternative way. First, notice that the parameterisation of the generalised metric in terms of $G,B$ implies that we can write it in the ``factorised form''
\begin{equation}
    \mathcal H=\left(\begin{array}{cc}
       \mathbf 1 & B \\
        \mathbf 0 & \mathbf 1
    \end{array}\right)\left(\begin{array}{cc}
       G & \mathbf 0 \\
        \mathbf 0 & G^{-1}
    \end{array}\right)\left(\begin{array}{cc}
       \mathbf 1 & \mathbf 0 \\
        -B & \mathbf 1
    \end{array}\right)=\mathcal B\mathcal G\mathcal B^t,
\end{equation}
where
\begin{equation}
    \mathcal B=\left(\begin{array}{cc}
       \mathbf 1 & B \\
        \mathbf 0 & \mathbf 1
    \end{array}\right)
    ,\qquad\mathcal G=\left(\begin{array}{cc}
       G & \mathbf 0 \\
        \mathbf 0 & G^{-1}
    \end{array}\right).
\end{equation}
We can understand this as the $O(D,D)$ matrix $\mathcal G$ (which is in fact of the type $\mathcal R$ with $\rho =G$) ``dressed'' by the $O(D,D)$ transformation by $\mathcal B$ (which is in fact of the type $\mathcal S$ with $s=B$).

Thanks to the above observation, and taking into account that the action of $\mathcal R$ does not change the block-diagonal or upper/lower-triangular forms of the above matrices, we can write
\begin{equation}
    \tilde {\mathcal H }= \mathcal R\mathcal H\mathcal R^t
    =(\mathcal R\mathcal B\mathcal R^{-1})(\mathcal R\mathcal G\mathcal R^t)(\mathcal R\mathcal B\mathcal R^{-1})^t.
\end{equation}
It is easy to check that this implies the transformation rules
\begin{equation}
    \tilde G=\rho G\rho^t,\qquad\qquad \tilde B=\rho B\rho^t.
\end{equation}
We recognise the transformation of the metric and Kalb-Ramond field under a coordinate transformation $x^m=\tilde x^n\rho_n{}^m$, with $\rho$ constant. This makes sense, because implementing this transformation on the double vector $ \Psi_M=\tilde \Psi_N\mathcal R^N{}_M$ we get $x'{}^m=\tilde x'{}^n\rho_n{}^m$ and $p_m=(\rho^{-1})_m{}^n\tilde p_n$.\footnote{Remember that in this presentation we want to discuss the $O(D,D)$ group, but eventually we will be interested only in an $O(d,d)$ subgroup, so that $\rho$ is of block form and is the identity for the $\mu$ indices. Similar comments apply also for the transformations that follow.}

Similarly, if we are interested to know the action of $\mathcal S$ we can write
\begin{equation}
    \tilde {\mathcal H}=\mathcal S\mathcal H\mathcal S^t=(\mathcal S\mathcal B)\mathcal G(\mathcal S\mathcal B)^t.
\end{equation}
In other words, we have the transformation rules
\begin{equation}
    \tilde G=G,\qquad \qquad \tilde B=B+s.
\end{equation}
We see that the only effect is that of shifting the $B$-field by the constant $s$. As a transformation of the phase-space variables, from $ \Psi_M=\tilde \Psi_N\mathcal S^N{}_M$ we read off
$x'{}^m=\tilde x'{}^m$ and $p_m=\tilde p_m+s_{mn}x'{}^n$.

To see the action of $\mathcal T_{(p)}$ let us just set $D=1$ and therefore consider the simplest possible case of $O(1,1)$. In this case we have only one such matrix
\begin{equation}
    \mathcal T=\left(\begin{array}{cc}
       0& -1 \\
       - 1 & 0
    \end{array}\right).
\end{equation}
If $D=1$ there is no $B$-field and the generalised metric is a $2\times 2$ matrix
\begin{equation}
    \mathcal H=\left(\begin{array}{cc}
       g& 0 \\
        0 & g^{-1}
    \end{array}\right).
\end{equation}
It is immediate to see that 
\begin{equation}
    \tilde{\mathcal H}=\mathcal T\mathcal H\mathcal T^t=\left(\begin{array}{cc}
       g^{-1}& 0 \\
        0 & g
    \end{array}\right).
\end{equation}
The transformation gives $g\to g^{-1}$, so that it implements the Buscher rules of T-duality in the case of $D=1$. Notice that as a transformation of the phase-space variables this T-duality map is in fact swapping the notion of (derivatives of) fields and conjugate momenta $x'=-\tilde p, p=-\tilde x'$. Checking that the Buscher rules are reproduced in the generic case $D\neq 1$ is left as the exercise~\ref{ex:Buscher-Odd} of section~\ref{sec:ex-Odd}. 

\vspace{12pt}

Having in mind the possible canonical transformations, we see that there is one still missing, namely
\begin{equation}
    x'{}^m=\tilde x'{}^m+\beta^{mn}\tilde p_n,\qquad \qquad p_m=\tilde p_m,
\end{equation}
with $\beta^{mn}=-\beta^{nm}$. One may directly check that this transformation is canonical, but an alternative way is to write it as a $2D\times 2D$ matrix and check that it is an element of $O(D,D)$. In fact, if we write $ \Psi_M=\tilde \Psi_N \mathcal U^N{}_M$ with 
\begin{equation}\label{eq:def-U}
    \mathcal U=\left(\begin{array}{cc}
       \mathbf 1&  \mathbf 0 \\
       \beta & \mathbf 1
    \end{array}\right),
\end{equation}
checking  that this is an $O(D,D)$ matrix is essentially the same calculation as the one for $\mathcal S$. Notice that this matrix is in fact equal to
\begin{equation}
    \mathcal U=\mathcal T\mathcal S\mathcal T,
\end{equation}
if we take $s=\beta$ and where we defined $\mathcal T=\mathcal T_{(1)}\mathcal T_{(2)}\cdots \mathcal T_{(D)}$,~i.e. we dualise all directions.

Notice that the composition of two ``$\beta$-deformations'' $\mathcal U_1\mathcal U_2$, say the first with $\beta_1$ and the second with $\beta_2$, is equivalent to a $\beta$-deformation with $\beta=\beta_1+\beta_2$. This is immediate to see if we consider the double matrices $\mathcal U$, but it can also be proved by looking at the formula for the transformation of $E$, as suggested in  exercise~\ref{ex:comp-TsT}  of  section~\ref{sec:ex-Odd}. Therefore, $\beta$-deformations form an abelian subgroup pf $O(D,D)$. 

Let us point out that we can always implement the canonical transformations induced by $\mathcal R,\mathcal S$. However, we see that we can implement those induced by $\mathcal T,\mathcal U$ only if the background is independent of the coordinates  on which the transformations are acting non-trivially. In those cases, therefore, one has to restrict to transformations to the indices $i,j$ and $I,J$ introduced earlier.

The $\beta$-deformations generated by $\mathcal U$ may be understood as sequences of TsT deformations, see the exercise~\ref{ex:TsT-Odd}  of section~\ref{sec:ex-Odd}. Although we do not show it explicitly, the deformations generated by $\mathcal{U}$ can also be recast as in the ``alternative construction'' of section~\ref{sec:TsT-alt}.  As we will see in the next section, the $\beta$-deformations are  the interesting $O(D,D)$ transformations to consider in order to describe the current-current deformations. 

\subsection{The dilaton and the Weyl invariance condition}\label{sec:dilaton}
The above discussion is enough if we are only interested in the \emph{classical} $\sigma$-model. Often, however, we want to impose more structure. An important example is when we want the $\sigma$-model to describe the dynamics of a string in a curved background, so that it must be described by 2-dimensional conformal field theory. Generic couplings $G,B$ will not necessarily give rise to a conformal theory on the worldsheet. At one loop, one finds that the scale invariance condition (i.e.~the conditions that the $\beta$-functions are 0) and the Weyl invariance condition (i.e.~the condition that the trace of the 2-dimensional stress-energy tensor is 0) imply 
\begin{equation}\label{eq:1loop-conf}
\begin{aligned}
    R_{mn}-\frac14 H_{mpq}H_n{}^{pq}+2\nabla_{(m}\nabla_{n)}\phi=0,\\
    \nabla^pH_{pmn}-2\nabla^p\phi H_{pmn}=0,\\
    R-\frac{1}{12}H^2+4\nabla^m\nabla_m\phi-4\partial_m\phi\partial^m\phi=0,
    \end{aligned}
\end{equation}
where $R_{mn},R$ are the Ricci tensor and Ricci scalar, $H_{mnp}=3\partial_{[m}B_{np]}$ is the field strength of the $B$-field,  $H^2=H_{mnp}H^{mnp}$ and $\nabla_m$ the covariant derivative. The above conditions involve a total of three fields: the metric $G$, the Kalb-Ramond field $B$ and the dilaton $\phi$. From the point of view of the $\sigma$-model action, the dilaton appears because it couples to the Ricci scalar of the 2-dimensional worldsheet metric at order $\hbar$ (order $\alpha'$ in the string-theory language).
The above equations may be derived also as equations of motion of a $D$-dimensional effective action.

The $O(D,D)$ transformations that we are considering will preserve the above one-loop conformal invariance conditions only if, together with the rules for $G$ and $B$, we assign the correct transformation rule also for the dilaton $\phi$. In other words, given a background $(G,B,\phi)$ that solves~\eqref{eq:1loop-conf}, we want to know what is the map $(G,B,\phi)\to (\tilde G,\tilde B,\tilde \phi)$ such that $(\tilde G,\tilde B,\tilde \phi)$ also solve ~\eqref{eq:1loop-conf}. The answer is given by demanding that the following combination is invariant under $O(D,D)$ transformations
\begin{equation}
\sqrt{\det G}e^{-2\phi}=\sqrt{\det \tilde G}e^{-2\tilde \phi},
\end{equation}
from which the transformation rules for $\tilde \phi$ may be derived.
The equations~\eqref{eq:1loop-conf} may be also obtained as the equations of motion of a ``doubled model'' living in $2D$ dimensions, the so-called Double Field Theory. The fundamental fields entering the action of Double Field Theory are the generalised metric $\mathcal H^{MN}$ and the generalised dilaton $d=\phi-\tfrac14 \det G$, which is in fact the $O(D,D)$ invariant combination written above.

\subsection{Exercises}\label{sec:ex-Odd}
\begin{enumerate}
\item \label{ex:simple}  \textbf{Simple checks ---} Check that the  matrices $\mathcal R,\mathcal S,\mathcal T_{(p)}, \mathcal U$ defined in this section are elements of $O(D,D)$.
   
\item \label{ex:transf-E} \textbf{The transformation of $E$ ---} Prove the transformation rule~\eqref{eq:tr-E} for $E$ under a generic $O(D,D)$ transformation. To do that, you may follow the following steps:
    \begin{enumerate}
   
    \item Start from $\tilde E=(aE+b)(cE+d)^{-1}$ and  compute $\tilde G=\tfrac12(\tilde E+\tilde E^t)$. The expressions that you will find suggest to write it in the form $\tilde G=M^{-1}NM^{-T}$ for matrices $M,N$ that you will find. Inspecting $N$ one can see that the $O(D,D)$ constraints~\eqref{eq:ODD-const-bl1} and~\eqref{eq:ODD-const-bl2} can be used to simplify the expression. Eventually, one finds
    \begin{equation}\label{eq:tr-G}
        \tilde G=(E^tc^t+d^t)^{-1}G(cE+d)^{-1}.
    \end{equation}
    \item Now use~\eqref{eq:tr-E} to compute $\tilde B=\tfrac12(\tilde E-\tilde E^t)$. The expressions that you will find suggest to write it again in the form $\tilde B=M^{-1}NM^{-T}$ for new matrices $M,N$ that you will find. Inspecting $N$ one can see that the $O(D,D)$ constraints~\eqref{eq:ODD-const-bl1} and~\eqref{eq:ODD-const-bl2} can be used to simplify the expression. Continuing to simplify the expression, eventually one finds
    \begin{equation}\label{eq:tr-B}
        \tilde B=[b+aE-(E^tc^t+d^t)^{-1}G](cE+d)^{-1}
    \end{equation}
        \item Now compute $\tilde{\mathcal H}=\mathcal O\mathcal H\mathcal O^t$ in block-form, using the blocks, $a,b,c,d$ to parameterise $\mathcal O$.
        Identify $\tilde G^{-1}$ by looking at the lower-right block, and $\tilde B\tilde G^{-1}$ by looking at the upper-right block.
        \item Compute $\tilde G^{-1}$  and $\tilde B\tilde G^{-1}$ from~\eqref{eq:tr-G} and~\eqref{eq:tr-B} and compare to the other expressions that you have found. Congratulations, you have checked~\eqref{eq:tr-E}.
         \item As a bonus, you may also work out equivalent formulas. Start from $\tilde E=(aE+b)(cE+d)^{-1}$ and argue that it is equivalent to \begin{equation}\label{eq:alt-tr-E}
        \tilde E=(Eb^t-d^t)^{-1}(c^t-Ea^t).
    \end{equation}
    This can be proved by considering the inverse transformation and by knowing how to write $\mathcal O^{-1}$ in block form.
    \end{enumerate}
   
\item \label{ex:Buscher-Odd} \textbf{Buscher rules ---} Use~\eqref{eq:tr-E} to check the Buscher rules given in~\eqref{eq:Buscher} when using the approriate $O(D,D)$ matrix to implement the T-duality transformation.
\item \label{ex:TsT-Odd} \textbf{TsT deformations ---} Use~\eqref{eq:tr-E} to check that you can reproduce the rules for the TsT deformation given in~\eqref{eq:TsT}, if you take $\beta$ such that $\beta^{12}=-\beta^{21}=-\lambda$.
\item \label{ex:comp-TsT} \textbf{Composition of TsT deformations ---} Use~\eqref{eq:tr-E} to check that a $\beta$-deformation with $\beta=\beta_1$ followed by another $\beta$-deformation with $\beta=\beta_2$ is equivalent to a $\beta$-deformation with $\beta=\beta_1+\beta_2$.
\end{enumerate}

\section{Finite current-current deformations from $O(d,d)$}\label{sec:finite}
The main claim that we will prove in this section is that the \emph{finite} version of an \emph{infinitesimal} current-current deformation is given by a $\beta$-deformation of $O(d,d)$. This is general, in particular there is no assumption regarding the conformal invariance nor the integrability of the original $\sigma$-model. Later we will also discuss what happens in the special case when the isometries are chiral.

The fact that $O(d,d)$ transformations can be used to ``integrate'' and construct a finite version of current-current deformations was first argued by Hassan and Sen in~\cite{Hassan:1992gi} by looking at concrete examples of Wess-Zumino-Witten models (as well as gauged versions of them) based on the $SU(2)$ group, and by working out explicit $O(2,2)$ and $O(3,3)$ transformations in concrete examples. The generic $O(d,d)$ case was left as a conjecture.
Shortly after,  in~\cite{Henningson:1992rn}, Henningson and Nappi proved the conjecture for generic $O(d,d)$ by considering a generic $\sigma$-model with chiral isometries. 
On the one hand, the proof of~\cite{Henningson:1992rn} is quite involved, possibly because of the choice for the  $O(d,d)$ parameterisation. On the other hand, it is not necessary, as done in~\cite{Henningson:1992rn}, to restrict to the case of $\sigma$-models with \emph{chiral} isometries: current-current deformations may be constructed even for $\sigma$-models with isometries that do not have a definite chirality, and $O(d,d)$ is still the answer to construct their finite version. For these reasons, in these lecture notes we construct an alternative proof that from our point of view is  simpler. In particular, we will use a nice representation of the non-geometric $O(d,d)$ transformation. Moreover, we first work out the generic case, without assuming the chirality of the isometries, and later discuss the extra features that appear in the presence of chiral symmetries.

\subsection{Deformations of $\sigma$-models with $d$ commuting isometries}
Let us start from a $\sigma$-model with the usual action
\begin{equation}
    S=-\int d^2\sigma\, \Pi^{\alpha\beta}_{(-)}\, \partial_\alpha x^m\partial_\beta x^n\, E_{mn}=-\int d^2\sigma\, \partial_+x^m\partial_-x^n\, E_{mn},
\end{equation}
where $E=G+B$.
Let's assume now that the action has a total of $d$-commuting isometries. We make sure to choose the coordinate system  so that these isometries are implemented simply as shifts of $d$ coordinates. In particular, as we did before, we split the total $D$ coordinates as 
\begin{equation}
    x^m=\{\phi^i,\hat x^\mu\},
    \qquad\qquad i=1,\ldots,d,\qquad \mu=1,\ldots,D-d.
\end{equation}
For convenience, let us rewrite the background field $E_{mn}$ in block form in terms of the $i,\mu$ indices as 
\begin{equation}
    E_{mn}=\left(\begin{array}{cc}
         E_{ij} & E_{i\nu}  \\
         E_{\mu j}& E_{\mu\nu}
    \end{array}\right)=\left(\begin{array}{cc}
         \mathsf E_{ij} & \mathsf F^R_{i\nu}  \\
         \mathsf F^L_{\mu j}& \hat {\mathsf E}_{\mu\nu}
    \end{array}\right).
\end{equation}
In the second equality we have assigned new names to the various blocks, which will be useful when introducing the index-free notation.
Obviously, the dimensions of the blocks in this decomposition  are
\begin{equation}\label{eq:dec-d-D-d}
    \left(\begin{array}{cc}
         d\times d & d\times (D-d)\\
         (D-d)\times d&(D-d)\times (D-d)
    \end{array}\right).
\end{equation}
The original action written in terms of this block decomposition of $E$ and with the index-free notation reads as
\begin{equation}\label{eq:S-d-iso}
    S=-\int d^2\sigma\left(\partial_+\phi^t \, \mathsf E\, \partial_-\phi+\partial_+\hat x^t \, \mathsf F^L\, \partial_-\phi+\partial_+\phi^t \, \mathsf F^R\, \partial_-\hat x+\partial_+\hat x^t\, \hat{ \mathsf E}\, \partial_-\hat x\right).
\end{equation}
By assumption, the action is invariant under the shifts $\phi^i\to \phi^i+c^i$, which implies a total of $d$ Noether currents $J_{i\alpha}$. In the light-cone worldsheet coordinates we may write them as
\begin{equation}\label{eq:Noether-indices}
  \begin{aligned}
        J_{i+}&=\mathsf E_{ji}\, \partial_+\phi^j+\mathsf F^L_{\mu i}\,  \partial_+\hat x^\mu,\\
        J_{i-}&=\mathsf E_{ij}\, \partial_-\phi^j+\mathsf F^R_{i\mu} \, \partial_-\hat x^\mu.
  \end{aligned}
\end{equation}
The above formulas may be obtained by the direct computation of the Noether currents or by using the known formula
\begin{equation}\label{eq:Noether-generic}
    J_{i\pm}=k_i^m(G_{mn}\mp B_{mn})\partial_\pm x^n\pm \omega_{in}\partial_\pm x^n,
\end{equation}
where $k_i^m$ are Killing vectors, so that $\mathcal L_{k_i}G=0$, and to remain general it is assumed that the $B$-field changes by an exact form under the Lie-derivative $\mathcal L_{k_i}B=d\omega_{i}$. In our case we may take $k_i^m=\delta_i^m$ and $\omega=0$. In the index-free notation, the formulas for the Noether currents would read
\begin{equation}
  \begin{aligned}
        J_{+}&=\partial_+\phi^t \, \mathsf E+\partial_+\hat x^t\,  \mathsf F^L,\\
        J_{-}&=\mathsf E\, \partial_-\phi+\mathsf F^R\, \partial_-\hat x,
  \end{aligned}
\end{equation}
where in the first line we multiply row vectors by matrices from the right, while in the second line we multiply column vectors by matrices from the left.

Thanks to the presence of these $d$ commuting isometries, we can implement a generic $O(d,d)\subset O(D,D)$ transformation of the background. Because of the discussion of the previous section, the $O(d,d)$ transformations that are \emph{not} interesting are those generated by $\mathcal R$ and $\mathcal S$, see~\eqref{eq:RST}. In fact, the first would correspond to a linear coordinate transformation on the $\phi^i$, while the second would simply shift $B_{ij}$ by a constant. The collection of  these transformations close into a subgroup that we call $\Lambda(d)$. We are therefore interested only in the coset $O(d,d)/ \Lambda(d)$. Taking into account that the $O(d,d)$ group has dimension $2d(2d-1)/2$, while the dimension of $\Lambda(d)$ is given by the sum of the number of coefficients needed to parameterise $\mathcal R$ and $\mathcal S$, namely $d^2+d(d-1)/2$, we see that the coset has dimension $d(d-1)/2$. This is in fact the number of parameters that enter the $d\times d$ antisymmetric matrix $\beta$ that defines the $\mathcal U$ transformation in~\eqref{eq:def-U}. Therefore, we may parameterise the component of the coset that is connected to the identity by the $\beta$-deformations. The T-duality transformations generated by $\mathcal T_{(p)}$ defined in~\eqref{eq:RST} are not part of the component connected to the identity.

\vspace{12pt}

Our first task is to identify how the background fields are transformed by the $\beta$-deformations, in other words what is the explicit expression for $\tilde E_{mn}$ in this case. To do that, we use the formula $\tilde E=(aE+b)(cE+d)^{-1}$ already given in~\eqref{eq:tr-E} after specifying what the blocks $a,b,c,d$ are in this case. It is clear that each of these $D\times D$ blocks can be written in a further block form because of the split of the $D=d+(D-d)$ indices. In particular, the transformation only acts non-trivially on the $i,j$ indices, while it is the identity on the $\mu,\nu$ indices and it does not mix indices of different type. Therefore, if we write the $D\times D$ blocks in the form of~\eqref{eq:dec-d-D-d} then we find
\begin{equation}
    \begin{aligned}
  &      a=\left(\begin{array}{cc}
       \mathbf 1 &  \mathbf 0\\
       \mathbf 0 & \mathbf 1
    \end{array}\right),\qquad
   && b=\left(\begin{array}{cc}
       \mathbf 0  &  \mathbf 0\\
       \mathbf 0 & \mathbf 0
    \end{array}\right),\\
  &      c=\left(\begin{array}{cc}
       \beta  &  \mathbf 0\\
       \mathbf 0  & \mathbf 0
    \end{array}\right),\qquad
   && d=\left(\begin{array}{cc}
       \mathbf 1 &  \mathbf 0\\
       \mathbf 0  & \mathbf 1
    \end{array}\right).
    \end{aligned}
\end{equation}
The formula for $\tilde E$ then reduces to $\tilde E=E(1+cE)^{-1}$ with $c$ given above. In particular, we have
\begin{equation}
    1+cE =\left(\begin{array}{cc}
      \mathbf 1+\beta \mathsf E  & \beta \mathsf F^R\\
       \mathbf 0  & \mathbf 1
    \end{array}\right),
\end{equation}
so that its inverse is easily calculated
\begin{equation}
  (  1+cE)^{-1} =\left(\begin{array}{cc}
      (\mathbf 1+\beta \mathsf E  )^{-1}& -(\mathbf 1+\beta \mathsf E  )^{-1}\beta \mathsf F^R\\
       \mathbf 0  & \mathbf 1
    \end{array}\right).
\end{equation}
Computing the expression for $\tilde E$ now amounts to just a simple matrix multiplication
\begin{equation}
\begin{aligned}
    \tilde E&=\left(\begin{array}{cc}
         \tilde{\mathsf E}&\tilde{ \mathsf F}^R  \\
         \tilde{\mathsf F}^L& \hat {\tilde{\mathsf E}}
    \end{array}\right)=E(1+cE)^{-1}\\
    &=\left(\begin{array}{cc}
         \mathsf E& \mathsf F^R  \\
         \mathsf F^L& \hat {\mathsf E}
    \end{array}\right)
    \left(\begin{array}{cc}
      (\mathbf 1+\beta \mathsf E  )^{-1}& -(\mathbf 1+\beta \mathsf E  )^{-1}\beta \mathsf F^R\\
       \mathbf 0  & \mathbf 1
    \end{array}\right),
\end{aligned}
\end{equation}
which allows us to identify
\begin{equation}
\begin{aligned}
    &     \tilde{\mathsf E}=\mathsf E(1+\beta \mathsf E)^{-1},\qquad
    && \tilde{ \mathsf F}^R  =[\mathbf 1-\mathsf E(1+\beta\mathsf E)^{-1}\beta]\mathsf F^R,\\ 
     &    \tilde{\mathsf F}^L=\mathsf F^L(\mathbf 1+\beta \mathsf E)^{-1},\qquad && \hat {\tilde{\mathsf E}}=\hat{\mathsf E}-\mathsf F^L(1+\beta\mathsf E)^{-1}\beta \mathsf F^R.
\end{aligned}
\end{equation}
At this point we have generated a $\beta$-deformed $\sigma$-model with background field $\tilde E_{mn}$, whose blocks are given above. Can we interpret the deformation as the ``integration'' of an infinitesimal current-current deformation? The answer is yes, and checking this is easy thanks to the intermediate results that we have already worked out. 

\vspace{12pt}

The action of the deformed model is still of the form of equation~\eqref{eq:S-d-iso}, simply by adding tildes everywhere. In particular, the $d$ isometries survive the deformation. It is therefore possible to construct an infinitesimal current-current deformation of the deformed model. The computation of the  Noether currents is formally the one of the undeformed case, it is simply enough to add tildes everywhere\footnote{In Section~\ref{sec:free} a tilde on the currents was used to denote the Hodge dual, while here the tilde refers to any object related to the transformed background. We are sure that this will not create confusion.}
\begin{equation}
  \begin{aligned}
       \tilde J_{+}&=\partial_+\phi \, \tilde{\mathsf E}+\partial_+\hat x\,  \tilde{\mathsf F}^L,\\
        \tilde J_{-}&=\tilde{\mathsf E}\, \partial_-\phi+\tilde{\mathsf F}^R\, \partial_-\hat x.
  \end{aligned}
\end{equation}
At this point we consider the deformed background and study the effect of an infinitesimal deformation of $\beta\to \beta+\delta \beta$. Using the obvious formula $\delta M^{-1}=-M^{-1}\delta M M^{-1}$ where $M$ is an invertible matrix, we find
\begin{equation}
\begin{aligned}
    &    \delta \tilde{\mathsf E}=-\mathsf E(1+\beta \mathsf E)^{-1}\, \delta\beta\, \mathsf E(1+\beta \mathsf E)^{-1}=-\tilde{\mathsf E}\, \delta\beta\, \tilde{\mathsf E},\\
    & \delta\tilde{ \mathsf F}^R  =-\mathsf E(1+\beta \mathsf E)^{-1}\, \delta\beta\, [\mathbf 1-\mathsf E(1+\beta\mathsf E)^{-1}\beta]\mathsf F^R=-\tilde{\mathsf E}\, \delta\beta\, \tilde{ \mathsf F}^R,\\ 
     &   \delta \tilde{\mathsf F}^L=-\mathsf F^L(\mathbf 1+\beta \mathsf E)^{-1}\, \delta\beta\, \mathsf E(1+\beta \mathsf E)^{-1}= - \tilde{\mathsf F}^L\, \delta\beta\, \tilde{\mathsf E},\\ 
     & \delta \hat {\tilde{\mathsf E}}=-\mathsf F^L(1+\beta\mathsf E)^{-1}\, \delta\beta\, [1-\mathsf E(1+\beta\mathsf E)^{-1}\beta] \mathsf F^R=-\tilde{\mathsf F}^L\, \delta\beta\, \tilde{ \mathsf F}^R.
\end{aligned}
\end{equation}
Therefore the infinitesimal deformation of the deformed action is
\begin{equation}
\begin{aligned}
   \delta\tilde S&=-\int d^2\sigma\left(\partial_+\phi^t \,\delta \tilde{ \mathsf E}\, \partial_-\phi+\partial_+\hat x^t \, \delta \tilde{ \mathsf F}^L\, \partial_-\phi+\partial_+\phi^t \, \delta \tilde{ \mathsf F}^R\, \partial_-\hat x+\partial_+\hat x^t\, \delta \hat{ \tilde{ \mathsf E}}\, \partial_-\hat x\right),\\
   &=\int d^2\sigma\Big(\partial_+\phi ^t\,\tilde{\mathsf E}\, \delta\beta\, \tilde{\mathsf E}\, \partial_-\phi+\partial_+\hat x^t \, \tilde{\mathsf F}^L\, \delta\beta\, \tilde{\mathsf E}\, \partial_-\phi\\
   &\qquad\qquad+\partial_+\phi^t \, \tilde{\mathsf E}\, \delta\beta\, \tilde{ \mathsf F}^R\, \partial_-\hat x+\partial_+\hat x^t\, \tilde{\mathsf F}^L\, \delta\beta\, \tilde{ \mathsf F}^R\, \partial_-\hat x\Big),
\end{aligned}
\end{equation}
but it is immediate to see that this is precisely the infinitesimal variation that we get if we perturb the action by the current-current bilinear
\begin{equation}
    \delta\tilde S=\int d^2\sigma\, \tilde J_+^t\, \delta\beta\, \tilde J_-=\int d^2\sigma\, \delta\beta^{ij}\,  \tilde J_{+i}\tilde J_{-j},
\end{equation}
where in the second step we have reinstated the indices that previously were omitted. We have therefore proved that the $\beta$-deformations are the ``integrated'' version of the infinitesimal current-current deformations.

\subsection{Deformations of $\sigma$-models with chiral isometries}\label{sec:chiral}
Let us now consider a $\sigma$-model with $d$ commuting isometries that we further split into $d_L$ chiral and $d_R$ antichiral isometries, so that $d_L+d_R=d$. If we split our coordinates as $\phi^i=\{\phi_L^{a},\phi_R^{\bar{a}}\}$ with $a=1,\ldots,d_L,\ \bar{a}=1,\ldots,d_R$, then  this means that the action is invariant under the transformations
\begin{equation}\label{eq:chiral-tr}
    \phi_L^a\to \phi_L^a+f_L^a(z),\qquad\qquad
    \phi_R^{\bar a}\to \phi_R^{\bar a}+f_R^{\bar a}(\bar z).
\end{equation}
In particular, the action is not only invariant under \emph{constant} shifts, it is invariant also when shifting by the chiral functions $f^a_L(z)$ and the antichiral functions $f^{\bar a}_R(\bar z)$.
As done before, from now on we will omit the explicit indices.
One can check that the $\sigma$-model action can be recast in the following form\footnote{As mentioned in section~\ref{sec:prel}, in this context it is natural to use the complex coordinates $z,\bar z$, but we continue to use the previous notation for the derivatives, so that $\partial_+\sim \partial_z$ and $\partial_-\sim \partial_{\bar z}$.}
\begin{equation}\label{eq:S-chiral}
\begin{aligned}
    S=-\int d^2\sigma\, \Big(&\partial_+\phi_L^t\partial_-\phi_L+\partial_+\phi_R^t\partial_-\phi_R +\partial_+\phi_R^t\, \mathsf e\, \partial_-\phi_L \\
    &+\partial_+\phi_R^t\mathsf G^R\partial_-\hat x+\partial_+\hat x^t \mathsf G^L\partial_-\phi_L+\partial_+\hat x ^t\hat{\mathsf E}\partial_-\hat x\Big).
\end{aligned}
\end{equation}
The verification that an action invariant under chiral isometries can be put into this form is left as the exercise~\ref{ex:action-chiral}  of section~\ref{sec:ex-def-act}.
In particular, matching the previous notation and identifying the $LL, LR, RL, RR$ blocks, we can set
\begin{equation}
    \mathsf E=\left(\begin{array}{cc}
       \mathbf 1_L  & \mathbf 0 \\
         \mathsf e & \mathbf 1_R
    \end{array}\right),\qquad
     \mathsf F^L=\left(\begin{array}{cc}
       \mathsf G^L  & \mathbf 0 
    \end{array}\right),\qquad
     \mathsf F^R=\left(\begin{array}{c}
        \mathbf 0 \\ \mathsf G^R
    \end{array}\right).
\end{equation}
When applying a $\beta$-deformation, we can  decompose also the $\beta$-matrix in terms of the $L/R$ blocks
\begin{equation}
    \beta^{ij}=\left(\begin{array}{cc}
        \beta_{LL}^{ab} &  \beta_{LR}^{a\bar b}\\
         \beta_{RL}^{\bar ab}&\beta_{RR}^{\bar a\bar b} 
    \end{array}\right),
\end{equation}
with the conditions $\beta_{LL}^{ab}=-\beta_{LL}^{ba}$, $\beta_{RR}^{\bar a\bar b}=-\beta_{RR}^{\bar b\bar a}$ and $\beta_{RL}^{\bar ab}=-\beta_{LR}^{b\bar a}$ to ensure antisymmetry of $\beta$. As remarked in Section~\ref{sec:actionOdd}, a $\beta$ that is given by the sum of other $\beta$'s can be understood as the composition of the various $\beta$-deformations. Therefore, we can individually switch on  $\beta_{LL}$ or $\beta_{RR}$ or $\beta_{LR}$ and study what their effects.

\subsubsection{Trivial deformations}
Let us start with $\beta_{LL}$. We are interested in knowing the transformations of the background fields. We therefore have to compute
\begin{equation}
    \begin{aligned}
        (1+\beta \mathsf E)^{-1}&=\left[\left(\begin{array}{cc}
           \mathbf 1  & \mathbf 0 \\
             \mathbf 0 & \mathbf 1
        \end{array}\right)+
        \left(\begin{array}{cc}
           \beta_{LL}  & \mathbf 0 \\
             \mathbf 0 & \mathbf 0
        \end{array}\right)\left(\begin{array}{cc}
           \mathbf 1  & \mathbf 0 \\
             \mathsf e & \mathbf 1
        \end{array}\right)\right]^{-1}\\
        &=
        \left(\begin{array}{cc}
           ( 1+\beta_{LL})^{-1}  & \mathbf 0 \\
             \mathbf 0 & \mathbf 1
        \end{array}\right).
    \end{aligned}
\end{equation}
Notice that, because of the form of $\mathsf E$ and of the choice for $\beta$, the above matrix is \emph{constant}. Computing the transformations of the background fields we find
\begin{equation}
    \begin{aligned}
   &     \tilde{\mathsf E}=\left(\begin{array}{cc}
           ( 1+\beta_{LL})^{-1}  & \mathbf 0 \\
             \mathsf e( 1+\beta_{LL})^{-1} & \mathbf 1
        \end{array}\right),\qquad\qquad
       &&  \tilde{\mathsf F}^R= \mathsf F^R=\left(\begin{array}{c}
        \mathbf 0 \\ \mathsf G^R
    \end{array}\right),\\
&        \tilde {\mathsf F}^L=\left(\begin{array}{cc}
       \mathsf G^L( 1+\beta_{LL})^{-1}  & \mathbf 0 
    \end{array}\right),\qquad\qquad
  &&  \hat{\tilde{\mathsf E}}=\hat{\mathsf E}.
    \end{aligned}
\end{equation}
Notice that $\mathsf F^R$ and $\hat{\mathsf E}$ remain unchanged, and that the transformations of $\mathsf E$ and $\mathsf F^L$ are quite simple. This is due to the fact that the matrices have various blocks that are 0, as a consequence of the invariance under chiral isometries.

We can write explicitly the transformed action as
\begin{equation}
\begin{aligned}
  \tilde  S=-\int d^2\sigma\, \Big(&\partial_+\phi_L^t( 1+\beta_{LL})^{-1}\partial_-\phi_L+\partial_+\phi_R^t\partial_-\phi_R \\
    &+\partial_+\phi_R^t\, \mathsf e( 1+\beta_{LL})^{-1}\, \partial_-\phi_L +\partial_+\phi_R^t\mathsf G^R\partial_-\hat x\\
    &+\partial_+\hat x^t \mathsf G^L( 1+\beta_{LL})^{-1}\partial_-\phi_L+\partial_+\hat x ^t\hat{\mathsf E}\partial_-\hat x\Big).
\end{aligned}
\end{equation}
At this point it is natural to reabsorb various factors of $( 1+\beta_{LL})^{-1}$ by redefining the left fields as $\phi_L\to ( 1+\beta_{LL})\phi_L$. The factors $( 1+\beta_{LL})^{-1}$ are reabsorbed everywhere yielding the same expressions as in the undeformed action, except in the first term. In particular we can write
\begin{equation}
\begin{aligned}
  \tilde  S=S+\int d^2\sigma\, \Big(\partial_+\phi_L^t\beta_{LL}\partial_-\phi_L\Big).
\end{aligned}
\end{equation}
It is easy to see that the difference $\partial_+\phi_L^t\beta_{LL}\partial_-\phi_L$ is in fact just a total derivative, and therefore we can drop it from the action. To conclude, the deformation by $\beta_{LL}$ is trivial, up to linear coordinate transformations and constant $B$-field shifts. See also the exercise~\ref{ex:trivial-left} of section~\ref{sec:ex-def-act}. Calculations for $\beta_{RR}$ are analogous and are left as the  exercise~\ref{ex:trivial-right}.

\subsubsection{Non-trivial deformations and the importance of being earnest chiral currents}
The only deformations that have a chance of being physical are those parameterised by $\beta_{LR}$. This is encouraging because it is very suggestive of the $LR$ coupling of the current bilinear. There is however a crucial subtlety.

When it comes to the currents of $\sigma$-models with chiral isometries, there is a very important comment to make. It turns out that the Noether currents corresponding to the $d_L+d_R$ isometries are \emph{not} chiral or antichiral. It is natural to worry about this point, because it means that the infinitesimal version of a $\beta$-deformation of a $\sigma$-model with chiral isometries is, as already proved, of the form of a current-current bilinear, but those currents are not chiral/antichiral as in the discussion of integrably marginal deformations of CFTs. 

Soon we will see the resolution of this puzzle. Before that, let us look more in detail at these conserved currents. Directly from Noether's theorem (or equivalently from~\eqref{eq:Noether-indices} or~\eqref{eq:Noether-generic}), we find the following expressions for the Noether currents
\begin{equation}
    \begin{aligned}        J_{a+}=\partial_+\phi_{La}+\partial_+\phi_R^{\bar{a}}\mathsf e_{\bar{a}a}+\partial_+\hat x^\mu\mathsf G^L_{\mu a},\qquad J_{a-}=\partial_-\phi_{La},\\
    J_{\bar a-}=\partial_-\phi_{R\bar a}+\mathsf e_{\bar{a}a}\partial_-\phi_L^{a}+G^R_{\bar a\mu}\partial_-\hat x^\mu\mathsf ,\qquad J_{\bar a+}=\partial_+\phi_{R\bar a}.
    \end{aligned}
\end{equation}
Here the indices $a,\bar a$ are lowered on $\phi_L,\phi_R$ with the Kronecker delta. The above currents are conserved, $\partial_+J_-+\partial_-J_+=0$, but they are not chiral or antichiral. Ideally, we would like to have left currents such that $J^L_+\neq 0, J^L_-=0$, and right currents such that $J^R_-\neq 0, J^R_+=0$, conserved in both cases. We see that we can indeed construct these chiral and antichiral currents from the above expressions. In fact, the conservation of $J_{a\alpha}$, for example, can be rewritten as
\begin{equation}
    0=\partial_+J_{a-}+\partial_-J_{a+}=
    \partial_+\partial_-\phi_{La}+\partial_-J_{a+}=\partial_-(J_{a+}+\partial_+\phi_{La}),
\end{equation}
which can be interpreted as the conservation of a chiral current. Similar considerations can be made for $J_{\bar a\alpha}$. Therefore we can define the chiral (left) and antichiral (right) currents
\begin{equation}
    \begin{aligned}
&J^L_{a+}\equiv J_{a+}+\partial_+\phi_{La},\qquad\qquad J^L_{a-}\equiv J_{a-}-\partial_-\phi_{La}=0,\\
&J^R_{\bar a-}\equiv J_{\bar a-}+\partial_-\phi_{R\bar a},\qquad\qquad J^R_{\bar a+}\equiv J_{\bar a+}-\partial_+\phi_{R\bar a}=0.
\end{aligned}
\end{equation}
The fact that the conservations of the Noether currents and of the chiral/antichiral currents are compatible can be made even more manifest rewriting these formulas as
\begin{equation}
    \begin{aligned}
J^{L\alpha}_{a}=J^\alpha_{a}+\epsilon^{\alpha\beta}\partial_\beta\phi_{La},\qquad\qquad J^{R\alpha}_{\bar a}=J^\alpha_{\bar a}-\epsilon^{\alpha\beta}\partial_\beta\phi_{R\bar a}.
\end{aligned}
\end{equation}
The shifts by $\epsilon^{\alpha\beta}\partial_\beta\phi_{La}$ or $\epsilon^{\alpha\beta}\partial_\beta\phi_{R\bar a}$ can be understood as ``improvement terms'' that, while not spoiling the conservation of the currents, they make them chiral or antichiral.

\vspace{12pt}

Let us now see how we can use these observations to make the infinitesimal version of the $\beta$-deformation of the wanted form. As already emphasized, we can turn on just the $\beta_{LR}$ block because it is the one with a chance of being non-trivial. We already know that the infinitesimal variation of the deformed action when considering $\beta\to \beta+\delta\beta$ is given by the Noether current bilinear, and using that result we find
\begin{equation}
\begin{aligned}
    \delta \tilde S&=\int d^2\sigma\, \delta\beta^{ij}\tilde J_{i+}\tilde J_{j-}=\int d^2\sigma\, (\delta\beta_{LR}^{a\bar a}\tilde J_{a+}\tilde J_{\bar a-}+\delta\beta_{RL}^{\bar aa}\tilde J_{\bar a+}\tilde J_{a-})\\
    &=\delta\beta_{LR}^{a\bar a}\int d^2\sigma\, (\tilde J_{a+}\tilde J_{\bar a-}-\tilde J_{\bar a+}\tilde J_{a-})\\
    &=\delta\beta_{LR}^{a\bar a}\int d^2\sigma\, (\tilde J^L_{a+}\tilde J^R_{\bar a-}-\tilde J^L_{a+}\partial_-\phi_{R\bar a}-\partial_+\phi_{La}\tilde J^R_{\bar a-}\\
    &\qquad\qquad\qquad+\partial_+\phi_{La}\partial_-\phi_{R\bar a}-\partial_+\phi_{R\bar a}\partial_-\phi_{La}),
\end{aligned}
\end{equation}
where in the last step we rewrote the Noether currents in terms of the chiral/antichiral currents. The last equality makes the problem evident: the first term with $\tilde J^L_{a+}\tilde J^R_{\bar a-}$ is the desired one at the infinitesimal level, and we should find a way  to get rid of all the remaining terms.\footnote{The last two terms add up to a total derivative and may be dropped. We will need to use this later. It would be tempting to get rid of the terms $-\tilde J^L_{a+}\partial_-\phi_{R\bar a}-\partial_+\phi_{La}\tilde J^R_{\bar a-}$ by integrating by parts and invoking the chirality/antichirality of the currents. However, that would prove the equivalence to the desired infinitesimal variation only on-shell, i.e.~on the $\sigma$-model equations of motion. Nevertheless, this observation makes it clear that there is a field redefinition that can help to remove the unwanted terms, as we will now see.} We will now show that we can remove these additional terms by a coordinate redefinition and a $B$-field gauge transformation.

First, consider the deformed action and implement the redefinition
\begin{equation}
    \phi_L\to \phi_L+\varepsilon_L,\qquad \qquad
    \phi_R\to \phi_R+\varepsilon_R,
\end{equation}
where $\varepsilon_L,\varepsilon_R$  depend on the worldsheet coordinates.
Given that the action is invariant under the above transformations if $\varepsilon_L,\varepsilon_R$ were constant, it is clear that the variation of the action produces terms with the Noether currents. We will now denote by $\delta'\tilde S$ the variation of the action as a consequence of this coordinate redefinition, to distinguish it from the variation $\delta S$ produced by taking $\beta\to\beta+\delta\beta$. We therefore find
\begin{equation}
\begin{aligned}
    \delta'\tilde S&=-\int d^2\sigma\, \eta^{\alpha\beta}(\partial_\alpha\varepsilon_L^aJ_{a\beta}+\partial_\alpha\varepsilon_R^{\bar a}J_{\bar a\beta})\\
    &=-\int d^2\sigma\, (\partial_+\varepsilon_L^aJ_{a-}+J_{a+}\partial_-\varepsilon_L^a+\partial_+\varepsilon_R^{\bar a}J_{\bar a-}+J_{\bar a+}\partial_-\varepsilon_R^{\bar a})\\
    &=-\int d^2\sigma\, (J^L_{a+}\partial_-\varepsilon_L^a+\partial_+\varepsilon_R^{\bar a}J_{\bar a-}^R\\
    & \qquad\qquad+\partial_+\varepsilon_L^a\partial_-\phi_{La}+\partial\phi_{R\bar a}\partial_-\varepsilon_R^{\bar a}-\partial_+\phi_{La}\partial_-\varepsilon_L^a-\partial_+\varepsilon_R^{\bar a}\partial_-\phi_{R\bar a}).
\end{aligned}
\end{equation}
At this point we see that we want  to set
\begin{equation}
    \varepsilon_L^a=-\delta\beta_{LR}^{a\bar a}\phi_{R\bar a},\qquad\qquad 
    \varepsilon_R^{\bar a}=-\delta\beta_{LR}^{a\bar a}\phi_{La},
\end{equation}
because then we get
\begin{equation}
\begin{aligned}
    \delta'\tilde S&=\delta\beta_{LR}^{a\bar a}\int d^2\sigma\, [J^L_{a+}\partial_-\phi_{R\bar a}+\partial_+\phi_{La}J_{\bar a-}^R\\
    &\qquad\qquad-2(\partial_+\phi_{La}\partial_-\phi_{R\bar a}-\partial_+\phi_{R\bar a}\partial_-\phi_{La})].
\end{aligned}
\end{equation}
Now we see that the combination of $\delta\tilde S$ and of the effect of the coordinate redefinition $\delta'\tilde S$ gives nice cancellations and leaves just
\begin{equation}
   \delta\tilde S+\delta'\tilde S =\delta\beta_{LR}^{a\bar a}\int d^2\sigma\, (\tilde J^L_{a+}\tilde J^R_{\bar a-}-\partial_+\phi_{La}\partial_-\phi_{R\bar a}+\partial_+\phi_{R\bar a}\partial_-\phi_{La}).
\end{equation}
It is easy to see that the last two terms add up to a total derivative, in other words they can be removed by a $B$-field gauge transformation. We are then left with the desired result. We conclude  that (up to coordinate redefinitions and $B$-field gauge transformations) the infinitesimal version of the $\beta$-deformation is given by a chiral-antichiral current bilinear
\begin{equation}
   \delta\tilde S+\delta'\tilde S =\delta\beta_{LR}^{a\bar a}\int d^2\sigma\, \tilde J^L_{a+}\tilde J^R_{\bar a-}.
\end{equation}

\subsection{Exercises}\label{sec:ex-def-act}
\begin{enumerate}
\item \label{ex:action-chiral} \textbf{Action of a $\sigma$-model with chiral isometries --- } Check that an action with $d_L+d_R$ chiral/antichiral commuting isometries can be put in the form~\eqref{eq:S-chiral}. Start from~\eqref{eq:S-d-iso} and find the conditions that the background should satisfy in order for the action to be invariant under~\eqref{eq:chiral-tr}. Implement coordinate redefinitions and $B$-field gauge transformations to put the action in the form of~\eqref{eq:S-chiral}.
\item \label{ex:trivial-left} \textbf{Triviality of the $\beta_{LL}$ deformation, alternative check --- } Consider a $\sigma$-model invariant under chiral isometries. Do an alternative check that the $\beta_{LL}$ transformation is trivial. In particular, construct the generalised metric and check that in this case the transformation induced by the particular $\mathcal U$ considered is equivalent to the composition of appropriate $\mathcal R,\mathcal S$ $O(D,D)$ transformations.

\item \label{ex:trivial-right} \textbf{Triviality of the $\beta_{RR}$ deformation --- } Consider a $\sigma$-model invariant under chiral isometries. Check that the deformation generated by $\beta_{RR}$ is trivial up to linear coordinate transformations of $\phi_R$ and to $B$-shifts.

\end{enumerate}


\section{Integrability and the equivalence to a twisted model}\label{sec:int}
In this section we want to discuss the relation between the deformations and the integrability in the case that the original $\sigma$-model is integrable.
TsT deformations and their relation to integrability were discussed in various papers in the literature of string theory and the AdS/CFT correspondence, see in particular~\cite{Frolov:2005ty,Frolov:2005dj,Alday:2005ww}. In these lecture notes we have decided to take a slightly different angle for the presentation compared to how the results were derived in the literature. In particular, the relation to integrability and the reformulation of the deformed models as undeformed $\sigma$-models with twisted boundary conditions on the worldsheet will be straightforward thanks to the discussion made in the previous sections.

\subsection{On-shell map and Lax connection}
The $O(d,d)$ transformations that we have been considering were originally motivated by demanding that they come from canonical transformations of the $\sigma$-models. In other words, they are maps that leave the Poisson brackets invariant, and therefore they relate the  equations of motion of the original model to those of the deformed model. 

Let us spend some more words on this point. Implementing the map $(x,p)\to (\tilde x,\tilde p)$ on the Hamiltonian $\mathcal H$ allows us to identify the Hamiltonian $\tilde{\mathcal H}$. To be more explicit, if the canonical transformation is given by $x=u(\tilde x,\tilde p), p=v(\tilde x,\tilde p)$ then we have the relation $\tilde{\mathcal H}(\tilde x,\tilde p)=\mathcal H(u(\tilde x,\tilde p),v(\tilde x,\tilde p))$.\footnote{ Remember  that the maps that we have been considering are actually of the form $(\phi'{}^i,p_i)\to (\tilde \phi'{}^i,\tilde p_i)$, where only the derivatives of the fields appear.}
In the Lagrangian formalism, however, if we want to obtain the transformed model from the original one, it is not enough to implement the substitution $(\phi'{}^i,\dot\phi^i)\to (\tilde \phi'{}^i,\dot {\tilde\phi}^i)$ in the Lagrangian. For example, one can check this starting from the Lagrangian for a collection of free bosons $\mathcal L=\partial_+\phi^t\partial_-\phi$ and substituting $\dot\phi=\dot{\tilde\phi}$, $\phi'=\tilde\phi'+\beta \dot{\tilde\phi}$. Indeed this does not yield the $\beta$-deformation of the Lagrangian. This is related to the fact that the canonical transformations that we are considering are understood as non-local maps of the fields. If, instead, we were considering a  canonical transformation implemented by the matrix $\mathcal R$, this would induce a  standard field redefinition that can be implemented directly at the Lagrangian level.

In general, therefore, the relation between the original Lagrangian $\mathcal L$ and the one of the $\sigma$-model after the canonical transformation $\tilde {\mathcal L}$ is only \emph{on-shell}. In other words, the map $(\phi'{}^i,\dot\phi^i)\to (\tilde \phi'{}^i,\dot {\tilde\phi}^i)$ makes sure that the \emph{equations of motion } of the two models are related to each other, and similarly for the \emph{solutions to the equations of motion}.

\vspace{12pt}

The above on-shell relation is the key that ensures the integrability of the deformed model. First, let us comment on how the Noether currents are mapped under this on-shell relation.
When computing the variation of the action with respect to $\delta\phi$, one obviously finds $\delta S=-\int d^2\sigma\, \eta^{\alpha\beta}\partial_\alpha(\delta\phi)J_\beta$ with $J$ the Noether currents. Exactly the same happens in the deformed model, so that the equations of motion for $\phi$ and $\tilde \phi$ imply the conservation of the corresponding Noether currents
\begin{equation}
    \partial_\alpha J^\alpha=0,\qquad\qquad
    \partial_\alpha \tilde J^\alpha=0.
\end{equation}
Let us now look in detail at the map in the case of the $\beta$-deformation. We have
\begin{equation}
    p=\tilde p,\qquad\qquad \phi'=\tilde\phi'+\beta\tilde p,
\end{equation}
where indices are omitted.
Recalling the definition of the momenta as $p=\delta S/\delta\dot \phi$, it is clear that they coincide with the time component of the Noether currents
\begin{equation}
    p_i=J_{i\tau}.
\end{equation}
Therefore, the on-shell map immediately implies that $J_0=\tilde J_0$. Although not obvious, an even stronger statement is true, namely that also the space components of the two currents agree, so that
\begin{equation}
    J_\alpha=\tilde J_{\alpha}.
\end{equation}
The proof of this last statement is a bit laborious, and it is left as the guided exercise~\ref{ex:Noether}  of section~\ref{sec:ex-int}.

Suppose, now, that the original model is integrable. In particular, there exists a Lax connection $\mathscr L(x,\partial x;\xi)$ depending on the spectral parameter $\xi$, such that its flatness
\begin{equation}
    \epsilon^{\alpha\beta}(\partial_\alpha \mathscr L_\beta +\mathscr L_\alpha \mathscr L_\beta )=0,
\end{equation}
is equivalent to the equations of motion of the $\sigma$-model. It follows immediately that the transformed model is also integrable, and that the corresponding Lax connection is obtained by 
\begin{equation}
    \tilde{\mathscr L}_\alpha(\partial_\alpha\tilde\phi,\hat x,\partial_\alpha\hat x;\xi)=
    \mathscr L_\alpha(u_\alpha(\partial\tilde\phi),\hat x,\partial_\alpha\hat x;\xi).
\end{equation}
In other words, the Lax connection of the deformed model may be obtained by implementing the on-shell map on the expression for the original $\mathcal L_\alpha$. In fact, the flatness of $\mathcal L$ implies the flatness of $\tilde {\mathcal L}$, and the latter is equivalent to the equations of motion of the deformed model thanks to the on-shell map. The same reasoning may be used to construct the Lax connection of models obtained by T-duality transformations.

Of course, we could have argued the conservation of integrability under the transformation from the very beginning in the Hamiltonian formulation, given that by construction the map is a canonical transformation.

\subsection{Twisted model}
It is natural to wonder whether the transformed models obtained by the canonical transformations are interesting at all. If they are on-shell equivalent to the original model, are we actually working with genuinely new models?

Of course, the geometric canonical transformations induced by $\mathcal R,\mathcal S$ lead to transformed models that are not genuinely new, as already explained. However, the T-duality transformations and the $\beta$-deformations do lead to new and interesting models. From the target-space point of view, we have already mentioned that they are not equivalent to simple coordinate redefinitions of $B$-field gauge transformations, but we can understand this also from a purely $\sigma$-model point of view. 

The important point is that,  in order to identify the dynamics of the model, the $\sigma$-model equations of motion alone are not enough, one also needs to specify boundary conditions. Suppose now that the spatial coordinate  $\sigma$ on the worldsheet goes from $r_-$ to $r_+$. If we wanted to work on the line, we would take $(r_-,r_+)=(-\infty,+\infty)$, while on the circle $(r_-,r_+)=(0,2\pi)$. Say that we fix the Dirichlet boundary condition $\tilde\phi(r_+)-\tilde\phi(r_-)=\Delta \tilde\phi$ in the deformed model, for some $\Delta \tilde\phi$ (for example zero, but not necessarily). Integrating the equality  $\phi'=\tilde\phi'+\beta J_\tau$ along $\sigma$ from $r_-$ to $r_+$ one finds
\begin{equation}
    \Delta\phi=\Delta\tilde\phi+\beta Q,\qquad \qquad Q\equiv \int_{r_-}^{r_+} d^2\sigma\, J_\tau,
\end{equation}
where we defined the Noether charge $Q$.
If, for example, we worked on the line, and we imposed falling-off boundary conditions for the deformed model $\Delta\tilde\phi=0$, then we would conclude that we must have \emph{kink} boundary conditions for the undeformed model that is equivalent to that, because $\Delta\phi\neq 0$. Configurations on the line are in fact called kinks if the value of the field at $\sigma=-\infty$ differs from the value of the field at $\sigma=+\infty$.  Similarly, if we worked on the circle and imposed periodic boundary conditions for the deformed model setting  $\Delta\tilde\phi=2\pi n$ with $n\in \mathbb Z$, then we would have \emph{twisted} boundary conditions for the undeformed model that is equivalent to that. Configurations on the circle such that the value of $\Delta\phi$ is not  $2\pi$ times an integer are in fact normally called twisted.

To conclude, it is certainly true that the deformed model can be reformulated in terms of an undeformed one, but we must be careful in identifying the correct boundary conditions on the two sides in order to make sure that we are describing the dynamics that are interesting for us. Actually, it is precisely because the boundary conditions are mapped to different ones under the on-shell map that these canonical transformations are non-trivial.

\subsection{The quantum theory and the S-matrix}
Until now our considerations have been made at the level of the classical model. Interestingly, one can understand the effect of the current-current deformations also at the quantum level. In particular, we will now assume that the seed model admits a formulation   of the scattering of asymptotic states at the quantum level. We therefore assume the existence of an S-matrix $\mathcal S$ for the theory at the deformation parameter $\lambda=0$. This S-matrix, then, encodes the effects of the interactions present in the model.

So far we have been discussing $\sigma$-models, in particular integrable ones, and it turns out that at least some of them do admit an S-matrix, which thanks to integrability can be even fixed exactly (i.e.~to all orders in the couplings of the theory, therefore beyond the standard perturbation theory). An example of this kind is the Principal Chiral Model (PCM), which we will present and deform at the \emph{classical} level in the next section. Importantly, in the case of the PCM the asymptotic states that are scattered by the exact S-matrix do not correspond to excitations of the fundamental fields of the theory, rather to solitonic configurations~\cite{Zamolodchikov:1978xm,Berg:1977dp,Wiegmann:1984ec}.

Other examples to keep in mind for this section are $\sigma$-models coupled to a worldsheet metric field $h_{\alpha\beta}$. This may be achieved by considering the projectors $\Pi^{\alpha\beta}_{(\pm)}$ of equation~\eqref{eq:Ppm} that are used to construct the $\sigma$-model action, and by replacing the flat Minkowski metric $\eta^{\alpha\beta}$ there with $\sqrt{|\det h|}h^{\alpha\beta}$.
 In this setup, the action of the 2-dimensional $\sigma$-model is invariant under reparameterisations of the worldsheet coordinates $\tau,\sigma$, so that this symmetry can be understood as a gauge freedom. At the same time, because the metric field $h_{\alpha\beta}$ appears without derivatives in the action, the energy momentum tensor $T_{\alpha\beta}=\delta S/\delta h^{\alpha\beta}$ is identically zero, $T_{\alpha\beta}=0$. These are the so-called Virasoro constraints. Solving the Virasoro constraints, one fixes the reparameterisation invariance on the worldsheet and ends up with a ``reduced'' model, because 2 unphysical modes have been gauged away. A useful choice is the so-called light-cone gauge. In general, the gauge-fixing procedure may result in a massive spectrum, and a notion of scattering for the fundamental excitations of the reduced model may be possible. A notable example that realises this scenario and that plays a fundamental role in the AdS/CFT correspondence is that of the superstring on the $AdS_5\times S^5$ background. We refer to~\cite{Arutyunov:2009ga} for a review on this particular case and on the more general setup. 

Finally, we point out that, as already mentioned,  current-current deformations  have an applicability that is wider than the scope of these lecture notes, a scope that is restricted by our particular motivations. In principle, one may consider generic field theories in 2 dimensions that are not necessarily $\sigma$-models and that from the start may have a natural formulation of the scattering problem for the fundamental fields. If the theory under consideration happens to have at least 2 commuting global internal symmetries, then it may be deformed by a current-current deformation, and the scattering problem may also be deformed in the way described in this section. Moreover, if this seed field theory can be understood as the reduced model resulting from the gauge-fixing procedure of a $\sigma$-model, then the currents generating the deformation will correspond to symmetries that act trivially on the unphysical fields that are gauged away in the gauge-fixing procedure.

\vspace{12pt}

We will now explain that it is possible to understand how the \emph{non-perturbative} S-matrix of the seed theory is deformed by the current-current deformation. The deformation of the S-matrix can be understood in terms of the so-called Drinfeld-Reshetikhin twists~\cite{drinfeld1983constant,Reshetikhin:1990ep}, which are known constructions in the context of Hopf algebras and quantum integrable models. See also~\cite{Giaquinto:1994jx,Kulish2009} for  useful references, and~\cite{Beisert:2005if} for the construction of  twisted worldsheet S-matrices corresponding to TsT deformations of the $AdS_5\times S^5$ superstring. To discuss the S-matrix, in these notes we closely follow~\cite{new-lcg}, see e.g.~\cite{Dubovsky:2023lza} for an alternative derivation.\footnote{This section explaining the effect of the current-current deformations at the level of the scattering formulation was missing at the time of the school in Durham, and it was added for completeness during the preparation of the paper~\cite{new-lcg}. Although the motivation in that paper is a bit different, the discussion on the twisted S-matrix is essentially the same. We  refer to that paper for more details on the discussion.} Although it may be possible to generalise the argument to S-matrices scattering solitonic objects, it will be useful to keep in mind the case of scattering of fundamental excitations of the theory.

First, to set the stage, let us give a brief recap on the construction of the S-matrix in a $1+1$-dimensional field theory. We assume that we are working on the line, so that $\sigma\in(-\infty,+\infty)$, and that the time evolution in the undeformed theory is dictated by the Hamiltonian $H$. A standard procedure is to separate the Hamiltonian into two parts
\begin{equation}
    H=H_0+V,
\end{equation}
where $H_0$ is interpreted as the ``free part'' and $V$ the one containing the interactions. The important point is that $H_0$ can be quantised exactly with standard methods. The simplest example to keep in mind is that of the Klein-Gordon field. In particular, we may consider a ``second quantisation'' procedure where fields are rewritten in terms of creation/annihilation operators $a^\dagger_k(p),a_k(p)$, where $k$ is a label to identify the quantum numbers of the operators, so that we can distinguish different ``flavours''. Particles with definite momentum $p$ are constructed acting with creation operators on the vacuum, e.g.~$\ket{p}_k=a^\dagger_k(p)\ket{0}$. We assume that the quantisation of the free part $H_0$ takes the simple form $H_0=\int dp\ \sum_k\ \omega_k(p) a^\dagger_k(p)a_k(p)$, where $\omega_k(p)$ is the dispersion relation (so that $\omega_k(p)=\sqrt{m_k^2+p^2}$ for relativistic theories).

The interacting part $V$ may be complicated. The standard procedure is to consider incoming and outgoing asymptotic states, defined at $\tau=-\infty$ and $\tau=+\infty$ respectively. Each asymptotic state may be composed of several particles, and these are understood as wave-packets that have well-defined momenta and are well separated from each other. If their distance is very large, interactions can be ignored, so asymptotic states only evolve with $H_0$. In particular, incoming states are constructed as the collection of $N$ particles as
\begin{equation}
\ket{p_1,p_2,\ldots,p_N}^{in}_{k_1,k_2,\ldots,k_N},
\end{equation}
where the momenta are taken to be ordered as $p_1>p_2>\ldots>p_N$. In other words, particles are ordered from the left to the right with decreasing order of velocity. This ensures that they will eventually meet and that scattering will occur. Similarly, outgoing states are constructed as the collection of $M$ particles as
\begin{equation}
\ket{p_M,p_{M-1},\ldots,p_1}^{out}_{l_M,k_{M-1},\ldots,l_1},
\end{equation}
where  the momenta are still ordered as $p_1>p_2>\ldots>p_M$, so that after scattering the fastest particles are to the right. In an integrable theory, $N=M$ and $p_i^{in}=p_i^{out}$.

Finally, the S-matrix is the object relating incoming and outgoing states, so that a particular entry may be denoted as
\begin{equation}
    \mathcal S_{k_1,k_2,\ldots,k_N}^{l_M,k_{M-1},\ldots,l_1}(\{p_i^{in}\},\{p_i^{out}\}).
\end{equation}
In particular, the S-matrix may be computed from the interacting part of the Hamiltonian as the time-ordered exponential of $V$
\begin{equation}
\mathcal S= \mathcal T\exp\left[-i\int_{-\infty}^{\infty}d\tau\ V\right].
\end{equation}

After these preliminaries, consider now a current-current deformation of the Hamiltonian density as \begin{equation}
    \delta \mathcal H=\frac{\lambda }{2}\epsilon^{\alpha\beta}\epsilon_{ij}J^i_\alpha J^j_\beta=-\lambda(J^1_\tau J^2_\sigma-J^1_\sigma J^2_\tau).
\end{equation}
Remember that the Hamiltonian is related to its density as $H=\int d\sigma\ \mathcal H$.
As already pointed out, the current-current deformations that we consider here form an abelian group, so that  compositions of different deformations can be easily considered.
The crucial point is that from the above expression one can prove the following relation for the infinitesimal deformation of the Hamiltonian 
\begin{equation}\label{eq:delta-H-dot-Q}
    \delta H=\frac{\lambda}{2}\partial_\tau\mathcal Q^{12},
\end{equation}
where we defined
\begin{equation}
\mathcal Q^{12}
=\int_{-\infty}^\infty d\sigma\ \int_{-\infty}^\sigma d\sigma'\ [J_\tau^1(\sigma)J_\tau^2(\sigma')-J_\tau^2(\sigma)J_\tau^1(\sigma')].
\end{equation}
To prove this relation, one may first define the field
\begin{equation}
\mathcal Q^i(\sigma)=\int^\sigma_{-\infty}d\sigma'\ J_\tau^i(\sigma'),
\end{equation}
which is related to the charge  $Q^i=\int^\infty_{-\infty}d\sigma J_\tau^i(\sigma)$  as $Q^i=\mathcal Q^i(\infty)$. Then, a simple computation shows that 
\begin{equation}
    J_\tau^1 J_\sigma^2-J_\sigma^1 J_\tau^2 = -\tfrac12 (J_\alpha^1\partial^\alpha \mathcal Q^2-J_\alpha^2\partial^\alpha \mathcal Q^1),
\end{equation}
which may be checked using just conservation of the currents and the fact that fields fall of at infinity. Integrating the above expression and using again these two properties, one arrives at the formula~\eqref{eq:delta-H-dot-Q}. See~\cite{new-lcg} for more details.

At the classical level, the time evolution is given by the Poisson bracket with the Hamiltonian, so that $\partial_\tau\mathcal Q^{12}=\{H,\mathcal Q^{12}\}$. At the quantum level, we have $\partial_\tau\mathcal Q^{12}=i[H,\mathcal Q^{12}]$. Therefore, we conclude that the quantum Hamiltonian $\tilde H$ of the deformed model satisfies the following differential equation 
\begin{equation}
\frac{d\tilde H}{d\lambda} = \frac{i}{2}[\tilde H,\mathcal Q^{12}].
\end{equation}
If we work in the Heisenberg picture, we may solve this equation by
\begin{equation}
\tilde H = e^{-\frac{i\lambda}{2}\mathcal Q^{12}} H e^{\frac{i\lambda}{2}\mathcal Q^{12}},
\end{equation}
where we used   that $\tilde H|_{\lambda=0}=H$.
This is the relation that will allow us to construct the exact deformation of the S-matrix.

To continue we need to understand the action of $\mathcal Q^{12}$ on the asymptotic states. First, we assume that the basis for the operators was chosen so that the charges $Q^i$ act diagonally
\begin{equation}
Q^i\ket{p}_k=Q^i_k\ket{p}_k,
\end{equation}
where $Q^i_k$ is the charge of the particle with flavour $k$. We may therefore represent the charge operators as $Q^i = \int dp\ \sum_{k} Q^i_k \ a^\dagger_k(p)a_k(p)$.
Importantly, when evaluating $\mathcal Q^{12}$ on asymptotic states, we may decompose the line parameterised by the spatial coordinate $\sigma$ as a collection of intervals $I_n$ in such a way that only the particle $n$ is contained in $I_n$. Therefore, although the definition of $\mathcal Q^{12}$ involves a double integral, thanks to the decomposition and the anti-symmetry one has
\begin{equation}
\mathcal Q^{12}\ket{p_1,\ldots,p_N}_{k_1,\ldots,k_N}=\sum_{n=1}^N \sum_{m=1}^{n-1} (Q^1_{k_n}Q^2_{k_m}-Q^2_{k_n}Q^1_{k_m})\ket{p_1,\ldots,p_N}_{k_1,\ldots,k_N}.
\end{equation}
The detailed proof of the above relation is left as an exercise, see also~\cite{new-lcg}.

At this point we have all the ingredients to construct the S-matrix of the deformed theory starting from that of the undeformed one. Having decomposed the undeformed Hamiltonian as the sum of a free part and an interacting part, we do the same for the deformed Hamiltonian $\tilde H$. In particular, we take
\begin{equation}
\tilde H_0 = e^{-\frac{i\lambda}{2}\mathcal Q^{12}} \ H_0\ e^{\frac{i\lambda}{2}\mathcal Q^{12}}.
\end{equation}
After noticing that both $H_0$ and $\mathcal Q^{12}$ act diagonally on asymptotic states, we conclude that 
\begin{equation}
    \tilde H_0 \ket{p_1,\ldots,p_N}_{k_1,\ldots,k_N}=H_0 \ket{p_1,\ldots,p_N}_{k_1,\ldots,k_N},
\end{equation}
so that the free part of the Hamiltonian in the deformed case is effectively the same as the one in the undeformed case.
The above definition of $\tilde H_0$ implies that the interacting part is $\tilde V = e^{-\frac{i\lambda}{2}\mathcal Q^{12}} V e^{\frac{i\lambda}{2}\mathcal Q^{12}}$, which in turn implies
\begin{equation}\label{eq:Sgamma}
\tilde{\mathcal  S} = e^{-\frac{i\lambda}{2}\mathcal Q^{12}}\ \mathcal S \ e^{\frac{i\lambda}{2}\mathcal Q^{12}}.
\end{equation}
 Specifying this relation to a particular S-matrix element, one finds
\begin{equation}
\tilde{\mathcal  S}_{k_1k_2\cdots k_N}^{l_Ml_{M-1}\cdots l_1} = e^{\frac{i\lambda}{2}\sum_{m>n}\epsilon_{ij}(Q^i_{l_m}Q^j_{l_n}+Q^i_{k_n}Q^j_{k_m})}\mathcal S_{k_1k_2\cdots k_N}^{l_Ml_{M-1}\cdots l_1},
\end{equation}
where $\epsilon_{12}=1$.
This is the desired relation between the deformed and undeformed S-matrices. In particular, this relation is exact in the deformation parameter $\lambda$, and may be valid even at the level of the non-perturbative S-matrix.

\subsection{An application: the Principal Chiral Model}

\subsubsection{The undeformed model}
To construct the action of the Principal Chiral Model (PCM) we start from a Lie group $G$ and we choose an element $g\in G$. In general, $g$ will depend on the coordinates that parameterise the group manifold of $G$. We will denote them by $x^m$. From the 2-dimensional point of view these are fields that depend on the worldsheet coordinates $\sigma^\pm$. Next, we construct the right-invariant Maurer-Cartan current as
\begin{equation}
    \ell=dgg^{-1}\in Lie(G),
\end{equation}
which is an element of the Lie algebra. Explicitly, we may write it as
\begin{equation}
    \ell_\alpha=\partial_\alpha g g^{-1}=\partial_\alpha x^m\ell_m{}^a T_a.
\end{equation}
Here $T_a$ is a basis of generators of $Lie(G)$, and the  explicit expression for the coefficients $\ell_m{}^a$ is found once the parameterisation of $g$ is fixed. The action of the PCM 
is given by 
\begin{equation}\label{eq:SPCM}
    S=-\int d^2\sigma\, \Trh[\ell_+\ell_-],
\end{equation}
where $\mathrm{Tr}$ stands for ``trace'', understood as the trace in the matrix realisation of $Lie(G)$.
We actually prefer to use a different normalisation, so that we write $\Trh=-\tfrac12 \mathrm{Tr}$.
More generally, one may replace $\Trh$ by a non-degenerate, ad-invariant, symmetric bilinear form on the Lie algebra. We will denote by 
\begin{equation}
    \kappa_{ab}=\Trh[T_aT_b]
\end{equation}
the components of the bilinear form on the Lie algebra.
The $\ell_m{}^a$ can be understood as a  ``vielbein'' because the action may be rewritten as
\begin{equation}
    S=-\int d^2\sigma\, \partial_+x^m\ell_m{}^a\kappa_{ab}\ell_n{}^b\partial_-x^n
    =-\int d^2\sigma\, \partial_+x^m\partial_-x^nG_{mn},
\end{equation}
where we have a target-space metric $G_{mn}=\ell_m{}^a\kappa_{ab}\ell_n{}^b$. Notice that for compact $G$ the metric on the Lie algebra may be chosen to be proportional to the identity $\kappa_{ab}\propto \delta_{ab}$. In the index-free notation we would write the metric as
\begin{equation}
    G=\ell\kappa\ell^t.
\end{equation}
It is clear that the PCM action is invariant under two copies of $G$-transformations. On the one hand we may consider the ``right'' multiplication of $g$ by a \emph{constant} $h_R\in G$, i.e.~$g\to gh_R$. Under these transformation $\ell$ remains invariant, and therefore also the action. On the other hand, we may also consider the ``left'' multiplication of $g$ by a \emph{constant} $h_L\in G$, i.e.~$g\to h_Lg$. Under these transformations $\ell$ is not invariant, but it transforms as $\ell\to h_L\ell h_L^{-1}$. After using the ad-invariance of the trace, one concludes that the PCM action is invariant also under these transformations. To summarise, the PCM action has a group of isometries given by $G_L\times G_R$.

Obviously, we could have constructed the model by exchanging left and right. In particular, one could consider the left-invariant (rather than right-invariant) Maurer-Cartan current $k=g^{-1}dg$ and use it to construct the action $S=-\int d^2\sigma\, \Trh[k_+k_-]$. The two definitions of the action coincide, because of the relation $\ell=gkg^{-1}$ and thanks to the ad-invariance of the trace.

It is not difficult to compute the Noether currents related to the $G_L\times G_R$ isometry.\footnote{Although we have left and right transformations and corresponding currents, in the case of the PCM this terminology does not correlate with chiral/antichiral currents. The PCM does not possess chiral isometries and is not a CFT.} One may do that by considering an infinitesimal transformation of the group element so that $\delta_L g=\epsilon_Lg$ in the case of left transformations, and $\delta_Rg=g\epsilon_R$ in the case of right transformations. In particular, one finds that $\ell$ is the Noether current for the $G_L$ symmetry, while $k$ is the Noether current for the $G_R$ symmetry. This is obviously compatible with the fact that $\ell$ is charged (i.e.~it transforms non-trivially) under $G_L$ transformations, while $k$ is charged under $G_R$.

The computation of the equations of motion of the PCM is analogous to the computation of the Noether currents. One considers the infinitesimal transformation $g\to g+ \delta g$ and then imposes the condition of extremality of the action. One may recast the form of the equations of motion in two equivalent ways
\begin{equation}
    \partial_\alpha \ell^\alpha=0,\qquad\qquad
        \partial_\alpha k^\alpha=0,
\end{equation}
which have the interpretation of the conservation of the left and right Noether currents.

Soon we will deform the PCM by a $\beta$-deformation that (partially) breaks the left symmetry and maintains the right one unbroken. Let's therefore focus on the right copy for a moment and think of the equations of motion as $\partial_\alpha k^\alpha=0$. Being $k$ a Maurer-Cartan current, it is not only conserved, it also satisfies the Maurer-Cartan identity
\begin{equation}\label{eq:MC-k}
    \epsilon^{\alpha\beta}(\partial_\alpha k_\beta+ k_\alpha k_\beta)=\partial_-k_+-\partial_+k_-+[k_-,k_+]=0.
\end{equation}
The conservation of the current and the Maurer-Cartan identity are enough to write down a Lax connection for the PCM as
\begin{equation}
    \mathscr L_\pm=\frac{k_\pm}{1\mp \xi}.
\end{equation}
Here $\xi\in \mathbb C$ is called the spectral parameter, and it is introduced because demanding that the Lax connection is flat at generic values of $\xi$ implies the two equations $\partial_\alpha k^\alpha=0$ and $\epsilon^{\alpha\beta}(\partial_\alpha k_\beta+k_\alpha k_\beta)=0$. The vice versa is also true, i.e. the two equations ensure that the Lax connection is flat at generic values of $\xi$. Obviously, we may have constructed an alternative Lax connection using the current $\ell$ instead.

\subsubsection{The $\beta$-deformation}\label{sec:beta-PCM}
After these initial considerations, we now want to implement a $\beta$-deformation of the PCM action. If the Lie group $G$ is $D$-dimensional, we consider a collection of $d\leq D$ generators $T_i\in Lie(G)$ that have the property of being mutually commuting, i.e. $[T_i,T_j]=0,\ \forall i,j$. To implement a $\beta$ deformation we need $d\geq 2$, and obviously we have $d\leq \mathrm{rank}(G)$. The remaining generators will be denoted with $T_{\hat a}$. Then, we also split the coordinates on $G$ as $x^m=\{\phi^i,\hat x^\mu\}$, and we construct the group element as
\begin{equation}
    g(x)=\exp(\phi^iT_i)\hat g(\hat x).
\end{equation}
In particular, we have factorised the dependence between the $\phi^i$ and $\hat x^\mu$ coordinates. It is clear that if we construct a constant group element
\begin{equation}
    h=\exp(c^iT_i),\qquad \qquad c^i\in \mathbb R,
\end{equation}
then the corresponding left multiplications $g\to hg$ correspond to the shifts of the $\phi^i$ coordinates
\begin{equation}
    \phi^i\to\phi^i+c^i.
\end{equation}
We conclude that, by construction, the action is invariant under these shifts. It is now interesting to compute the expression for $\ell$ with this choice of parameterisation of $g$
\begin{equation}
\ell=d\phi^iT_i+e^{\phi^iT_i}d\hat g\hat g^{-1}e^{-\phi^iT_i}=d\phi^iT_i+d\hat x^\mu \ell_\mu{}^aT_a.
\end{equation}
If we now want to think of $\ell_m{}^a$ as a $D\times D$ matrix, we may decompose it into a block form according to the split of the indices $m=\{i,\mu\}$ and $a=\{j,\hat a\}$. Doing that we may write\footnote{Notice that we are using the same notation $i,j,\ldots$ for the ``curved indices'' of the target space coordinates $\phi^i$ and for the indices of the Lie algebra generators $T_i$. In the latter case, we may think of the indices as ``frame-like''. In principle, we should use a different notation for the two, and use a vielbein to translate from one to the other. However, in this case we may choose this vielbein to be the Kronecker delta, so we prefer to keep the notation simple and not introduce a new set of indices.}
\begin{equation}\label{eq:ell-mat}
    \ell_m{}^a=\left(\begin{array}{cc}
         \delta_i{}^j & \mathbf 0  \\
         \ell_\mu{}^j& \ell_\mu{}^{\hat a}
    \end{array}\right).
\end{equation}
We are now in the position of looking at a $\beta$-deformation of the PCM. Thanks to the preliminary considerations made above, we will be able to recast it in an algebraic form as a deformation of the action~\eqref{eq:SPCM}. First, we consider the transformation of the $\beta$-deformation that we have obtained from section~\ref{sec:actionOdd}
\begin{equation}
    \tilde E=\tilde G+\tilde B=G(1+ \beta G)^{-1},
\end{equation}
where we are using the information that before the deformation $E=G$ because $B=0$. Notice that in the presence of the deformation the $B$-field is not vanishing anymore, as it is easy to see for example by expanding in powers of $\beta$ and remembering that $\beta$ is antisymmetric. Using now the geometric series expansion we may write
\begin{equation}
    \tilde E=\sum_{p=0}^\infty(-1)^p G(\beta G)^p
    =\sum_{p=0}^\infty(-1)^p  \ell\kappa\ell^t(\beta \ell\kappa\ell^t)^p
    =\sum_{p=0}^\infty(-1)^p  \ell\kappa(\ell^t\beta \ell\kappa)^p\ell^t.
\end{equation}
We see that we have to calculate $\ell^t\beta \ell$. However, taking into account that $\beta^{mn}$ is
\begin{equation}
    \beta^{mn}=\left(\begin{array}{cc}
         \beta^{ij}&\mathbf 0  \\
        \mathbf 0 & \mathbf 0
    \end{array}\right),
\end{equation}
and that the upper-right block of $\ell$ in~\eqref{eq:ell-mat} is zero and the upper-left one is just the identity, it is not difficult to convince ourselves that $\beta^{ab}=\ell_m{}^a\beta^{mn}\ell_n{}^b$ is given again just by
\begin{equation}
    \beta^{ab}=\left(\begin{array}{cc}
         \beta^{ij}&\mathbf 0  \\
        \mathbf 0 & \mathbf 0
    \end{array}\right).
\end{equation}
We may therefore write
\begin{equation}
    \tilde E   =\sum_{p=0}^\infty(-1)^p \ell\kappa(\beta \kappa)^p\ell^t=\ell\kappa\left(1+ \beta\kappa\right)^{-1}\ell^t.
\end{equation}
At this point we can reinterpret $\left(1+ \beta\kappa\right)^{-1}$ as (the inverse of) a linear operator that acts on the Lie algebra of $G$. In particular, let us define the linear operator $R:Lie(G)\to Lie(G)$ such that
\begin{equation}
    R(T_a)=T_b\beta^{bc}\kappa_{ca}.
\end{equation}
Then we can recast the action of the $\beta$-deformation of the PCM in the following form
\begin{equation}
   S=-\int d^2\sigma\, \Trh\left[\ell_+\frac{1}{1+R}(\ell_-)\right].
\end{equation}
This form of the action makes it manifest that, on the one hand, the deformation does not break the $G_R$ copy of the isometries, since it is constructed out of $\ell$ which is invariant under right transformations. On the other hand, the $G_L$ copy of isometries is in general broken because of the presence of the $R$-operator: $\ell$ is not invariant under left transformations, and the action would be invariant only if the $R$-operator were ``transparent'' to the adjoint action by the constant Lie-group element. In other words, the group of left isometries is broken to a subgroup $H_L\subset G_L$ identified by all the elements $h_L$ such that $\Trh[Xh_L^{-1}(R(h_LYh_L^{-1}))h_L]$ for generic $X,Y\in Lie(G)$. Writing $h_L=\exp(\epsilon_L)$ with $\epsilon_L\in Lie(H_L)$ we may therefore write this condition as
\begin{equation}
    \Trh[X[\epsilon_L,R(Y)]]=
    \Trh[XR([\epsilon_L,Y])].
\end{equation}
Nevertheless, given that the $G_R$ symmetry is unbroken, we may compute the corresponding Noether current $K$, and an explicit computation gives
\begin{equation}\label{eq:def-K-YB}
    K_\pm = g^{-1}\left[\frac{1}{1\mp R}\left(\ell_\pm\right)\right]g.
\end{equation}
This current, then, is conserved and, similarly to what was happening in the undeformed setup, its conservation is equivalent to the equations of motion of the deformed PCM
\begin{equation}
    \partial_\alpha K^\alpha=0.
\end{equation}
The current $K_\alpha$ is not of Maurer-Cartan form, and therefore it does not satisfy the Maurer-Cartan identity written in~\eqref{eq:MC-k}. However, it does satisfy this equation if, at the same time, we impose the equations of motion of the $\sigma$-model. To see this, let us first invert~\eqref{eq:def-K-YB} and write
\begin{equation}
    k_\pm=g^{-1}[(1\mp R)(gK_\pm g^{-1})]g=(1\mp R_g)(K_\pm),
\end{equation}
where we defined a $g$-dependent linear operator $R_g=\Ad_g^{-1}\circ R\circ \Ad_g$ with $\Ad_g$ the adjoint action by $g$, so that $R_g(x)=g^{-1}(R(gxg^{-1}))g$.
Now, using the fact that $k_\pm$ is of Maurer-Cartan form we can write
\begin{equation}\label{eq:MC-K}
    \begin{aligned}
        0&=\partial_+ k_--\partial_-k_++[k_+,k_-]\\
        &=\partial_+ K_--\partial_-K_++[K_+,K_-]\\
        &-R_g(\partial_+K_-+\partial_-K_+)\\
        &+([R_g (K_+),R_g(K_-)]-R_g[R_g(K_+),K_-]-R_g[K_+,R_g(K_-)]).
    \end{aligned}
\end{equation}
The derivation of this equation is left as the exercise~\ref{ex:Maurer} of section~\ref{sec:ex-int}. In the second line we recognise the action of $R_g$ on the equations of motion, therefore this expression does not vanish in general, but it is zero \emph{on shell}. The third line, on the other hand,  is identically zero, and one way to see that is to define \begin{equation}
    r=\beta^{ij}T_i\otimes T_j\in Lie(G)\otimes Lie(G).
\end{equation}
In fact, one can prove that the equation
\begin{equation}\label{eq:CYBE-R}
    [R(T_a),R(T_b)]-R([R(T_a),T_b]+[T_a,R(T_b)])=0,\qquad \forall a,b,
\end{equation}
is equivalent to \begin{equation}\label{eq:CYBE-r}
    [r_{12},r_{13}]+[r_{12},r_{23}]+[r_{13},r_{23}]=0,
\end{equation}
where the subscripts in the second equation denote the spaces on the triple tensor product, so that
\begin{equation}
    r_{12}=\beta^{ij}\, T_i\otimes T_j\otimes T_k,\qquad
    r_{23}=\beta^{jk}\, T_i\otimes T_j\otimes T_k,\qquad
    r_{13}=\beta^{ik}\, T_i\otimes T_j\otimes T_k.
\end{equation}
As it should be clear when doing the explicit calculation (proposed later as the exercise~\ref{ex:CYBE} of section~\ref{sec:ex-int}), to prove the equivalence between the two equations~\eqref{eq:CYBE-R} and~\eqref{eq:CYBE-r} one needs to assume invertibility of the metric $\kappa_{ab}$ on the algebra, as well as that it is ad-invariant. The two equations~\eqref{eq:CYBE-R} and~\eqref{eq:CYBE-r} go under the name of ``classical Yang-Baxter equation'', and the latter is the more familiar form. 
It is easy to see that~\eqref{eq:CYBE-r} is identically zero for our choice of $R$ or $r$, because by assumption $[T_i,T_j]=0$. As anticipated, then, the third line in~\eqref{eq:MC-K} is identically zero. 

To summarise, $K_\pm$ does not satisfy the Maurer-Cartan identity, but it satisfies this equation if, at the same time, we assume the equations of motion. In other words, even for the \emph{deformed} model we can claim that the dynamics is identified by the equations
\begin{equation}
\partial_\alpha K^\alpha=0,\qquad \quad \epsilon^{\alpha\beta}(\partial_\alpha K_\beta+ K_\alpha K_\beta)=0, \quad \text{imposed simultaneously}.
\end{equation}
These are essentially the equations for the \emph{undeformed} PCM with the formal replacement $k_\alpha\to K_\alpha$.
For this pair of equations, we already know that we can write the Lax pair
\begin{equation}
    \mathscr L_\pm=\frac{K_\pm}{1\mp \xi}.
\end{equation}
The above discussion is a more explicit way to see the integrability of the deformed model, at least  in the PCM case. It also allows us to obtain an explicit construction of the corresponding Lax connection.

\subsubsection{An example: the $U(2)$ PCM and its $\beta$-deformation}
Let's look at an explicit example, and let's choose in particular $G=U(2)=SU(2)\times U(1)$. As for the basis generators $T_a$ with $a=1,\ldots,4$  we may choose
\begin{equation}
    T_a=\sigma_a,\ a=1,2,3,\qquad T_4=\mathbf 1,
\end{equation}
where $\sigma_a$ are the Pauli matrices and $\mathbf 1$ the $2\times 2$ identity matrix. With this choice the metric on the Lie algebra induced by the trace is $\kappa_{ab}=\Trh[T_aT_b]=-\delta_{ab}$. We can parameterise the group element as
\begin{equation}
    g=e^{i\theta \mathbf 1}\, e^{\frac{i}{2}(\varphi_1+\varphi_2)\sigma_1}\, e^{i(\xi-\frac{\pi}{4})}\, e^{\frac{i}{2}(\varphi_1-\varphi_2)\sigma_2}.
\end{equation}
As an explicit $2\times 2$ matrix it would read
\begin{equation}
    g=\sqrt{i}e^{i\theta}
    \left(
\begin{array}{cc}
  \cos
   \varphi_1 \sin \xi -i \cos \xi  \cos \varphi_2 & \sin \xi  \sin \varphi_1+i \cos \xi  \sin \varphi_2 \\
 i \sin \xi  \sin \varphi_1+\cos \xi  \sin \varphi_2 &  \cos \xi  \cos \varphi_2-i \cos \varphi_1 \sin \xi  \\
\end{array}
\right).
\end{equation}
One can check that the above matrix is unitary and has determinant equal to $e^{2i\theta}$. We can then compute $k=g^{-1}dg=k^aT_a$ finding
\begin{equation}
    \begin{aligned}
        k^1&=\frac{i}{2}  (2 d\xi  \sin (\varphi_1-\varphi_2)+\sin (2 \xi )
   (d\varphi_1+d\varphi_2) \cos (\varphi_1-\varphi_2)),\\
   k^2&=-i \left(\cos ^2(\xi ) d\varphi_2-\sin ^2(\xi ) d\varphi_1\right),\\
   k^3&=\frac{i}{2}  (2 d\xi  \cos (\varphi_1-\varphi_2)-\sin (2 \xi )   (d\varphi_1+d\varphi_2) \sin (\varphi_1-\varphi_2)),\\
   k^4&=i\, d\theta.
    \end{aligned}
\end{equation}
With these results we can compute the action of the PCM and find that the corresponding target-space metric is
\begin{equation}
    ds^2=dx^mdx^nG_{mn}=d\xi^2+d\varphi_1^2\sin^2\xi+d\varphi_2^2\cos^2\xi+d\theta^2.
\end{equation}
We recognise the metric of an $S^3$ (which is the group manifold of $SU(2)$) parameterised by $\xi,\varphi_1,\varphi_2$, together with a circle $S^1$ (because of the extra $U(1)$) parameterised by $\theta$.

\vspace{12pt}

We may now use the above results to construct a $\beta$-deformation of the PCM. We will consider just a single TsT. It will be a TsT on $\theta$ and $\varphi_+=(\varphi_1+\varphi_2)/2$. To implement it, we define the $R$-operator as
\begin{equation}
    R(T_1)=\lambda T_4,\qquad
    R(T_4)=-\lambda T_1,\qquad R(T_a)=0, \text{ for } a=2,3,
\end{equation}
where we have introduced an explicit deformation parameter $\lambda$. Notice that we are selecting $T_1$ and $T_4$ as the two commuting generators. Other choices would be related to this by an $SU(2)$ transformation.
With this definition, we can construct the linear operator $O\equiv 1+R$. We can think of it as a matrix $O_a{}^b$ if we write $O(T_a)=O_a{}^bT_b$. We find
\begin{equation}
    O_a{}^b=\left(
\begin{array}{cccc}
 1 & 0 & 0 & \lambda  \\
 0 & 1 & 0 & 0 \\
 0 & 0 & 1 & 0 \\
 -\lambda  & 0 & 0 & 1 \\
\end{array}
\right),
\end{equation}
whose inverse is
\begin{equation}
   (O^{-1})_a{}^b= \left(
\begin{array}{cccc}
 \frac{1}{\lambda ^2+1} & 0 & 0 & -\frac{\lambda }{\lambda ^2+1} \\
 0 & 1 & 0 & 0 \\
 0 & 0 & 1 & 0 \\
 \frac{\lambda }{\lambda ^2+1} & 0 & 0 & \frac{1}{\lambda ^2+1} \\
\end{array}
\right).
\end{equation}
With this result we can construct the action of the deformed model. After the deformation, the metric is modified and a $B$-field is generated. The explicit results are
\begin{equation}
    \begin{aligned}
        ds^2&=d\xi^2+(1+\lambda^2)^{-1}\Big(d\varphi_1^2\sin^2\xi\left(1+\lambda^2\cos^2\xi\right)\\
&        +d\varphi_2^2\cos^2\xi\left(1+\lambda^2\sin^2\xi\right)
        -2\lambda^2d\varphi_1d\varphi_2\sin^2\xi\cos^2\xi\Big)\\
       & +(1+\lambda^2)^{-1}d\theta^2,\\
        B&=\tfrac12 B_{mn} dx^m\wedge dx^n =\lambda(1+\lambda^2)^{-1}\left(\sin^2\xi d\theta\wedge d\varphi_1+\cos^2\xi d\theta\wedge\varphi_2\right).
    \end{aligned}
\end{equation}
Looking at the metric, we see that now the space is the product of a deformation of $S^3$ and an $S^1$ (with a rescaled radius). The two mix at the level of the $B$-field. 

\vspace{12pt}

\begin{figure}
    \centering
   \includegraphics[width=0.7\textwidth]{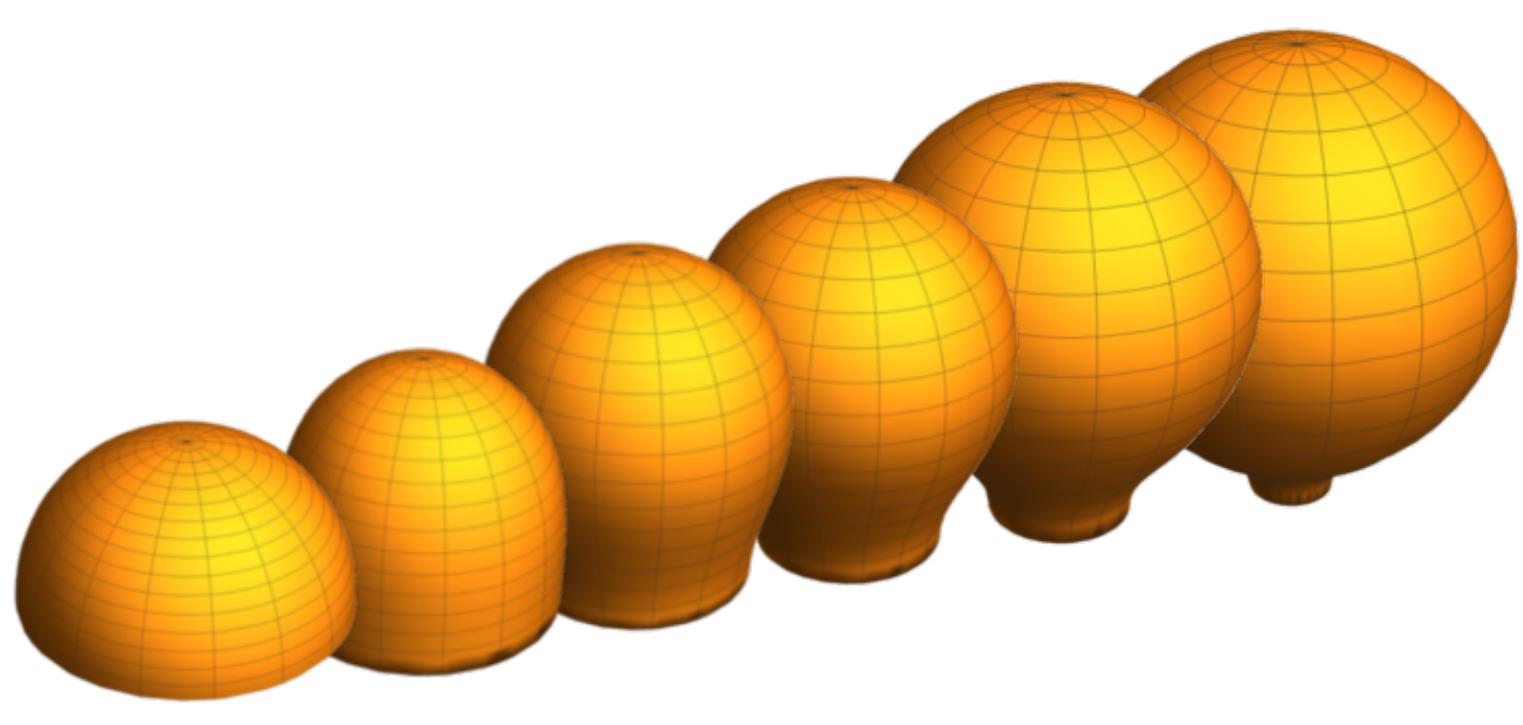}
    \caption{Visualisation of the deformation of the half-sphere according to the metric given in~\eqref{eq:def-S2}. The leftmost picture corresponds to $\lambda=0$. We then slightly increase the value of the deformation parameter until we show the picture for $\lambda=10$.}
    \label{fig:def-S2}
\end{figure}

Let us now play a bit to visualise the deformation. In order to draw pictures,  we want to focus on a 2-dimensional sub-manifold. In particular, we set for example $\theta=\varphi_2=0$, and we only look at the reduced metric
\begin{equation}\label{eq:def-S2}
        d\bar s^2=d\xi^2+(1+\lambda^2)^{-1}\sin^2\xi\left(1+\lambda^2\cos^2\xi\right)d\varphi_1^2.
\end{equation}
In the limit $\lambda\to 0$ this reduces to 
\begin{equation}\label{eq:ds2-S2}
        d\bar s^2\to d\xi^2+\sin^2\xi d\varphi_1^2,
\end{equation}
which is the metric of a 2-sphere. Actually, taking into account how this submanifold is embedded in the original one, we want to take $\xi\in [0,\pi/2[$ and $\varphi_2\in [0,2\pi[$, so that we actually have a half-sphere.
In the undeformed case, we can embed a 2-dimensional sphere in a 3-dimensional Euclidean space, because we can identify the sphere as the locus $X_1^2+X_2^2+X_3^2=1$ (if the radius of the sphere is 1). In fact, if we parameterise this constraint as 
\begin{equation}
    X_1=\cos\xi\cos \varphi_2,\qquad 
    X_2=\cos\xi\sin \varphi_2,\qquad 
    X_3=\sin\xi,
\end{equation}
then the flat metric $d\bar s^2=dX_i^2$ reproduces equation~\eqref{eq:ds2-S2}. Our task, therefore, is to find a modification of the embedding coordinates, now depending on the deformation parameter $\lambda$, such that  $dX_i^2$ reproduces the \emph{deformed} metric~\eqref{eq:def-S2}. We will not write the explicit expression here, because it is a bit complicated. But we can use the result to plot some pictures of the deformed space with Mathematica. In Figure~\ref{fig:def-S2} we have a collection of pictures of the deformed space at different values of the deformation parameter $\lambda$. On the left we start with $\lambda=0$, then we slowly turn on $\lambda$ until we reach the value of $\lambda=10$. We see that the effect is indeed that of a deformation of the (half) sphere in the standard sense: the manifold starts to elongate in the vertical direction, until it becomes a round sphere in the strict $\lambda\to\infty$ limit. This can be easily checked also at the level of the metric~\eqref{eq:def-S2}.

Following the discussion presented in this section, one may further construct the explicit Lax connection for this deformed PCM. Moreover, one can also explicitly reformulate it as an undeformed twisted model by writing down the twisted boundary conditions for the fields.

\subsection{Exercises}\label{sec:ex-int}

\begin{enumerate}
    \item \label{ex:Noether} \textbf{Noether currents under the on-shell map ---} The map between the original undeformed model and the new $\beta$-deformed model implies that $J_0=\tilde J_0$. Prove that
\begin{equation}\label{eq:beta-map}
    p=\tilde p,\qquad\qquad \phi'=\tilde\phi'+\beta\tilde p.
\end{equation}
implies also the equality $J_1=\tilde J_1$.
Try to follow these steps:
\begin{enumerate}
        \item Compute the conjugate momenta $p$ in your favourite way (by direct calculation or by computing $J_0$). You should find
    \begin{equation}\label{eq:p-phi}
        p=\mathsf G\dot\phi-\mathsf B\phi'+\frac{1}{\sqrt 2}(\mathsf F^L)^t\partial_+\hat x-\frac{1}{\sqrt 2}\mathsf F^R\partial_-\hat x,
    \end{equation}
    where $\mathsf G=(\mathsf E+\mathsf E^t)/2$, $\mathsf B=(\mathsf E-\mathsf E^t)/2$.
    Obviously, an analogous formula for $\tilde p$ holds if we add tildes everywhere.
    \item Using~\eqref{eq:p-phi}, solve the equations~\eqref{eq:beta-map} in terms of $\dot \phi$ and $\phi'$. You should find
    \begin{equation}
    \begin{aligned}
        \dot\phi&=\mathsf G^{-1}(1+\mathsf B\beta)\tilde {\mathsf G}\dot{\tilde\phi}+\mathsf G^{-1}(\mathsf B-(1+\mathsf B\beta)\tilde{\mathsf  B})\tilde\phi'\\
        &+\frac{1}{\sqrt 2}\mathsf G^{-1}[(1+\mathsf B\beta)(\tilde {\mathsf F^L})^t-(\mathsf F^L)^t]\partial_+\hat x\\
        &-\frac{1}{\sqrt 2}\mathsf G^{-1}[(1+\mathsf B\beta)\tilde {\mathsf F^R}-\mathsf F^R]\partial_-\hat x,\\
        \phi'&=\beta\tilde{\mathsf G}\dot{\tilde \phi}+(1-\beta\tilde{\mathsf B})\tilde\phi'
        +\frac{1}{\sqrt 2}\beta(\tilde{\mathsf F}^L)^t\partial_+\hat x
        -\frac{1}{\sqrt 2}\beta\tilde{\mathsf F}^R\partial_-\hat x
    \end{aligned}
    \end{equation}
    \item Now compute $J_1$ using your favourite definition of the Noether currents. You should find
    \begin{equation}
        J_1=-\mathsf B\dot\phi+\mathsf G\phi'+\frac{1}{\sqrt 2}(\mathsf F^L)^t\partial_+\hat x+\frac{1}{\sqrt 2}\mathsf F^R\partial_-\hat x,
    \end{equation}
    \item Substitute the above expressions for $\dot\phi$ and $\phi'$ into $J_1$. Collect the expressions proportional to $\dot{\tilde \phi}$, $\tilde\phi'$, $\partial_\pm\hat x$ and check that they match with the above formula after adding tildes everywhere. If you can do that, then it means that you have proved $J_1=\tilde J_1$. 
    
    \textbf{Hint:} instead of using the formulas for $\tilde{\mathsf G}, \tilde{\mathsf B}$, it is actually simpler to derive relations from the transformation of the generalised metric $\tilde{\mathcal  H}=\mathcal U\mathcal H\mathcal U^t$. For example, you will find these useful relations
    \begin{equation}
    \begin{aligned}
      \tilde{\mathsf G}-\tilde{\mathsf B}\tilde{\mathsf G}^{-1}\tilde{\mathsf B}&=
        \mathsf G-\mathsf B\mathsf G^{-1}\mathsf B,\\
        \tilde{\mathsf B}\tilde{\mathsf G}^{-1}&=\mathsf B\mathsf G^{-1}-(\mathsf G-\mathsf B\mathsf G^{-1}\mathsf B)\beta.
    \end{aligned}
    \end{equation}
\end{enumerate}

\item \label{ex:Maurer} \textbf{Maurer-Cartan identity and the deformed PCM ---} Derive equation~\eqref{eq:MC-K}. To obtain the result, first prove that $d\Ad_g=\Ad_g \ad_{k}$ and $dR_g=[R_g,\ad_k]$. Here $\ad_kx=[k,x]$.

\item \label{ex:CYBE} \textbf{Classical Yang-Baxter equation ---} Prove the equivalence of equations~\eqref{eq:CYBE-R} and ~\eqref{eq:CYBE-r}.

\end{enumerate}

\section{Beyond these notes}\label{sec:concl}
To conclude these notes, let us summarise other aspects that, albeit they may not have been discussed here, they have nevertheless motivated  the choice of the content of the notes. This brief conclusion is meant to give a perspective on how the  discussion in the previous sections is related or is embedded into other research,  in particular in relation to more recent developments. As a disclaimer, the following presentation is strongly influenced by the taste of the author. It is of course not meant to be a complete review of the corresponding fields, nor a complete list of the possible connections between the current-current deformations of these notes and other topics. It may be seen simply as a first step to delve into the literature, with a biased preference for certain directions.

\vspace{12pt}

\textbf{T-duality and ``double formulations'' --- }
We briefly touched upon the subject of T-duality because of its relation to the deformations constructed here. There is a vast literature on T-duality, see for example the standard review~\cite{Giveon:1994fu}. More recently, T-duality transformations have been used as a guiding principle to construct a ``duality covariant'' language that was called ``Double Field Theory'', see for example~\cite{Zwiebach:2011rg,Aldazabal:2013sca,Berman:2013eva,Hohm:2013bwa} for some reviews. Double Field Theory builds on previous observations that T-duality transformations are part of a larger $O(d,d)$ group, see for example~\cite{Giveon:1990era,Meissner:1991zj,Giveon:1991jj}, and on the double $\sigma$-model of Tseytlin~\cite{Tseytlin:1990nb,Tseytlin:1990va}.

\vspace{12pt}

\textbf{Current-current deformations of WZW models --- }
Current-current deformations of Wess-Zumino-Witten (WZW) models have been investigated in various works. We mention in particular~\cite{Forste:2003km} and~\cite{Forste:2003yh}. The discussion of D-branes in a deformation of $SU(2)$ WZW was given in~\cite{Forste:2001gn}, see also~\cite{Forste:2002uv}. A deformation in a non-compact WZW along a null direction was discussed in~\cite{Forste:1994wp}.

\vspace{12pt}

\textbf{The $O(d,d)$ group and string theory --- }
The deformations considered here are related to Narain's $O(d,d)$ transformations relating different toroidal compactifications~\cite{Narain:1985jj,Narain:1986am}.
Original arguments that $O(d,d)$ transformations  give rise to solutions of the string theory effective equations were later generalised to all orders in the $\alpha'$-expansion, see for example~\cite{Sen:1991zi,Kiritsis:1993ju}.
The compatibility of the $O(d,d)$ transformations with worldsheet supersymmetry, and consequently the fact that they can give rise to superconformal deformations, was presented in~\cite{Hassan:1994mq}.

\vspace{12pt}

\textbf{Recent developments on integrable deformations --- }
Recently, the topic of current-current deformations has received a renewed interest because of new developments.
In fact, an important direction of research has been that of integrable deformations of $\sigma$-models. We refer to~\cite{Hoare:2021dix} for a recent review on this topic. Various families of integrable deformations have been constructed. Let us start by focusing on two kinds of integrable deformations: the so-called ``$\lambda$-deformation'' and the ``Yang-Baxter deformation''. 

\vspace{12pt}

\textbf{The $\lambda$-deformation --- }
The  $\lambda$-deformation, in its original construction of~\cite{Sfetsos:2013wia}, may be understood as a perturbation of a WZW model by an ``isotropic'' current bilinear, meaning that the Lie-algebra indices of the conserved currents are paired with the Kronecker delta. This deformation does not satisfy the marginality condition of Chaudhuri and Schwartz and is not exactly marginal. It is in fact a relevant deformation. The construction of~\cite{Sfetsos:2013wia} is also referred to as the $\lambda$-deformation \emph{of the PCM}. It is in fact naturally understood as a transformation obtained by starting from the PCM action on the Lie group $G$, that results in a model that interpolates between the WZW model on $G$ and the non-abelian T-dual of the PCM. Although in general it breaks the conformal invariance, the $\lambda$-deformation is an integrable deformation. It corresponds to a $q$-deformation with $q$ being a root of unity.  Later, the construction was extended to coset and supercoset models~\cite{Hollowood:2014rla,Hollowood:2014qma}. Various works discussed the embedding of the $\lambda$-deformation in string theory and supergravity~\cite{Sfetsos:2014cea,Lunin:2014tsa,Demulder:2015lva,Chervonyi:2016bfl,Chervonyi:2016ajp,Borsato:2016zcf,Borsato:2016ose}.
The RG flows of  $\lambda$-deformed models were discussed also in~\cite{Itsios:2014lca,Appadu:2015nfa,Sagkrioti:2018rwg,Georgiou:2018vbb,Delduc:2020vxy}.
Generalisations of the construction of the $\lambda$-deformation were worked out for example in~\cite{Sfetsos:2015nya,Georgiou:2016urf,Georgiou:2018gpe,Driezen:2019ykp}.

\vspace{12pt}

\textbf{Yang-Baxter deformations --- }
The family of Yang-Baxter deformations may be  divided into two distinct constructions: the \emph{inhomogeneous} and the \emph{homogeneous} Yang-Baxter deformation. The name Yang-Baxter is justified by the central ingredient that enters the construction, namely an $r$-matrix that satisfies the classical Yang-Baxter equation. This equation is ``modified'' in the inhomogeneous case and not modified in the homogeneous one. In both cases, the deformation can be understood as being generated at the infinitesimal level by a current-current bilinear, where the pairing of the Lie-algebra indices is now controlled by the $r$-matrix. 

The \textbf{inhomogeneous Yang-Baxter deformation} also corresponds to a $q$-deformation of the underlying symmetry algebra, where now $q$ is real. It was originally constructed in~\cite{Klimcik:2002zj,Klimcik:2008eq} for the case of the PCM, and later generalised to coset and supercoset models~\cite{Delduc:2013fga,Delduc:2013qra,Delduc:2014kha}. Further generalisations were also constructed, like the deformation of the PCM with a WZ term~\cite{Hoare:2020mpv}, and multi-parameter deformations, see for example \cite{Klimcik:2008eq,Klimcik:2014bta,Delduc:2015xdm,Delduc:2017fib,Delduc:2018xug,Seibold:2019dvf}. The study of the inhomogeneous Yang-Baxter deformation in the context of superstring backgrounds~\cite{Arutyunov:2015qva} was pivotal for the construction of the ``generalised supergravity equations''~\cite{Arutyunov:2015mqj}, and for the understanding of the relation between these equations and the constraints imposed by kappa-symmetry in the Green-Schwartz formulation~\cite{Wulff:2016tju}. Later it was shown that the inhomogeneous Yang-Baxter deformation can give rise to backgrounds that are solutions to the standard equations of type IIB supergravity~\cite{Hoare:2018ngg}.
Various aspects of this kind of deformation have been investigated, including the application of resurgence methods~\cite{Demulder:2016mja}.

The \textbf{homogeneous Yang-Baxter deformation} is the one that is more closely related to these notes. It started to receive a lot of attention after the works~\cite{Kawaguchi:2014qwa,Matsumoto:2015jja}, and an important influential work was also~\cite{vanTongeren:2015soa}. In section~\ref{sec:beta-PCM} we have already seen how the $\beta$-deformation of the PCM can be recast into the language of Yang-Baxter deformations, with an $r$-matrix satisfying the classical Yang-Baxter equation. In particular, we refer to the papers~\cite{Matsumoto:2014nra,Osten:2016dvf} where it was first shown that ``abelian'' Yang-Baxter deformations are equivalent to sequences of TsT transformations. In the case of \emph{compact} Lie groups, this is the end of the story: the classical Yang-Baxter equation only admits ``abelian'' solutions as the ones of section~\ref{sec:beta-PCM}. However, if the Lie group of isometries is non-compact, then the classical Yang-Baxter equation may have also ``non-abelian solutions''. In other words, the $r$-matrix is constructed out of generators of the Lie algebra that do not necessarily commute, but rather close into a non-abelian subalgebra of the Lie algebra of isometries. As a generalisation of some of the aspects that we have seen in the case of $\beta$-deformations, homogeneous Yang-Baxter deformations were shown to be related to \emph{non-abelian} T-duality~\cite{Hoare:2016wsk,Borsato:2016pas,Borsato:2018idb}. Homogeneous Yang-Baxter deformations were also shown to have a meaningful interpretation in string theory given that (when an extra condition for $r$ is satisfied) they map solutions of the supergravity equations of motion to other solutions~\cite{Borsato:2016ose,vanTongeren:2019dlq}.
As already argued for the smaller class of $\beta$-deformations, also homogeneous Yang-Baxter deformations may be seen as arising from canonical transformations on the worldsheet, see for example~\cite{Osten:2019ayq,Borsato:2021gma}.

Certain Yang-Baxter deformations are  current-current deformations that might be integrably marginal despite eluding the marginality condition of Chaudhuri and Schwartz. The reason for this possibility lies in the fact that the underlying Lie algebra is non-compact, and therefore some of the assumptions that we made in Section~\ref{sec:ChSch} do not hold. The relation between non-abelian Yang-Baxter deformations and the marginality condition was explored in~\cite{Araujo:2018rho} and in~\cite{Borsato:2018spz}, but more work is needed to fully understand this issue.

\vspace{12pt}

\textbf{AdS/CFT aspects of the deformations and their integrability --- }
Certain TsT deformations have been given an AdS/CFT interpretation, see in particular~\cite{Lunin:2005jy,Maldacena:1999mh,Imeroni:2008cr}. There are proposals for AdS/CFT interpretations of more general homogeneous Yang-Baxter deformations, see in particular~\cite{vanTongeren:2015uha,vanTongeren:2016eeb,Araujo:2017jkb,Meier:2023kzt,Meier:2023lku}. At present, there is no proposal for a gauge theory realisation of the duals of the inhomogeneous Yang-Baxter or $\lambda$-deformations of the $AdS_5\times S^5$ superstring.

The formulation of the integrability methods for certain TsT deformations of $AdS_5\times S^5$ is very well understood, see for example the review~\cite{vanTongeren:2013gva}.
Nevertheless, other TsT deformations~\cite{Guica:2017mtd} and more general homogeneous Yang-Baxter deformations remain elusive.  Recently,   general arguments were provided to justify the already known construction of  worldsheet S-matrices of TsT-deformed models~\cite{Dubovsky:2023lza}, see also~\cite{new-lcg}. These results may be seen also as a further confirmation of the previous successful constructions of the integrability methods for TsT deformations in AdS/CFT. More work, however, still needs to be done to understand the cases in which the deformation is not compatible with the standard ``gauge fixing'' procedure, see~\cite{new-lcg} for the first step in that direction.

The integrability description of the spectrum of the $q$-deformation of $AdS_5\times S^5$ with $q$ real was also carried out, see~\cite{Arutyunov:2014wdg,Klabbers:2017vtw} and references therein. The same level of result was not obtained for the $\lambda$-deformation.

\vspace{12pt}

\textbf{Double aspects of the deformed model --- } Given their close relations to various notions of T-duality (including non-abelian T-duality), it is not too surprising that the deformations admit a natural formulation in terms of ``double languages''. From the point of view of the $\sigma$-model, this double structure gets exposed once one goes to the Hamiltonian formalism, see for example~\cite{Vicedo:2015pna}.
The relation to Poisson-Lie duality and the E-models~\cite{Klimcik:1995dy,Klimcik:1996nq,Klimcik:1996np} was explored in various works like~\cite{Vicedo:2015pna,Hoare:2017ukq,Demulder:2018lmj}. 
This interpretation of the deformed models actually offers a resolution to an obvious puzzle, that appears as soon as one considers deformations that are more general than the $\beta$-deformations considered in these notes: the deformations are defined by current bilinears related to symmetries that are, however, broken by the deformations themselves, so how is it possible to ``iterate'' the infinitesimal deformation and construct the finite one? Relaxing our notion of symmetries seems to be the key to answer this question: although the \emph{isometries} are broken by the deformations, they are traded for hidden symmetries that are  called \emph{Poisson-Lie symmetries}.
In~\cite{Delduc:2019whp,Lacroix:2020flf} E-models were also recast in the language of the 4d Chern-Simons theory of~\cite{Costello:2019tri,Vicedo:2019dej}, see also the lecture notes~\cite{Lacroix:2021iit}.
A closely related program was that of reformulating Poisson-Lie duality and the deformations in the language of Double Field Theory and generalised geometry~\cite{Hassler:2017yza,Sakamoto:2017cpu,Sakamoto:2018krs,Hassler:2019wvn,Demulder:2019vvh,Sakatani:2019jgu,Borsato:2021vfy,Butter:2022iza}.
The double formulation has been particularly useful to understand certain aspects like the
$\alpha'$-corrections, see for example~\cite{Borsato:2020bqo,Borsato:2020wwk,Hassler:2020tvz,Codina:2020yma}.

\vspace{12pt}

\textbf{The $T\bar T$ and $JT$ deformation --- }
Finally, we cannot conclude without mentioning a famous current-current deformation that has recently attracted a lot of attention in the literature, the so called ``$T\bar T$-deformation''~\cite{Smirnov:2016lqw,Cavaglia:2016oda}. This infinitesimal deformation is, in fact, controlled by the components of the worldsheet energy-momentum tensor, which encodes the conserved currents for the invariance under translations on the worldsheet. The difference with the deformations considered in these notes is that here we were considering deformations generated by the currents of internal global symmetries, rather than spacetime symmetries. Nevertheless, there are several points in common between the two constructions. A complete discussion on $T\bar T$ cannot fit here, and we refer for example to the lecture notes~\cite{Jiang:2019epa}.
A closely related construction is that of the $JT$ deformation of~\cite{Guica:2017lia}.

\section*{Acknowledgements}
I thank Alejandra Castro and Linus Wulff for useful discussions, and Linus Wulff for comments on the draft.
I am supported by the grant RYC2021-032371-I (funded by MCIN/AEI/10.13039/501100011033 and by the European Union ``NextGenerationEU''/PRTR) and by the grant 2023-PG083 with reference code ED431F 2023/19 funded by Xunta de Galicia. I also acknowledge AEI-Spain (under project PID2020-114157GB-I00 and Unidad de Excelencia Mar\'\i a de Maetzu MDM-2016-0692),  Xunta de Galicia (Centro singular de investigaci\'on de Galicia accreditation 2019-2022, and project ED431C-2021/14), and the European Union FEDER. 

\vspace{12pt}

These notes were prepared for the Young Researchers Integrability School and Workshop (YRISW) held in Durham from 17 to 21 July 2023, see \url{https://indico.cern.ch/e/YRISW23} for the website. I wish to thank the organisers of the school (Patrick Dorey, Ben Hoare, Ana Retore, Fiona Seibold and Alessandro Sfondrini) for the opportunity to give the lectures, and for all the help and support before and during the school. I also thank the students for their engagement during the lectures and for the interesting questions. The YRISW 2023 was realised with the financial support of Durham University, the University of Padova, Imperial College London, GATIS+, UKRI and the European Union.

\input{Current_current_deformation_notes.bbl}


\end{document}